\documentclass[twocolumn,english,showpacs,amsmath,amssymb,floatfix]{revtex4-1}
\usepackage{amsmath}
\usepackage{amssymb}
\usepackage{bbold}
\usepackage{graphicx}
\usepackage[usenames,dvipsnames]{color}
\usepackage{babel}
\usepackage{times}
\usepackage{array}
\usepackage{hyperref}

\allowdisplaybreaks[3]

\newcommand{\citationtitle}[1]{{\color{Gray}\small #1}}

\newcommand{\ba}{{\bf a}}
\newcommand{\bk}{{\bf k}}
\newcommand{\bq}{{\bf q}}
\newcommand{\br}{{\bf r}}
\newcommand{\kk}{{\bf k}_0}
\newcommand{\kt}{\tilde{\bf k}_0}
\newcommand{\id}{\ensuremath{\mathbb{1}}}

\newcommand{\Z}{Z$_{\rm 2}$}

\newcommand{\NaIrO}{Na$_{\text 2}$IrO$_{\text 3}$}
\newcommand{\aLiIrO}{$\alpha$-Li$_{\text 2}$IrO$_{\text 3}$}
\newcommand{\bLiIrO}{$\beta$-Li$_{\text 2}$IrO$_{\text 3}$}
\newcommand{\cLiIrO}{$\gamma$-Li$_{\text 2}$IrO$_{\text 3}$}
\newcommand{\RuCl}{RuCl$_{\text 3}$}

\newcommand{\ie}{{i.e.~}}
\newcommand{\eg}{{e.g.~}}

\newcolumntype{L}[1]{>{\raggedright\let\newline\\\arraybackslash\hspace{0pt}}m{#1}}
\newcolumntype{C}[1]{>{\centering\let\newline\\\arraybackslash\hspace{0pt}}m{#1}}

\graphicspath{{Figures/}}

\begin{document}

\title{Classification of gapless \Z\ spin liquids in three-dimensional Kitaev models}

\author{Kevin O'Brien}
\author{Maria Hermanns}
\author{Simon Trebst}
\affiliation{Institute for Theoretical Physics, University of Cologne, 50937 Cologne, Germany}

\date{\today}

\begin{abstract}
  Frustrated quantum magnets can harbor unconventional spin-liquid ground states in which the elementary magnetic moments fractionalize into new emergent degrees of freedom.
  While the fractionalization of quantum numbers is one of the recurring themes in modern condensed matter physics,
  it often remains a challenge to devise a controlled analytical framework  tracking this phenomenon.
  A notable exception is the exactly solvable Kitaev model, in which spin degrees of freedom fractionalize into Majorana fermions and a \Z\ gauge field. 
  Here we discuss the physics of fractionalization in {\it three-dimensional} Kitaev models and demonstrate that the itinerant Majorana fermions generically form a (semi)metal which, depending on the underlying lattice structure, exhibits Majorana Fermi surfaces, nodal lines, or topologically protected Weyl nodes.
  We show that the nature of  these Majorana metals can be deduced from an elementary symmetry analysis of the projective time-reversal and inversion symmetries for a given lattice.
  This allows us to comprehensively classify the gapless spin liquids of Kitaev models for the most elementary tricoordinated lattices in three dimensions. 
  We further expand this classification by addressing the effects of time-reversal symmetry breaking and additional interactions.  
\end{abstract}

\maketitle



\section{Introduction}
The low-temperature collective physics of interacting quantum many-body systems often calls for a novel description in terms of emergent degrees of freedom that are not only distinct from those of the original constituents of the system, but describe certain ``fractions'' thereof. 
Familiar examples include the spin-charge separation in one-dimensional metals \cite{SpinChargeSeparation}, the electron fractionalization in fractional quantum Hall states of two-dimensional electron gases \cite{FQH}, as well as the emergence of monopoles in spin ice \cite{MonopolesSpinIce} or chiral magnets \cite{MonopolesChiralMagnets}.
Quantum spin liquids in frustrated quantum magnets \cite{SpinLiquids} provide another important venue for such quantum number fractionalization.
For these spin liquids the theoretical formulation of this phenomenon is often closely linked to a lattice gauge theory description of the quantum magnet; the spin degrees of freedom typically decompose into spinons coupled to an emergent $U(1)$ or \Z\ gauge field whose elementary excitations remain deconfined \cite{ReadSachdev,SenthilFisher,XiaoGang}.
One of the paradigmatic examples of a model harboring a \Z\ spin liquid ground state is Kitaev's exactly solvable honeycomb model \cite{kitaev2006}.
It describes a spin-$\frac{1}{2}$ quantum magnet subject to strong exchange frustration arising from bond-directional interactions of the form
\begin{equation}
  H_{\rm Kitaev} = -\sum_{\gamma \rm-bonds}  J_{\gamma} \,\, \sigma_i^\gamma \sigma_j^\gamma\,,  
  \label{eq:spinH}
\end{equation}
where $\gamma = x,y,z$ labels the three different bond directions of the honeycomb lattice.
The low-energy physics of this spin model can be captured in terms of Majorana degrees of freedom and a \Z\ gauge field.
Crucially, the gauge field remains static for the pure Kitaev model \eqref{eq:spinH}, and identifying the ground state configuration of the gauge field reduces to an essentially classical problem.
Typically this yields a unique ground state with a finite gap for the elementary vison excitations  of the \Z\ gauge field.
Fixing the gauge structure then allows to recast the original spin model as a free Majorana fermion model and thus paves the way to a full analytical solution.
The phase diagram of the Kitaev model generically exhibits two types of spin-liquid phases.
Around the limits where one of the three couplings dominates over the other two one finds a gapped spin liquid which, for the two-dimensional honeycomb model, is known to exhibit Abelian topological order \cite{kitaev2006}.
The second phase, which is found for roughly isotropic couplings (\ie, $J_x \sim J_y \sim J_z$) is gapless and can generically be understood as a metallic state of the itinerant Majorana fermions.
For the two-dimensional honeycomb model the itinerant Majorana fermions form a graphene-like band structure with two Dirac cones \cite{kitaev2006}.

\begin{table*}
  \begin{tabular}{L{13mm} | C{70mm} C{15mm} C{20mm} C{20mm} c c}
    lattice 			& alternative  			& sites in 	& sublattice	& inversion  	& \multicolumn{2}{c}{space group} \\
    	   			    & names 				& unit cell   &   symmetry  	&   symmetry  	& symbol & No. \\
    \hline \hline
    (10,3)a 	&    hyperoctagon \cite{hermanns14}, Laves graph \cite{Laves}, K$_4$ crystal \cite{K4}  		& 4 		& $\kk \neq 0$ 		& chiral 		& I$4_132$ 		& 214 \\
    (10,3)b 	&  hyperhoneycomb \cite{beta} 	& 4	& \checkmark	& \checkmark 		& Fddd 			& 70 \\
    (10,3)c 	& ---					& 6		& \checkmark 	& chiral		& P$3_112$		& 151 \\
    \hline
    (9,3)a 	& ---					& 12		& --- 			& \checkmark		& R$\bar{3}$m		& 166 \\
    (9,3)b 	& ---					& 24		& --- 			& \checkmark		&P4$_2$/nmc		&137	 \\
    \hline
    (8,3)a 	& ---					& 6		& $\kk  \neq 0$		& chiral		& P$6_222$		& 180 \\
    (8,3)b 	& ---					& 6		& $\kk  \neq 0$ 		& \checkmark		& R$\bar{3}$m		& 166 \\
    (8,3)c 	& ---					& 8		& \checkmark		& \checkmark		& P$6_3$ / mmc 	& 194 \\
    (8,3)n 	& ---					& 16		& \checkmark		& $\tilde{\mathbf{k}}_0  \neq 0$ 		& I4 / mmm		& 139 \\
    \hline\hline
    (6,3)  	& honeycomb			& 2		& \checkmark & \checkmark	& & 
  \end{tabular}
  \caption{Overview of elementary tricoordinated lattices in three spatial dimensions. Following the classification of A. F. Wells 	
		\cite{Wells77}, we only consider lattices of fixed polygonality $p$ (\ie, a fixed length of all elementary closed 
 	loops) and vertex coordination $c=3$ using the Schl\"afli symbol $(p,c)$ followed by a letter.
 	For each lattice, we list alternative names used in the literature along with some basic lattice information including 
 	the number of sites $Z$ in the unit cell, whether the lattice exhibits a (non-trivial) sublattice symmetry (see also the 
 	discussion in the main text), whether the lattice exhibits (non-trivial) inversion symmetry and provide the space-group information.
 	More technical details, such as precise unit cell definitions including Wyckoff positions, can be found in an extensive
 	appendix.
 	}
\label{tab:lattice_overview}
\end{table*}
 
In this paper, we comprehensively classify the nature of the gapless spin liquids and their underlying Majorana metals for {\it three-dimensional} Kitaev models.
Our motivation has been rooted in the mounting experimental evidence that spin-orbit entangled Mott insulators can provide solid-state realizations of the Kitaev model following the theoretical guidance by Khaliullin and coworkers \cite{Khaliullin}.
This materials-oriented search \cite{honeycomb-iridates,RuCl3} has produced various candidate 4d and 5d compounds, most notably \NaIrO, \aLiIrO\ and \RuCl, which realize hexagonal arrangements of local, spin-orbit entangled $j=1/2$ moments that are indeed subject to strong bond-directional exchanges as indicated by recent experiments \cite{BondDirectionalExchange}.
A byproduct of this experimental search has been the discovery \cite{beta,gamma} of the polymorphs \bLiIrO\ and \cLiIrO, which realize three-dimensional arrangements of the spin-orbit entangled moments which retain the {\it tricoordination} familiar from the hexagonal lattice.
This has sparked a surge of interest in three-dimensional variants of the Kitaev model which, hitherto, had evaded the attention of the broader community \cite{FootnoteEarlier3D}.
It was quickly recognized that the analytical tractability of the two-dimensional Kitaev model largely carries over to the three-dimensional variants, and it has recently been demonstrated that such three-dimensional Kitaev models harbor a rich variety of gapless \Z\ spin liquids in which the emergent Majorana metals form nodal structures which include Majorana Fermi surfaces \cite{hermanns14}, nodal lines \cite{Mandal09} as well as topologically protected Weyl nodes \cite{wsl2014}.
The purpose of this paper is to go beyond these initial examples and to impart a more systematic classification of gapless Kitaev spin liquids in three spatial dimensions.
In particular, we comprehensively discuss how the nature of the emergent Majorana metal depends on the underlying lattice geometry.
We do so by considering Kitaev models for the most elementary three-dimensional, tricoordinated lattices, \ie, lattices that have elementary loops of only one fixed length \cite{FootnoteElementaryLoop}.
For instance, the well-known honeycomb lattice is the only tricoordinated lattice with elementary loops of length 6.
However, there are multiple lattice structures with elementary loops of lengths 7, 8, 9 or 10 (and possibly higher), which are all three-dimensional.
In fact, such three-dimensional, tricoordinated structures have been comprehensively classified in the work of Wells in the 1970's \cite{Wells77}.
Here, we focus on those lattice structures that exhibit equidistant bonds and approximately 120$^\circ$ bond angles at every vertex \cite{equidistantbond}.
An overview of the so-identified family of three dimensional, tricoordinated lattice structures and their basic properties is provided in Table \ref{tab:lattice_overview}. 
A convenient way to systematically label the individual lattices is to use the so-called Schl\"afli symbol $(p,c)$ followed by a letter, where $p$ is the fixed polygonality (or elementary loop length) of the lattice, $c=3$ refers to the tricoordination of the vertices, and the additional letter simply enumerates the lattices for a given Schl\"afli symbol.
It should be noted that some of these lattices are well known in the literature under alternative names.
These include the (10,3)a lattice, which has long been known as the Laves graph \cite{Laves} in the crystallographic literature or as $K_4$ crystal \cite{K4} in the mathematical literature.
It has also been renamed hyperoctagon lattice \cite{hermanns14} by some of the authors of this manuscript in an earlier study.
Similarly, the (10,3)b lattice has recently gained some attention under the name hyperhoneycomb lattice \cite{beta} after it had been discovered as the iridium-sublattice in the iridate \bLiIrO. 

It is precisely this family of tricoordinated lattice structures that serves as principal input in our quest to comprehensively discuss three-dimensional Kitaev models in the following.
We show that these Kitaev models harbor a plethora of gapless spin liquids that can be cast as  different incarnations of Majorana metals  
whose precise nature can be systematically understood from a basic symmetry analysis.

\subsection*{Overview of results}
 
\begin{table}
  \begin{tabular}{L{13mm} | C{3cm} | C{3cm}}
    Lattice &  Majorana metal  &  TRS breaking   \\
    \hline \hline
    (10,3)a  	&   Fermi surface 	&  Fermi surface 	 \\
    (10,3)b  	&  Nodal line		& Weyl nodes 		 \\
    (10,3)c  	&  Nodal line		& Fermi surface  \\
    \hline
    (9,3)a$^*$ 	& Weyl nodes		& Weyl nodes	 \\
    \hline
    (8,3)a 	& Fermi surface 	& Fermi surface	  \\
    (8,3)b 	& Weyl nodes 		& Weyl nodes		  \\
    (8,3)c$^*$ 	& Nodal line		& Weyl nodes 		 \\
    (8,3)n 	& Gapped		
    &  Weyl nodes	 \\
    \hline
    (6,3)  	& Dirac cones		& Gapped		 
  \end{tabular}
  \caption{Overview of Majorana metals in three-dimensional Kitaev models. 
		Shown is a characterization of the nodal structure of the metallic states formed by the itinerant
 	Majorana fermions in the gapless spin-liquid phase of three-dimensional Kitaev models defined 
 	on tricoordinated lattices of Table \ref{tab:lattice_overview}. 	
 	Results for the pure Kitaev model \eqref{eq:spinH} are given in the second column. 
 	The third column provides information on how the nodal structure changes if the Kitaev model 
 	is augmented by an explicit time-reversal symmetry (TRS) breaking magnetic field term (\ie, a magnetic
 	field pointing along the 111-direction). 
 	The asterisk indicates that for these two lattices we are providing results for the lowest-energy flux sector that does not
 	 break any point group symmetries of the lattice.  
  }
  \label{tab:majorana_metals}
\end{table}

Before going into a detailed discussion of the Kitaev models for these individual lattices, we provide a brief overview of our main results.
For all but one lattice, \ie, (8,3)n, we find that there is an extended gapless spin-liquid phase around the point of isotropic coupling, \ie, $J_x \sim J_y \sim J_z$.
This gapless phase is best described as a Majorana metal (or semimetal), since it is the band structure of the itinerant Majorana fermions that exhibits gapless excitations, while the vison excitations of the static \Z\ gauge field remain gapped for all lattices \cite{FootnoteGappedVisons}. 
A summary of our results characterizing the various Majorana metals for different lattice geometries is provided in Table \ref{tab:majorana_metals}.
We do so by listing the nodal manifold of gapless excitations in the band structure of the itinerant Majorana fermions.
As can be seen from the table, the various three-dimensional lattice geometries realize multiple examples for the emergence of  Fermi surfaces, nodal lines, or Weyl nodes.
As we will lay out in detail in the remainder of the paper, an understanding of the systematics in this table is closely linked to a symmetry analysis of the Kitaev models for the respective lattice geometries (see in particular Sec.~\ref{sec:symmetries}).
For instance, the occurrence of Majorana Fermi surfaces for the two lattices (10,3)a and (8,3)a is closely linked to a non trivial sublattice symmetry for these lattices.
Similarly, the emergence of Weyl nodes can be understood from a close inspection of time-reversal and inversion symmetry.
For instance, for a lattice with an {\it odd} number of bonds in the elementary loop, such as the (9,3)a lattice, time-reversal symmetry has to be broken spontaneously for the emergent Majorana degrees of freedom, which in turn allows for the occurrence of Weyl nodes.
For the lattice (8,3)b it is a more intricate interplay of time-reversal symmetry, inversion symmetry, and non-trivial sublattice symmetry that allows for the emergence of Weyl nodes in the Majorana band structure without breaking time-reversal symmetry {\it nor} inversion symmetry, a situation that cannot occur for electronic band structures.
We will discuss these different incarnations of  Weyl physics in a broader context in Sec.~\ref{sec:Weyl}.

For the two-dimensional Kitaev honeycomb model, it is well known that the Dirac spin-liquid of the pure Kitaev model \eqref{eq:spinH} gaps out into a massive phase with non-Abelian topological order when perturbing the spin system with a magnetic field pointing along the 111-direction,
\ie, by considering a Hamiltonian
\begin{equation}
  H = -\sum_{\gamma \rm-bonds}  J_{\gamma} \,\, \sigma_i^\gamma \sigma_j^\gamma - \sum_j \vec{h}\cdot\vec{\sigma}_j  \,.
  \label{eq:spinH2}
\end{equation}
It may, thus, be a natural question to ask whether the breaking of time-reversal symmetry can give rise to gapped phases with non-trivial topological order also for three-dimensional Kitaev models.
Bearing in mind that the Kitaev model can be recast as a free Majorana fermion system, one can immediately answer this question in the negative by considering the classification of topological insulators \cite{TopoClassification} rooted in the symmetry classification of free-fermion systems \cite{AltlandZirnbauer}.
In this free-fermion classification scheme, the pure Kitaev model \eqref{eq:spinH} falls into symmetry class BDI, while the one with broken time-reversal symmetry of Eq.~\eqref{eq:spinH2} belongs to symmetry class D.
As can readily be seen from the classification tables of Refs.~\onlinecite{TopoClassification}, there are no topologically non-trivial band insulators in either symmetry class BDI or D for three spatial dimensions -- in contrast to two spatial dimensions where symmetry class D allows for this possibility [and as realized for Hamiltonian \eqref{eq:spinH2} on the honeycomb lattice]. 
It is, of course, nevertheless an interesting question to ask what effect the breaking of time-reversal symmetry has on the Majorana metals for  the three-dimensional Kitaev models of Table \ref{tab:majorana_metals}.
The answer is provided in the third column of that table and is discussed in much detail in the remainder of the paper.
Generically, we find that the metallic nature per se remains untouched by the breaking of time-reversal symmetry, \ie, for no system do we observe a transition into a (topologically trivial) massive phase.
Instead, we find that the nodal structure remains robust under this perturbation for systems that exhibit Majorana Fermi surfaces or Weyl nodes.
The only effect of a (small) magnetic field is to either deform the Majorana Fermi surface or move the positions of the Weyl nodes.
A different picture emerges for those lattices where the pure Kitaev model \eqref{eq:spinH} exhibits nodal lines.
Here, the magnetic field does alter the nodal structure.
For the lattices (10,3)b and (8,3)c the magnetic field gaps out the nodal line with the exception of two or six nodes, respectively.
These nodes turn out to be one or three pairs of Weyl nodes.
A symmetry analysis again indicates that another symmetry plays a crucial role in stabilizing this unusual band structure.
It is the presence of inversion symmetry that fixes these Weyl nodes to the Fermi energy (at zero energy) for these lattices.
Inversion symmetry is absent for the chiral lattice (10,3)c, which also exhibits nodal lines for the unperturbed Kitaev model.
Upon applying a magnetic field the nodal lines vanish and the system creates six pairs of Weyl nodes, which in the absence of inversion symmetry are no longer fixed to the Fermi energy and move away from it.
The result is the emergence of 12 pockets of Majorana Fermi surfaces, each of which encapsulates a Weyl node.
As a result, these Fermi surfaces also acquire some non-trivial topological properties from the Weyl nodes, as we discuss in further detail in Sec.~\ref{ssec:topFermiSurface}.
As such, the Kitaev model for lattice (10,3)c stands out as the only system where the effect of breaking time-reversal symmetry is to {\it increase} the nodal structure and the associated density of states of the Majorana metal.

Finally, we briefly comment on the \Z\ gauge structure of these models.
As we will discuss in more detail in Sec.~\ref{sec:kitaev}, the assignment of local \Z\ gauges (on the bonds of the lattice) corresponds to an assignment of \Z\ fluxes through the elementary loops in the lattice.
For the two-dimensional Kitaev honeycomb model, this correspondence can be used to readily fix the gauge structure of the ground state via a theorem by Lieb \cite{Lieb}, which states that the ground state of the honeycomb model has no fluxes through any of the hexagonal plaquettes (corresponding to the elementary loops).
Lieb's theorem can also be applied to three-dimensional Kitaev models if the lattice structure exhibits certain mirror symmetries.
As it turns out, only one of the lattices in our family, namely the lattice (8,3)b, fulfills this criterion.
It is thus the only lattice for which we can rigorously assign the flux configuration of the ground state.
For all other lattices, we have to resort to alternative ways to identify precisely this ground state configuration of the (static) \Z\ gauge field.
In this study, we have resorted to numerical simulations of this essentially classical problem for finite systems.
In general, we find that the result of this numerical procedure is a flux configuration that precisely corresponds to the one indicated by Lieb's theorem if it were to apply to the lattice structure at hand. 
Specifically, all elementary loops of length 2 (mod 4) carry zero flux, while elementary loops of length 0 (mod 4) carry $\pi$ flux.
Non-bipartite lattices with elementary loops of odd length are not covered by Lieb's theorem at all.
For these lattices  we assign the flux configuration using symmetry arguments. 
Notably, however, for the lattice (9,3)a our numerical checks indicate the possibility of low-energy flux assignments that break at least one of the point-group symmetries of the lattice.
Such an exotic scenario might also be relevant to lattice (8,3)c where the $\pi$-fluxes are subject to geometric frustration.
We have not explored this possibility in full detail in the paper at hand, but instead have provided results for the lowest-energy \Z\ flux structure that does not break any point-group symmetries.
We revisit this point in an outlook at the end of the paper.

The paper is structured as follows.
In Sec.~\ref{sec:kitaev}, we briefly review the general framework to analytically solve the Kitaev model in arbitrary spatial dimensions.
A detailed analysis of the relevant symmetries of three-dimensional Kitaev models is presented in Sec.~\ref{sec:symmetries}.
We then go through all lattices of Table~\ref{tab:lattice_overview} one by one in Sec.~\ref{sec:3d} and provide further details on the underlying lattices, the definition of the Kitaev model, and its solution for each of these lattices.
This includes the overall phase diagram of the model along with a detailed discussion of the gapless phase around the point of isotropic couplings.
In the subsequent Sec.~\ref{sec:Weyl}, we take a step back and discuss the different scenarios for the emergence of Weyl physics in these three-dimensional Kitaev models.
This also includes a discussion of the topological properties of some of the observed Majorana Fermi surfaces.
Section~\ref{sec:spin-peierls} focuses on a discussion of Majorana Fermi surfaces in general and their BCS-type spin-Peierls instabilities.
We round off the paper with an outlook in Sec.~\ref{sec:outlook} that touches on the possibility of realizing some of the Kitaev models of interest here in spin-orbit entangled Mott insulators.
We further lay out some future directions to be pursued for this family of three-dimensional Kitaev models.
The main paper is complemented by an extensive appendix that provides many of the technical details on the lattice structures.

\section{Solving the Kitaev model}
\label{sec:kitaev}

We start our discussion by briefly reviewing the main traits of Kitaev's original solution \cite{kitaev2006} of the honeycomb model and lay out how it can be adapted to the three-dimensional model systems of interest here.
At the heart of Kitaev's exact solution is the existence of a macroscopic number of conserved quantities, which are associated with closed loops in the underlying two- or three-dimensional lattice.
For each of these loops, we can define a corresponding loop operator by 
\begin{align}
  \label{eq:loop}
  W_\ell&=\prod_{s\in \ell}K_{s,s-1} \,,
\end{align}
where $s$ labels the sites within the loop $\ell$ and $K_{i,j}$ is given by
\begin{align}
  K_{i,j}=\left\{
  \begin{array}{cc} 
    \sigma_i^x\sigma_{j}^x, & \mbox{if }\langle i,j\rangle  \mbox{ is a }x\mbox{-link}\\
    \sigma_i^y\sigma_{j}^y,& \mbox{if }\langle i,j \rangle \mbox{ is a }y\mbox{-link}\\
    \sigma_i^z\sigma_{j}^z, & \mbox{if }\langle i,j \rangle \mbox{ is a }z\mbox{-link}. 
  \end{array}\right.
\end{align}
For even-length loops, the operator has eigenvalue $\pm 1$, where eigenvalue $-1$ is identified with the presence of  a \Z\ flux through the enclosed plaquette and eigenvalue $+1$ implies that there is no flux through the plaquette.
If the operator contains an odd number of bonds/sites, its eigenvalue is instead $\pm i$.
This case is treated in more detail in the discussion of lattice (9,3)a in Sec.~\ref{ssec:9a}. 
Note that the definition \eqref{eq:loop} is chosen in order to be consistent with the conventions chosen by Lieb in discussing his flux-theorem \cite{Lieb} for the ground state.

\begin{figure}
  \includegraphics[width=\columnwidth]{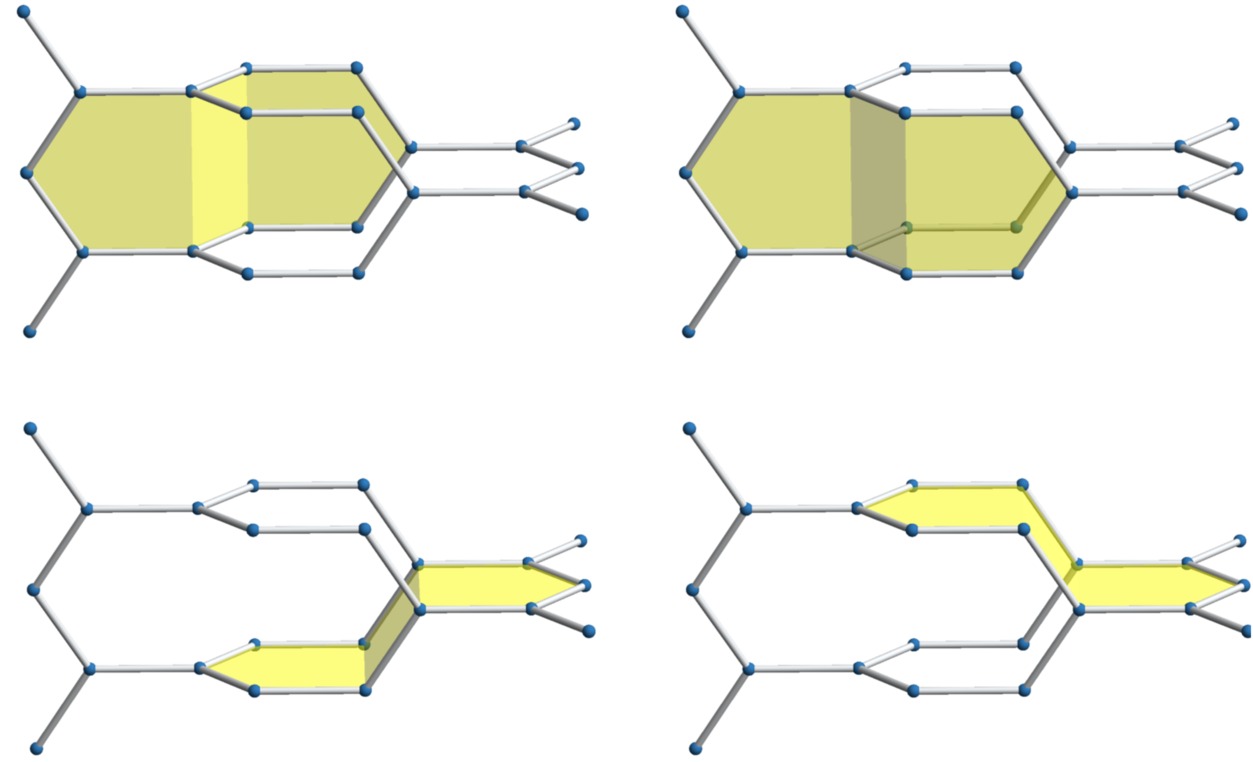}
  \caption{(Color online) Example of a minimal volume arising for the (10,3)b hyperhoneycomb lattice, for which four elementary loops of length 10 are needed to enclose a closed volume.
     		We provide illustrations of such minimal volumes for all lattices in  Appendix \ref{sec:appendix3DKitaev}.}
  \label{fig:MinimalVolumes}
\end{figure}
The set of loop operators is generically linearly dependent for three-dimensional (3D) lattices, due to the presence of volume constraints as exemplified in Fig.~\ref{fig:MinimalVolumes}.
In particular, if a set of loop operators forms the boundary of a closed volume, the product of their eigenvalues is fixed to $+1$.
Thus, we are going to restrict the discussion to a linearly independent subset of these.
For each of the lattices considered here, we can identify $M/2$ fundamental loop operators (per unit cell), where $M$ is the number of sites per unit cell, from which all other loop operators can be built by successive multiplication.
The Hilbert space factorizes into eigenstates of these loop operators (in the following called "flux sectors") and we will usually restrict our analysis to a single such flux sector.
Another consequence of the volume constraint is that flux excitations, also called visons,  always form closed loops; there are no magnetic monopoles in the corresponding \Z\ gauge field. 

Following Kitaev's original solution \cite{kitaev2006}, we proceed to represent each spin in terms of four Majorana fermion operators 
\begin{align}
  \label{eq:spinMaj}
  \sigma_j^\gamma(\mathbf R)=ia_j^\gamma(\mathbf R) c_j(\mathbf R)
\end{align}
with $\gamma=x,y,z$, and $j$ denoting the site within the unit cell at position $\mathbf R$.
The Majorana fermion operators obey the usual anti-commutation relations 
\begin{align}
  \{a_j^\alpha(\mathbf R),a_k^\beta(\mathbf R ')\}&=2\delta_{j,k}\delta_{\alpha,\beta}\delta_{\mathbf R,\mathbf R'},\nonumber\\
  \{c_j(\mathbf R), c_k(\mathbf R')\}&=2\delta_{j,k}\delta_{\mathbf R,\mathbf R'},\nonumber\\
  \{c_j(\mathbf R),a_k^\alpha(\mathbf R')\}&=0. 
\end{align}
This representation faithfully reproduces the spin algebra within the physical Hilbert space defined by 
\begin{align}
  \hat D_j |\mbox{phys}\rangle\equiv a_j^x a_j^y a_j^z c_j  |\mbox{phys}\rangle&=+1|\mbox{phys}\rangle. 
\end{align}
As a second step, we regroup the Majoranas into bond operators $\hat u_{jk}=ia_j^\gamma a_k^\gamma$, with $\gamma$ being the label of the nearest-neighbor bond $\langle j,k\rangle$.
The bond operators  have eigenvalues $\pm 1$, which therefore can be identified with an emergent \Z\ gauge field.
Note that, in contrast to the loop operators, these \Z\ gauge fields are \emph{not} physical, but merely a consequence of enlarging the Hilbert space in Eq.~\eqref{eq:spinMaj}.
In particular, we can identify the loop operators \eqref{eq:loop} as the gauge-invariant quantities of this emergent \Z\ gauge field.  
It turns out that gauge transformations  play an important role when classifying the possible Majorana metals, as they affect how symmetries are implemented in the Majorana system.
This is discussed in more detail in Sec.~\ref{sec:symmetries}.

\begin{table}
  \begin{tabular}{L{13mm} | C{15mm} C{15mm} C{15mm} C{15mm}}
    lattice & flux sector 	& Lieb  		& vison gap 	& vison loop  \\
    	   & 			&  theorem 	&  			& length  \\
    \hline \hline
    (10,3)a 	& $0$-flux  		& --- 	& 0.09(1) 	& 10  \\
    (10,3)b  	& $0$-flux 		& --- 	& 0.13(1)	& 6    \\
    (10,3)c  	& $0$-flux 		& ---	& 0.13(1)	& 3    \\
    \hline
    (9,3)a$^*$ 	& $\pi/2$-fluxes& ---	& ---	& 4 \\
    \hline
    (8,3)a 	& $\pi$-flux 	& ---	& 0.07(1) 	& 2 \\
    (8,3)b 	& $\pi$-flux 	& yes & 0.16(1) 	& 2 \\
    (8,3)c$^*$ 	& $0$-flux  	& ---	& ---	& 4 \\
    (8,3)n	& $\pi$-flux 	& ---	& 0.16(1) 	& 2 \\
    \hline
    (6,3) 	& $0$-flux  		& yes & 0.27	& 2
    \end{tabular}
    \caption{Overview of the physics of the \Z\ gauge field for three-dimensional Kitaev models.
    		The second column provides the flux sector assignment of the elementary loops in the ground state of the Kitaev model defined
     	 for the various tricoordinated lattice geometries of Table \ref{tab:lattice_overview}. 
     	 The third column indicates whether the ground-state flux sector can be assigned via a theorem by Lieb \cite{Lieb}, 
     	 which can only be applied if the underlying lattice has certain mirror symmetries (see the discussion in the main text).
     	 The additional columns provide information on the physics of the vison excitations of the \Z\ gauge field, in particular, the
     	 size of the vison gap and the length of an elementary vison loop in the ground state (\ie, the number of flipped loop operators of  minimal length).
     	 The asterisk indicates that for these two lattices we are providing results for the lowest-energy flux sector that does not
     	 break any point-group symmetries of the lattice.  
      	 }
  \label{tab:Z2}
\end{table}

\begin{figure}
  \includegraphics[width=\columnwidth]{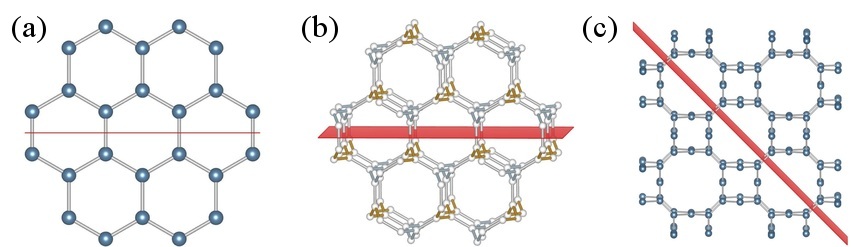}
  \caption{(Color online) (a) Mirror line for the honeycomb lattice (6,3).  (b) One of the three mirror planes for (8,3)b. The other two are obtained by 120$^{\circ}$ rotations around the $\hat z$-direction. The combination determines the eigenvalues of  all the fundamental loop operators.  (c) One of the mirror planes for (8,3)n. Another is obtained by a 90$^{\circ}$ rotation around $\hat z$. A third mirror plane (not shown) lies in the $xy$-plane. These determine the eigenvalues  of all but one loop operator per unit cell.   }
  \label{fig:Lieb}
\end{figure}

It can easily be shown that all bond operators commute with each other, as well as with all of the loop operators and with the Hamiltonian. 
Thus, we can fix their eigenvalues $u_{j,k}$ and solve the resulting quadratic Hamiltonian for the $c$ Majoranas 
\begin{align}
  H&=i\sum_{\gamma-\mbox{\tiny bonds}} J_\gamma u_{j,k}  \, c_j c_k .
\end{align}
Note that when assigning eigenvalues to the bond operators, we need to pick a directionality as $\hat u_{j,k}=-\hat u_{k,j}$ \cite{FootnoteDirectionality}. 
The remaining difficulty is to decide on how to assign the bond eigenvalues. 
One guiding principle is Lieb's theorem \cite{Lieb}.
This theorem determines the ground-state flux configuration for any plaquette that is invariant under mirror symmetry, as long as the mirror line (2D) or plane (3D) does not cut through lattice sites, but only bonds.
The flux per plaquettes is 0 if the loop length equals 2 (mod 4), whereas it is $\pi$ for loop lengths 0 (mod 4). 
For the honeycomb lattice, we show the relevant mirror line in  Fig.~\ref{fig:Lieb} (a). The bond length 6 implies that the ground state sector has 0 flux per plaquette, \ie all loop operators have eigenvalue $+1$.

For most of the 3D lattices discussed here, Lieb's theorem cannot be applied.
Notable exceptions are the lattices (8,3)b and (8,3)n. 
For the former, Lieb's theorem determines the flux of all the fundamental loop operators; for the latter, Lieb's theorem determines seven of the eight fundamental loop operators per unit cell. 
Examples of the relevant mirror planes are shown in Fig.~\ref{fig:Lieb} (b) for the lattice (8,3)b and in Fig.~\ref{fig:Lieb} (c) for (8,3)n. 
For all the other lattices, one needs to resort to numerical simulations  to determine the flux configuration of the ground state.
However, Lieb's theorem still provides a good educated guess.
In particular, choosing the flux configuration as explained above and preserving the symmetries of the underlying lattices often yields the correct ground state sector.
Notable exceptions are (9,3)a and (8,3)c, where numerical studies indicate that the system may prefer to spontaneously break lattice symmetries in the ground state. 

\begin{figure}
  \includegraphics[width=\columnwidth]{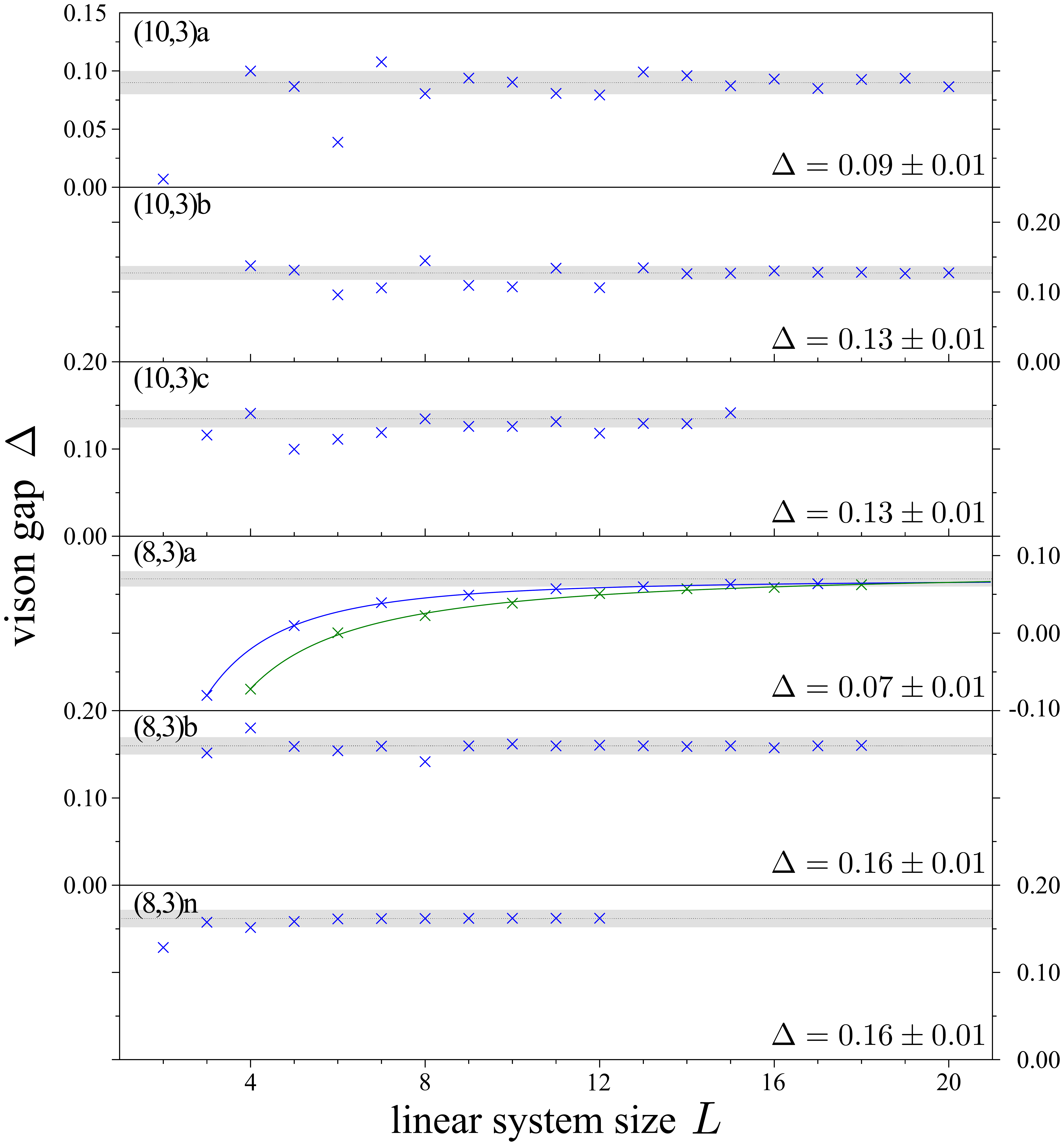}
  \caption{(Color online) Vison gap obtained for the smallest vison loop as a function of system size. The dotted line marks the extrapolation of the gap for infinite system size, and the gray bar denotes the error of the extrapolation. Details on the vison loops can be found in Appendix~\ref{sec:supplemental}.}
  \label{fig:visongap}
\end{figure}

Note that the elementary excitations of the \Z\ gauge field remain gapped for all lattices.
We calculated this vison gap for all lattices [but lattices (8,3)c and (9,3)a, see the discussion above] for the smallest possible vison loop. This can usually be obtained by flipping a single $z$-type bond, except for the lattices (10,3)b and c, where instead a single $x$- or $y$-type bond is flipped. 
Results for the vison gap are given in Table \ref{tab:Z2} and Fig.~\ref{fig:visongap}, where the explicit system-size dependence of the vison gap is illustrated. Details on the smallest vison loop and which bond is flipped can be found in Appendix~\ref{sec:supplemental}.

\paragraph*{Magnetic field.--}
When discussing the effect of symmetries on possible Majorana metals, we are particularly interested in the effect of time reversal and what consequences breaking time-reversal symmetry can have.
For simplicity, the time-reversal symmetry breaking term we consider is chosen as an external magnetic field in the 111 direction.
The particular direction is not essential as long as it couples to all spin components, but it ensures that all other (lattice) symmetries remain intact.
As $\vec h\cdot \vec \sigma$ does not commute with the loop operators \eqref{eq:loop}, the model is no longer exactly solvable.
However, as long as the visons are gapped and the strength of $\vec h$ sufficiently small as not to excite them, we can treat the magnetic field term perturbatively.
Following Ref.~\cite{kitaev2006}, one finds that the first non-trivial contribution arises in third-order perturbation theory and can be written as a three-spin interaction,  
\begin{align}
  \label{eq:kappa}
  H_{\rm eff}&=-\kappa \sum_{ (j,l,k)} \sigma_j^\alpha \sigma_k^\beta \sigma_l^\gamma,
\end{align} 
where the summation is over all triples of adjacent sites $j$, $l$, and $k$ such that the bond $\langle j,l\rangle$ ($\langle k,l\rangle$) is of $\alpha$- ($\beta$-) type and $\gamma$ is chosen as the remaining bond type.
The prefactor $\kappa\sim(h_x h_y h_z)/\Delta^2$ depends on the vison gap $\Delta$ and the strength of the magnetic field. 
Using the representation in terms of Majorana fermions \eqref{eq:spinMaj}, we can rewrite this as an effective next-nearest neighbor hopping of the $c$ Majoranas by
\begin{align}
  \sigma_j^\alpha \sigma_k^\beta \sigma_l^\gamma &= i\epsilon^{\alpha\beta\gamma}\hat D_l \hat u_{jl}\hat u_{lk}c_j c_k , 
\end{align}
where we can neglect the operator $\hat D_l$ as it will act as the identity on the physical subspace.

 
\section{Projective symmetries}
\label{sec:symmetries}

While the general framework of Kitaev's original solution of the honeycomb model can be applied to arbitrary tricoordinated lattices, including three-dimensional ones, important differences between individual lattice structures arise when considering the most elementary symmetries of the system.
In particular, it is important to note that the symmetries of the original spin Hamiltonian can manifest themselves in distinct ways when considering their effect in the physical Majorana subspace.
These projective symmetries turn out to be sensitive to the underlying lattice geometry.
The deeper origin of this lattice dependence can be traced back to the somewhat subtle incarnation of the \Z\ gauge theory description of the Kitaev model.
All calculations are done in a fixed gauge, \ie, one chooses a specific set of bond operator eigenvalues that is compatible with the flux sector. 
In order for the symmetries to act within this fixed gauge sector, they often need to be supplemented by a gauge transformation.
As we detail in the following, such a supplemental gauge transformation may lead to an additional shift in momentum space in the projective symmetry relations for the Majorana fermions.
As we will argue in the following, a careful analysis of these projective symmetries, in particular particle-hole, time-reversal, and inversion symmetry, will allow us to systematically classify the Kitaev Hamiltonian in its Majorana representation and infer the nature of the emergent Majorana metal from it.

\paragraph*{Particle-hole symmetry \emph{(PHS)}.--}
Particle-hole symmetry is, strictly speaking, not a symmetry of our systems, but rather a consequence of describing the spins in terms of Majorana fermions (instead of complex fermions) which over-counts the degrees of freedom by a factor of 2. 
The Majorana condition $c_\alpha(\mathbf r)=c^\dagger_\alpha(\mathbf r) $ immediately leads to the condition $c_\alpha^\dagger (\mathbf k)=c_\alpha(-\mathbf k)$ in momentum space. 
For the Hamiltonian and its eigenenergies, this implies 
\begin{align}
  \hat h(\bk)&=-\hat h^\star(-\bk),\nonumber\\ 
	\epsilon(\bk) &= - \epsilon(- \bk),
	\label{eq:PH}
\end{align}
where the asterisk indicates complex conjugation. 
The over-counting of degrees of freedom can be taken care of in two ways: either by restricting the allowed momenta to half the Brillouin zone or  by discarding  the lower half of the energy bands. 

\begin{figure}
  \includegraphics[width=\columnwidth]{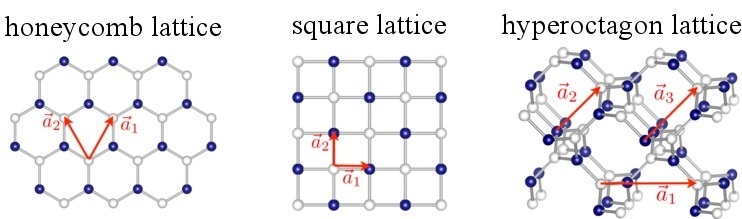}
  \caption{(Color online) Visualization of the A and B sublattices (in white and blue, respectively) for the honeycomb, square, and hyperoctagon lattices. While the sublattices of the honeycomb lattice have the same translation vectors as the original lattice, the same is not true for the square and hyperoctagon lattices, which leads to a finite value for $\kk$. }
  \label{fig:sublattice}
\end{figure}

\paragraph*{Sublattice symmetry \emph{(SLS)}.--}
All lattices that are considered here, with the exception of (9,3)a, are bipartite lattices, \ie, the lattice sites can be partitioned into two sublattices (referred to as sublattices A and B in the following) such that nearest neighbors are always from  different sublattices.
The pure Kitaev Hamiltonian \eqref{eq:spinH} has only nearest neighbor interactions.
Consequently, a sublattice transformation,
\begin{align}
  c_\alpha(\br)&\rightarrow \left\{ \begin{array} {rll} & c_\alpha(\br)& \mbox{ for  sublattice A} \\ -&c_\alpha(\br) &\mbox{ for sublattice B}\end{array}\right.
\end{align}
changes the overall sign of the Hamiltonian and implies that 
\begin{align}
  \hat h(\bk)&= - U_{\mbox{\tiny SLS}} \, \hat h(\bk+\kk)\, U_{\mbox{\tiny SLS}}^{-1},\nonumber\\
  \epsilon(\bk)& = - \epsilon(\bk+ \kk),	
	\label{eq:sublattice}
\end{align}
where $U_{\mbox{\tiny SLS}}$ is a unitary matrix and $\kk$ is a reciprocal lattice vector of the sublattice. 
Note that $\kk=0$, if the sublattice has the same translation vectors as the full lattice.
Examples for this case are the honeycomb lattice in 2D and the (10,3)b (hyperhoneycomb) lattice in 3D. 
If the translation vectors connect the two sublattices, as is the case, \eg, for the square lattice in Fig.~\ref{fig:sublattice}, then $\kk$ has a finite value; in particular,  $\kk=(\frac \pi 2,\frac \pi 2)$ for the square lattice.
In 3D, $\kk$ is non-vanishing for the lattices (10,3)a (hyperoctagon), (8,3)a, as well as for (8,3)b. 
 
\paragraph*{Time-reversal symmetry \emph{(TRS)}.--}
The importance of the sublattice symmetry becomes apparent when trying to implement time-reversal symmetry for the Majorana system.
As in the Kitaev honeycomb model \cite{kitaev2006}, we consider a time-reversal transformation $T$ that squares to one, $T^2=1$, which implies symmetry class BDI. 
Time-reversal symmetry flips the spin eigenvalues, and can be implemented in the Majorana language by 
\begin{align}
  T c_j  (\br )T^{-1}&=c_j (\br ), &T a_j^\alpha  (\br )T^{-1}&=a_j^\alpha  (\br )
\end{align}
and complex conjugation.
However, due to the complex conjugation it also flips the eigenvalue of all the bond operators, $T \hat u_{jk} T^{-1}=-\hat u_{jk}$ and, thus, needs to be supplemented by a gauge transformation so as to remain in the same (fixed) gauge sector.
The required gauge transformation is, in fact, simply the sublattice transformation discussed above, and we can implement time-reversal by requiring 
\begin{align}
\tilde T c_j  (\br )\tilde T^{-1}&=\mu_j(\br) c_j (\br ), &\tilde T a_j^\alpha  (\br )\tilde T^{-1}&=\mu_j(\br)a_j^\alpha (\br ) \,,  
\end{align}
where $\mu_j(\br)=1$ or $-1$ depending on whether the site is in the A- or B-sublattice.
Thus, the gauge-invariant time-reversal $\tilde T$ inherits the $\kk$ vector from the sublattice transformation, and we obtain
\begin{align}
  \hat h(\bk)&=  U_{\mbox{\tiny T}} \, \hat h^\star(-\bk+\kk)\, U_{\mbox{\tiny T}}^{-1}\nonumber\\
  \epsilon(\bk)& =  \epsilon(- \bk + \kk)	. 
	\label{eq:TR}
\end{align}
Note that if $\kk=0$, the only stable zero-mode manifolds are nodal lines independent of other symmetries, as was shown in \cite{hermanns14}.
If $\kk\neq 0$ or time-reversal symmetry is broken, then other symmetries become important for determining the stable zero-energy modes, see also Table~\ref{tab:majorana_metals}. 

\paragraph*{Inversion symmetry \emph{(I)}.--}
Analogously to time-reversal invariance discussed above, inversion symmetry also needs to be supplemented by a gauge transformation in order to act within a fixed gauge sector.
However, the exact form of the gauge transformation depends on the details of the lattice and the flux configuration.
In general, inversion symmetry will act as 
\begin{align}	
  \label{eq:inversion}
  \hat h(\bk)   &=  U_{\mbox{\tiny I}} \, \hat h(-\bk+\tilde{\mathbf k}_0)\, U_{\mbox{\tiny I}}^{-1}\nonumber\\
  \epsilon(\bk) &= \epsilon(- \bk + \tilde{\mathbf k}_0)	\,,
\end{align}
where $\tilde{\mathbf k}_0$ is either  half a reciprocal lattice vector or zero, depending on whether the necessary gauge transformation enlarges the unit cell or not.
In particular, $\tilde{\mathbf k}_0 $ may or may not be equal to $\kk$, and various different possibilities are realized in the models we discuss here. 
Of particular interest is the lattice (8,3)b, where $\kk\neq 0$ and $\tilde{\mathbf k}_0=0$, thus allowing for the presence of Weyl nodes with both inversion and time-reversal symmetries unbroken. 
An example for $\kk= 0$ and $\tilde {\mathbf{k}}_0  \neq0$ is the lattice (8,3)n, where the time-reversal broken model allows for Majorana Fermi surfaces, even in the presence of inversion symmetry. 
Note that other lattice symmetries, such as rotations, may also have such an additional translation in momentum space.


\section{3D Kitaev models}
\label{sec:3d}

We now turn to a detailed discussion of the 3D Kitaev models for the various three-dimensional, tricoordinated lattices of Table~\ref{tab:lattice_overview}.
For each lattice, we go through a similar line of arguments where we (i) provide some elementary information about the lattice structure such as its unit
cell, associated lattice vectors and the assignment of Kitaev couplings to the bonds of the lattice, (ii) discuss the structure of the elementary loops and
the assignment of \Z\ fluxes in the ground state of the Kitaev model, (iii) determine the manifestation of the projective symmetries, and (iv) discuss the
nature of the emergent Majorana metal.

This discussion is supplemented by two appendices where we provide more detailed technical information for each lattice. 
In Appendix~\ref{sec:appendix3DKitaev} we present visualizations of the lattice structures along various high-symmetry projections 
along with detailed information on their space group and the Wyckoff positions for the unit cell. 
This lattice information is further supplemented by VESTA \cite{VESTA,supplemental} visualization files, which are provided in the Supplemental Material of this paper.
In Appendix~\ref{sec:supplemental}, we provide additional information on the 3D Kitaev models for these lattices.
In particular, we provide a detailed summary
of the gauge structure of the ground state and give an explicit expression for the Kitaev Hamiltonian in its Majorana representation in this gauge.

\subsection{(8,3)a}
\label{ssec:8a}

\begin{figure}
  \includegraphics[width=\linewidth]{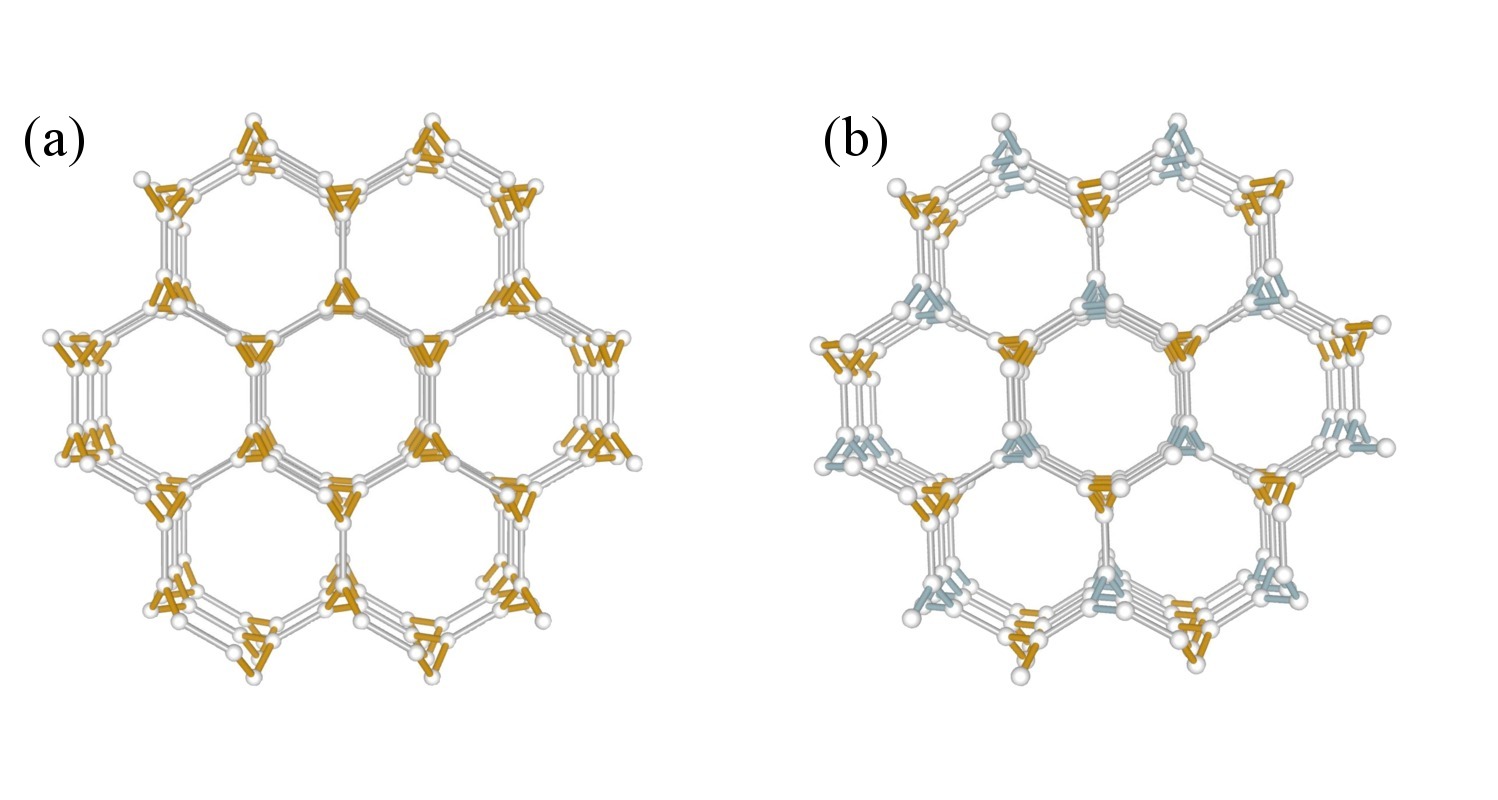}
  \caption{(Color online)  The distinct features of the (8,3)a lattice [shown in (a)] are co-rotating spirals, while the (8,3)b lattice [shown in (b)] has counter-rotating spirals. 
  The two different rotation directions are indicated by orange and blue, respectively. }
  \label{fig:comparison_8_3_a_b}
\end{figure}
\begin{figure*}
  \includegraphics[width=\linewidth]{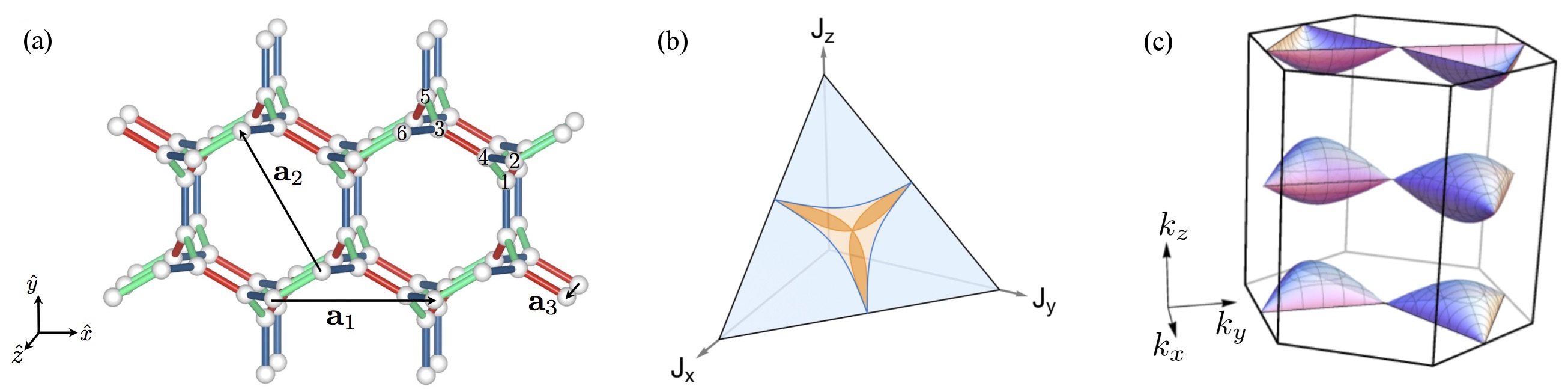}
  \caption{(Color online) (a) Unit cell and translation vectors for the Kitaev model on the (8,3)a lattice.  (b) Phase diagram for (8,3)a. The parameter regions shaded darker orange have topological Fermi surfaces, while the lighter orange regions have trivial Fermi surfaces. (c) Visualization of the four Majorana Fermi surfaces for isotropic Kitaev couplings.}
  \label{fig:phase_diagram_8_3a}
\end{figure*}

We start the discussion with lattices of elementary loop length 8.
Of the 15 nets reported on  by Wells \cite{Wells77}, only four have equal-length bonds and 120$^\circ$ bond angles and will be discussed in detail in the coming sections.
The first two lattices, (8,3)a and (8,3)b, are in fact close cousins of each other. 
Both of them can be viewed as a three-dimensional version of the 3-12-12 lattice \cite{KitaevMosaicModels} or, alternatively, the Yao-Kivelson lattice \cite{YaoKivelson}, where the triangles are replaced by triangular spirals.
While these spirals are co-rotating in (8,3)a, resulting in a chiral lattice, they are counter-rotating in (8,3)b, thus leading to an inversion-symmetric lattice.
A comparison of the two lattices is given in Fig.~\ref{fig:comparison_8_3_a_b}, where the two different rotation directions are marked by different colors. 

\begin{figure}
  \includegraphics[width=\linewidth]{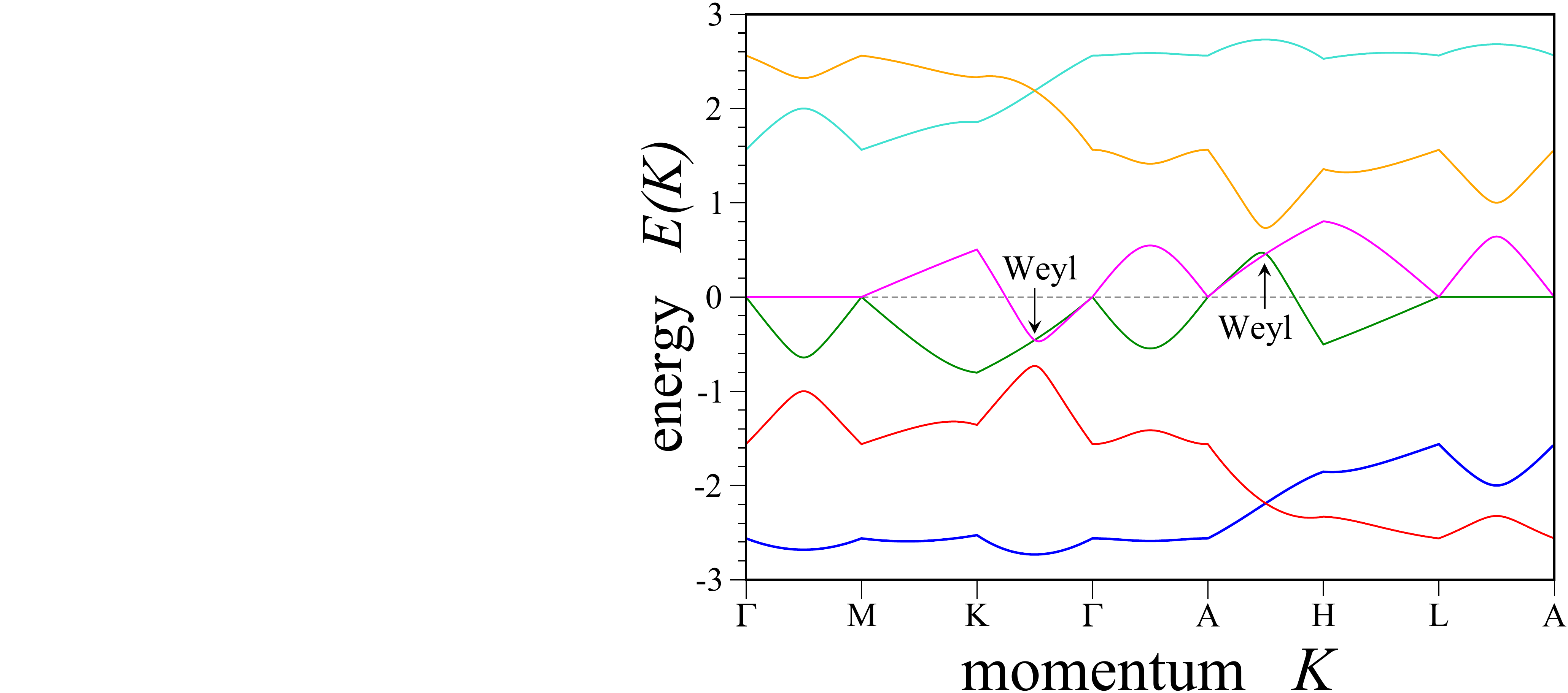}
  \caption{(Color online) 
    Left-hand side: Brillouin zone with Majorana Fermi surfaces and high-symmetry points for (8,3)a. 
    Right-hand side: Energy dispersion along the high-symmetry lines. 
	  The Weyl points are indicated at the band crossings between the green and pink bands between $K$ and $\Gamma$, as well as between $A$ and $H$.}
  \label{fig:energydispersion8a}
 \end{figure}
%

\paragraph*{Lattice structure.--}
More formally, the (8,3)a lattice is a hexagonal lattice with six sites per unit cell at positions 
\begin{align}
  \br_1 &= \left( \frac{1}{2}, \frac{\sqrt{3}}{10}, 0 \right)\,,
  &\br_2 &= \left( \frac{3}{5}, \frac{\sqrt{3}}{5}, \frac{2\sqrt{2}}{5} \right)\,, \nonumber\\
  \br_3 &= \left( \frac{1}{10}, \frac{3\sqrt{3}}{10}, \frac{\sqrt{2}}{5} \right)\,,
  &\br_4 &= \left( \frac{2}{5}, \frac{\sqrt{3}}{5}, \frac{\sqrt{2}}{5} \right)\,, \nonumber\\
  \br_5 &= \left( 0, \frac{2\sqrt{3}}{5}, 0 \right)\,,
  &\br_6 &= \left( -\frac{1}{10}, \frac{3\sqrt{3}}{10}, \frac{2\sqrt{2}}{5} \right)\,.
\end{align}
We choose the lattice vectors as
\begin{align}
  \ba_1 &= \left( 1, 0, 0 \right),&
  \ba_2 &= \left( -\frac{1}{2}, \frac{\sqrt{3}}{2}, 0 \right),&\nonumber\\
  \ba_3 &= \left( 0, 0, \frac{3\sqrt{2}}{5} \right),
\end{align}
and the corresponding reciprocal lattice vectors become
\begin{align}
  \bq_1 &= \left( 2\pi, \frac{2\pi}{\sqrt{3}}, 0 \right),&
  \bq_2 &= \left( 0, \frac{4\pi}{\sqrt{3}}, 0 \right),&\nonumber \\
  \bq_3 &= \left( 0, 0, \frac{5\sqrt{2}\pi}{3} \right).
  \label{eq:recip_vec_8_3a}
\end{align}
The unit cell and the translation vectors are illustrated in Fig.~\ref{fig:phase_diagram_8_3a} (a).
The color coding of bonds in this figure indicates the assignment of the bond-directional
coupling along $x$-, $y$-, and $z$-type bonds with colors green, red, and blue, respectively.
This particular assignment of the bonds is chosen such as to retain as many of the lattice symmetries as possible, and is {\em unique} up to an overall permutation of the three bond types. 
Note that the sets of all $x$-, $y$-, and $z$-bonds are related to one another by the threefold screw-rotation symmetry around the  $\hat z$ axis; consequently, the phase diagram, shown in Fig.~\ref{fig:phase_diagram_8_3a} (b), has to be symmetric in all couplings.

\begin{figure*}
  \includegraphics[width=\linewidth]{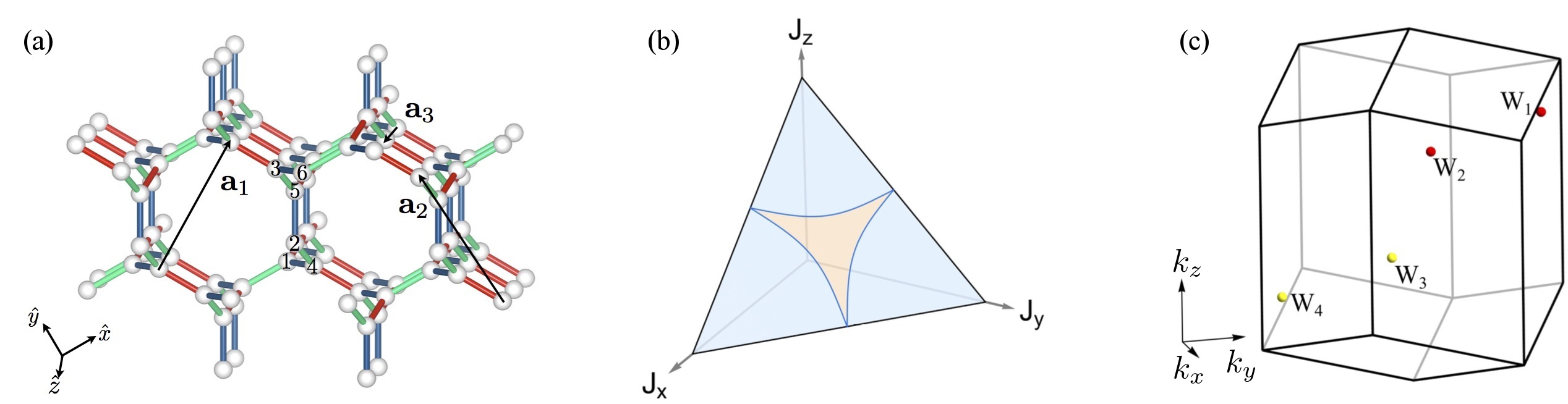}
  \caption{(Color online) (a) Unit cell and translation vectors, (b) phase diagram, and (c) Brillouin zone and position of Weyl points for isotropic couplings  for the lattice (8,3)b. Yellow/red denote Weyl points with negative/positive charge.  }
  \label{fig:phase_diagram_8_3b}
\end{figure*}
%

\paragraph*{Gauge structure.--}
The (8,3)a lattice has three loop operators of length 8 and three of length 14 per unit cell.
These six loop operators form three closed volumes which leads to only three linearly independent loop operators per unit cell (see Appendix \ref{ssec:supplemental_8_a} for details).
In what follows, we consider the flux sector where all loop operators of length 8 have eigenvalue $-1$ and all loop operators of length 14 have eigenvalue $+1$.
This configuration of fluxes respects all the lattice symmetries and, although this lattice does not possess the symmetries required for rigorous application of Lieb's theorem \cite{Lieb}, is consistent with the flux assignments one would expect were Lieb's theorem to hold.
In addition, we have also checked numerically that this flux sector is indeed the ground-state sector. 
The vison gap for this lattice, reported in Fig.~\ref{fig:visongap}, is computed by flipping a single $z$-bond operator, which changes the signs of four loop operators.
For further details, we refer the reader to  Appendix \ref{ssec:supplemental_8_a}.

\paragraph*{Projective symmetries.--}
This lattice has the property that the translation vector $\ba_3$ maps the two sublattices onto each other.
Therefore, sublattice symmetry and time-reversal symmetry involve a non-vanishing translation in momentum space by $\kk = \bq_3/2 = (0, 0, \frac{5\pi}{3\sqrt{2}})$.
As the lattice is chiral, the relevant energy relations are given by particle-hole and time-reversal symmetry
\begin{align}
  \epsilon(\bk) &= -\epsilon(-\bk)& \mbox{ and }&&
  \epsilon(\bk) &= \epsilon(-\bk + \kk) \,.
\end{align}
%

\paragraph*{Majorana metal.--}
Following the symmetry analysis of Sec.~\ref{sec:symmetries}, the projective symmetries of lattice (8,3)a indicate that the only stable zero-energy manifolds are {\em surfaces}.
Indeed, we find an extended gapless phase around the point of isotropic coupling [see the  phase diagram shown in Fig.~\ref{fig:phase_diagram_8_3a} (b)],
 where the gapless modes sit on four Majorana Fermi surfaces, visualized in Fig.~\ref{fig:phase_diagram_8_3a} (c). 
The darker shaded orange region of the phase diagram denotes the parameter space where these Majorana Fermi surfaces are topological, \ie, they contain a Weyl point at finite energy. 
The Weyl points can be seen in the energy dispersion in Fig.~\ref{fig:energydispersion8a} as the band crossings between the middle two bands. 
For isotropic coupling strengths, the Weyl points are located  at  $\bk= \pm (\frac {\pi}{ 3} , \frac {\pi}{\sqrt{3}},0)$ and their translations by $\kk$, while pairs of (oppositely charged) Weyl nodes are located at the touching points of the surfaces [see Fig.~\ref{fig:phase_diagram_8_3a}(c)].
The latter are located at $\bk =(0,0,0),\, (\pi,\frac \pi {\sqrt 3} ,0)$, and  $(0,\frac{2\pi}{\sqrt 3} ,0)$, as well as their translations by $\kk$. 
Note that combining time-reversal with particle-hole symmetry implies that the spectrum is anti-symmetric under translation of $\kk$, \ie, $\epsilon_\alpha(\bk)=-\epsilon_{7-\alpha}(\bk+\kk)$ [assuming energies are sorted $\epsilon_1(\bk)>\epsilon_2(\bk)>\ldots >\epsilon_6(\bk)$], which is clearly visible in the dispersion plot in Fig.~\ref{fig:energydispersion8a}. 
For a more detailed description on topological Fermi surfaces, as well as the evolution of the position of Weyl points throughout the phase diagram, we refer the reader to the discussion on topological Fermi surfaces in Sec.~\ref{ssec:topFermiSurface}.

As can be seen from Fig.~\ref{fig:phase_diagram_8_3a} (c), pairs of Majorana Fermi surfaces are mapped onto each other by the perfect nesting vector $\kk$. 
This suggests a  Fermi surface instability as soon as the Majorana fermions become interacting, which happens naturally when adding additional interactions to the pure Kitaev Hamiltonian \eqref{eq:spinH}.
A very similar situation occurs in the (10,3)a hyperoctagon lattice (for details, see Sec.~\ref{ssec:10a}) and was studied in Ref.~\cite{SpinPeierlsHyperoctagon}. 
It was shown that generic interactions always induce a BCS-type instability of the Majorana Fermi surface, albeit with the important difference that it is translation symmetry that is spontaneously broken, not $U(1)$ symmetry.
Thus, the instability was dubbed spin-Peierls BCS instability. 
The resulting phase is still a quantum spin liquid, but with a nodal line instead of a full Fermi surface. 
As the arguments in Ref.~\cite{SpinPeierlsHyperoctagon} are very general and only rely on the perfect nesting condition, we expect the same type of behavior for (8,3)a in the presence of interactions. 
One important difference to the (10,3)a lattice lies in the larger number of surfaces for the (8,3)a lattice. 
As a consequence, time-reversal symmetry does not guarantee the presence of nodal lines, and interactions might thus gap the system completely. 
For further details on the spin-Peierls instability, as well as the effect of time-reversal breaking interactions, the reader is referred to Section \ref{sec:spin-peierls}.

\subsection{(8,3)b}
\label{ssec:8b}
 
\paragraph*{Lattice structure.--}
The (8,3)b lattice has a lattice structure very similar to the lattice (8,3)a discussed in the previous section.
It can again be viewed as coupled triangular spirals, however, in contrast to (8,3)a, the rotation directions alternate between nearest-neighbor spirals in (8,3)b , leading to an inversion symmetric lattice.

It has six sites per unit cell that are located at  
\begin{align}
  \br_1 &= \left( \frac{1}{10}, \frac{1}{2\sqrt{3}}, \frac{1}{5}\sqrt{\frac{2}{3}} \right) \,,  &\br_2 &= \left( \frac{1}{5}, \frac{\sqrt{3}}{5}, \frac{\sqrt{6}}{5} \right) \,, \nonumber\\ 
  \br_3 &= \left( \frac{3}{10}, \frac{11}{10\sqrt{3}}, \frac{4}{5}\sqrt{\frac{2}{3}} \right) \,, &\br_4 &= \left( \frac{1}{5}, \frac{2}{5\sqrt{3}}, \frac{2}{5}\sqrt{\frac{2}{3}} \right) \,, \nonumber\\ 
  \br_5 &= \left( \frac{3}{10}, \frac{3\sqrt{3}}{10}, \frac{\sqrt{6}}{5} \right) \,, &\br_6 &= \left( \frac{2}{5}, \frac{1}{\sqrt{3}}, \sqrt{\frac{2}{3}} \right)  \,,
\end{align}
and the lattice vectors are chosen as
\begin{align}
  \ba_1 &= \left( \frac{1}{2}, \frac{1}{2\sqrt{3}}, \frac{1}{5}\sqrt{\frac{2}{3}} \right),&
  \ba_2 &= \left( 0, \frac{1}{\sqrt{3}}, \frac{2}{5}\sqrt{\frac{2}{3}} \right),&\nonumber\\
  \ba_3 &= \left( 0, 0, \frac{\sqrt{6}}{5} \right).
\end{align}
The corresponding reciprocal lattice vectors are given by
\begin{align}
  \bq_1 &= \left( 4\pi, 0, 0 \right),&
  \bq_2 &= \left( -2\pi, 2\sqrt{3}\pi, 0 \right),&\nonumber \\
  \bq_3 &= \left( 0, -\frac{4\pi}{\sqrt{3}}, 5\sqrt{\frac{2}{3}}\pi \right).
  \label{eq:recip_vec_8_3b}
\end{align}

The unit cell and translation vectors are illustrated in Fig.~\ref{fig:phase_diagram_8_3b} (a).
The assignment of  the three bond types is chosen analogously to (8,3)a; again, this is the most symmetric assignment of the different bond types and unique up to overall permutations of $x$, $y$, and $z$. 
Note that also for this lattice, the sets of all $x$-, $y$-, and $z$-bonds are related to one another by the threefold rotation symmetry around the  $\hat z$ axis, and the phase diagram, consequently, is symmetric in all couplings. 

\paragraph*{Gauge structure.--}
The lattice (8,3)b takes a special role among the lattices considered in this paper, as it is the only three-dimensional lattice that allows for a direct application of 
Lieb's theorem \cite{Lieb} to determine the ground-state flux sector {\em rigorously}.
The mirror planes used for determining the flux are illustrated in Fig.~\ref{fig:Lieb}.  
There are three linearly independent loop operators per unit cell, which all have length eight and, thus, $\pi$ flux in the ground state.
Flipping a single $z$-type bond changes the sign of four plaquettes.
The corresponding vison gap $\Delta\sim 0.16$ (measured in units of the Kitaev coupling at the isotropic point $J_x=J_y=J_z=1$) is the largest found for the lattices considered here (see Table \ref{tab:Z2} and Fig.~\ref{fig:visongap}).

\begin{figure}
  \includegraphics[width=\linewidth]{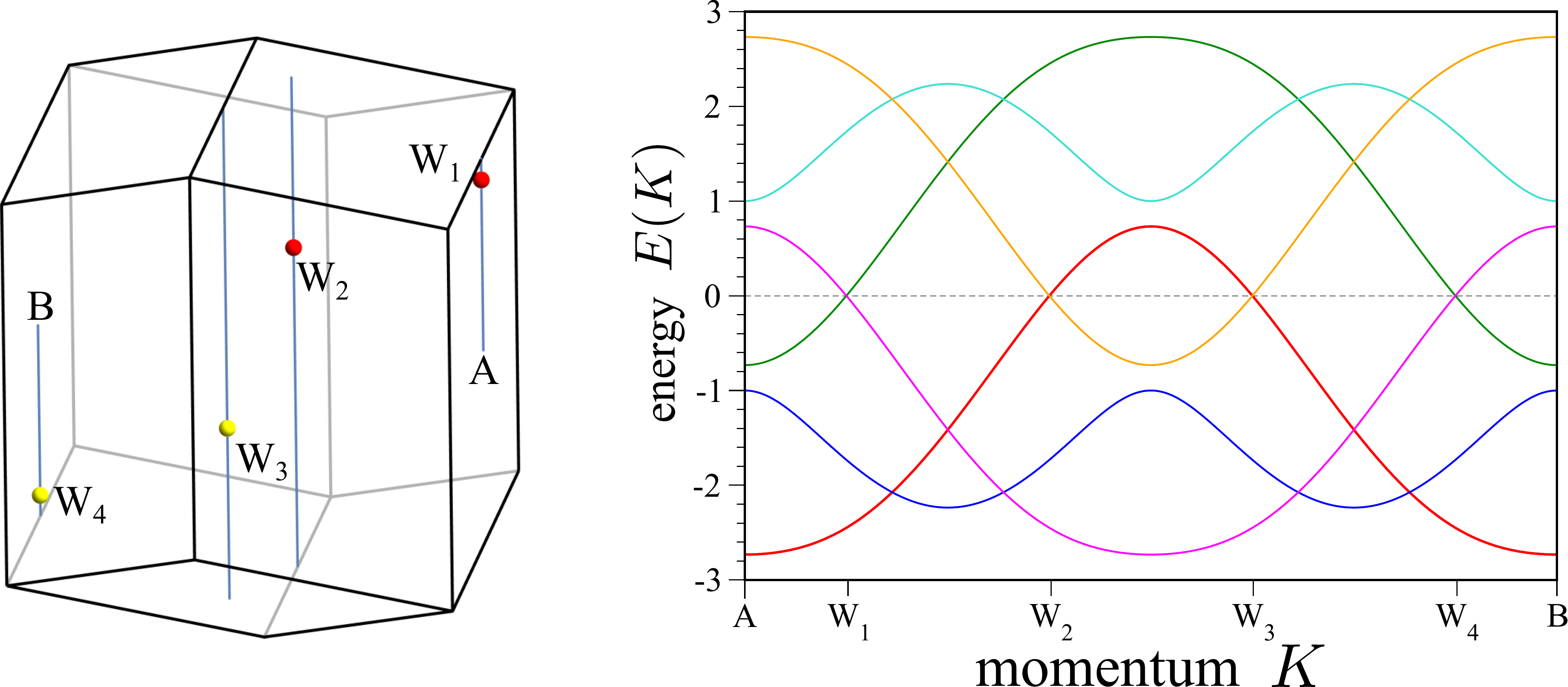}
  \caption{(Color online) Left-hand side: At the isotropic point, the Weyl points are located on the 120$^\circ$ rotation invariant line, marked in blue. The right-hand side shows the energy dispersion along this high-symmetry line.}
  \label{fig:dispersion_8_3b}
\end{figure}
%

\paragraph*{Projective symmetries and Majorana metal.--}
This lattice has a non vanishing $\kk=\mathbf q_1/2+\mathbf q_3/2$ vector for time-reversal symmetry which, in the absence of other symmetries, would imply that the system exhibits stable Majorana Fermi surfaces.
However, as (8,3)b is also inversion symmetric with $\kt=0$,  the energy dispersion is particle-hole symmetric at every momentum which, in turn, prohibits stable Fermi surfaces.
Instead, the system exhibits gapless {\em Weyl points} in a finite parameter region around the isotropic point, as shown in the phase diagram of Fig.~\ref{fig:phase_diagram_8_3b} (b) and the dispersion plot of Fig.~\ref{fig:dispersion_8_3b}.
Note that the Weyl points are fixed to zero energy due to inversion symmetry. 
\begin{figure}
  \includegraphics[width=\linewidth]{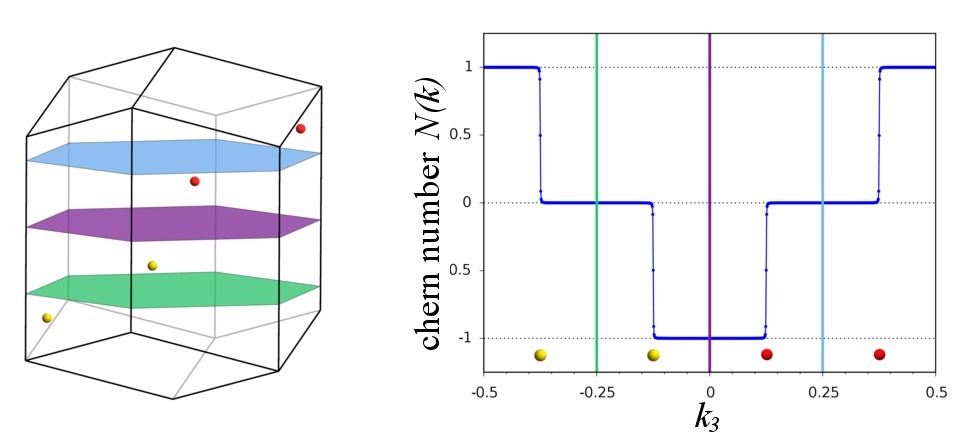}
  \caption{(Color online) The left-hand side shows the Brillouin zone with the positions of the Weyl points, the right-hand side shows the corresponding Chern number. The positions of the Weyl points, as well as the three example planes (defined by  $\bk_3=0,\pm 1/4$), are indicated as a guide to the eye.  }
  \label{fig:chern_numbers_8_3b}
\end{figure}
\begin{figure*}
  \includegraphics[width=\linewidth]{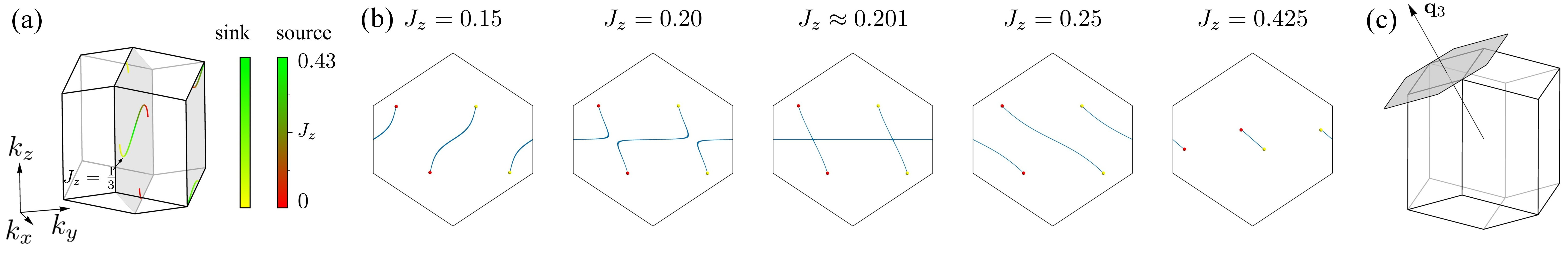}
  \caption{(Color online) (a) Evolution of Weyl points for varying the coupling constants  $0 \leq J_z \lesssim 0.43$ and $J_x = J_y = (1-J_z)/2$ for (8,3)b.  (b) Corresponding Fermi arcs. The Fermi arcs touch and reconnect at $J_z\approx 0.201$. (c) Surface Brillouin zone for the 001 surface.}
  \label{fig:fermi_arcs_8_3b}
\end{figure*}
\begin{figure*}
  \includegraphics[width=\linewidth]{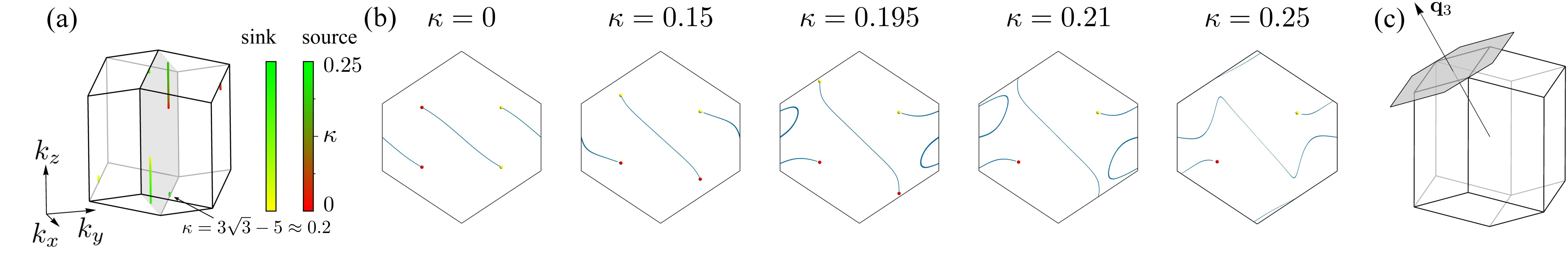}
  \caption{(Color online) (a) Evolution of the Weyl nodes  for (8,3)b in presence of a magnetic field for varying $\kappa$ from $0,\ldots,0.25$. (b) Corresponding Fermi arc evolution. (c) Visualization of the surface Brillouin zone for open boundary conditions in the 001 direction.
  \label{fig:weyl_points_8_3b}}
\end{figure*}

Before discussing the manifestation of Weyl physics in the 3D Kitaev model for the (8,3)b lattice, we want to briefly recapitulate some elementary aspects of Weyl semi-metals
as they are typically discussed in the context of electronic band structures \cite{WeylSM}. 
Weyl points are, in fact, a very common phenomenon in three-dimensional band models, as band touching points generically show a linear dispersion. 
Projected onto the two relevant bands, the low-energy band Hamiltonian can (to leading order) be expanded as 
\begin{align}
  \hat H_{2\times 2}&=\mathbf{v}_0\cdot \mathbf q \,\id +\sum_{j=1}^3 \mathbf{v}_{j}\cdot \mathbf q \,  \sigma_j \,,
\end{align}
where $\mathbf q$ denotes the displacement (in momentum space) from the band-touching point.  
The "velocities" $\mathbf v_j$ are, in general,  all non-zero and linearly independent, in which case we  call the band touching a Weyl point (WP). 
Each WP carries a charge, or chirality, that is defined by  sgn$[\mathbf v_1 \cdot(\mathbf v_2 \times \mathbf v_3)]$.  
More mathematically, we can identify WPs with monopoles of "Berry flux", defined by 
\begin{align}
  \label{eq:BerryFlux}
  \mathcal F&=\nabla_{\mathbf k}\times \mathcal{A}(\mathbf k)\nonumber\\
  \mathcal{A}&=\sum_{n} i\langle u_{n,\mathbf k}|\nabla_{\mathbf k}|u_{n,\mathbf k}\rangle, 
\end{align}
where the summation in the Berry connection $\mathcal A$ runs over all occupied bands and $|u_{n\mathbf k}\rangle$ denotes the Bloch state of band $n$ at momentum $\mathbf k$.
As such, WPs are topological objects and can only be gapped out in pairs of opposite chirality. 
In absence of scattering between the WPs, this is only possible by bringing a pair of them with opposite chirality to the same point in momentum space, where they then can annihilate each other. 
Throughout this paper, we will mark Weyl points of positive/negative chirality by red/yellow dots, respectively. 

Due to the overall (Berry) charge neutrality, Weyl points always occur in pairs.
For (8,3)b, however, we find that Weyl points need to occur in multiples of four as long as time-reversal remains intact.
As particle-hole symmetry maps a Weyl node at $\bk$ to a Weyl node of \emph{opposite} chirality at $-\bk$, while time-reversal maps it to a Weyl node of \emph{the same} chirality at $-\bk +\kk$. 
Thus, there are in total four Weyl points, located at $\pm\bk$ and $\pm \bk+\kk$. 
At the isotropic point, we find positively charged Weyl points at $W_1 = (5/8)\mathbf q_1 + (3/4)\mathbf q_2 +(3/8)\mathbf q_3$ and $W_2 = -\mathbf q_1/8+\mathbf q_2/4+\mathbf q_3/8$
and negatively charged Weyl points at $W_3 = \mathbf q_1/8-\mathbf q_2/4-\mathbf q_3/8$ and $W_4 = -(5/8)\mathbf q_1 - (3/4)\mathbf q_2 -(3/8)\mathbf q_3$.
These four Weyl points are visualized in Fig.~\ref{fig:phase_diagram_8_3b}(c).

\begin{figure*}
  \includegraphics[width=\linewidth]{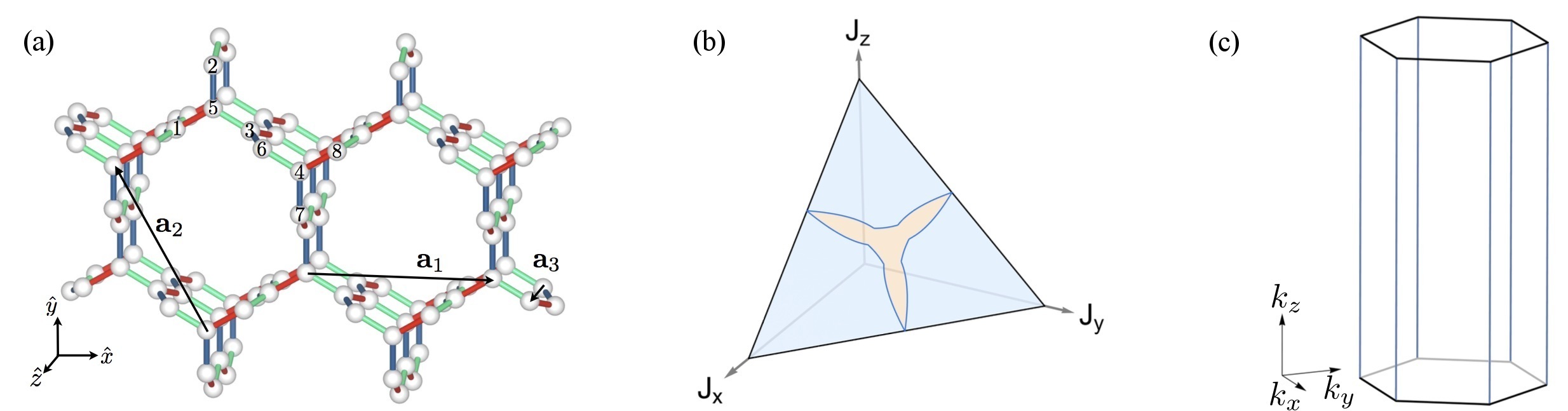}
  \caption{(Color online) (a) Unit cell and translation vectors for the lattice (8,3)c. (b) Phase diagram of the Kitaev model on (8,3)c. Around the isotropic point, there is a gapless phase with line nodes. (c) The line nodes (marked in blue) are located precisely at the edge of the Brillouin zone for  isotropic couplings. }
  \label{fig:phase_diagram_8_3c}
\end{figure*}

The charge or chirality of a WP can be measured by computing the Chern number on an arbitrary closed surface (in momentum space) around it. 
An alternative setup is computing the Chern number on parallel planes and observing its jump when moving the plane "through" the WP. 
That these two methods give the same information on the WP, can readily be seen by noting that a closed surface can be smoothly deformed into  a pair of planes, albeit with  normal vectors that point in opposite directions. 
Reversing the normal vector of one of the planes is equivalent to changing its Chern number, thus identifying the Chern number of the closed surface to the \emph{difference} in  Chern number of two planes on either side of the WP. 
The latter method is slightly easier to implement numerically and will be the method of choice in this paper. 
For the lattice (8,3)b, the Chern number as a function of $\bk_3$  is shown in Fig.~\ref{fig:chern_numbers_8_3b}. 
The left-hand side shows the Brillouin zone with the position of the Weyl points and planes at positions $\bk=0,\pm 1/4$, the right-hand-side shows the corresponding Chern number. 
As a guide to the eye, we marked the positions of the Weyl points and the three example planes in the plot. 
Note that at each position of the Weyl points, the Chern number jumps by its respective charge. 

Figure~\ref{fig:fermi_arcs_8_3b} (a)  shows the evolution of the Weyl nodes in the 3D Brillouin zone as the exchange couplings are varied with $J_x = J_y = (1-J_z)/2$ and $0 \leq J_z \lesssim 0.43$.
The position of Weyl nodes of \textit{negative} chirality is marked by colors changing from yellow to green as $J_z$ is increased,  while Weyl nodes of \textit{positive} chirality are marked by colors changing from red to green.
As $J_z$ is increased, Weyl nodes of opposite chirality are seen to move towards each other, ultimately meeting and mutually annihilating for $J_z \approx 0.43$ at $\bq=(0,0,0)$ and $\kk$.
For decreasing $J_z$ the Weyl nodes move in the Brillouin zone, but rather than meeting and annihilating at isolated momenta, the velocity vectors of the isolated Weyl points approach zero, collapsing the bulk gap.
Figure~\ref{fig:fermi_arcs_8_3b} (b) shows the associated Fermi arc surface states in the 001-surface Brillouin zone, visualized in Fig.~\ref{fig:fermi_arcs_8_3b} (c), for a slab geometry. 
Fermi arcs are exact zero-energy surface modes that connect Weyl points of opposite chirality.
Similar to the Weyl points themselves, the Fermi arcs are also topologically protected. 
As long as the Weyl points remain intact, no disorder or any other type of perturbation can gap these surface states. 
As the Weyl nodes move around, the Fermi arcs are seen to deform, ultimately shrinking to nothing as the Weyl nodes meet and annihilate for $J_z \approx 0.43$.
For $J_z \approx 0.201$, the Fermi arcs are seen to cross each other in the surface Brillouin zone. 
As $J_z$ is increased further, the Fermi arcs split once again. 
While they still connect the same pairs of Weyl nodes as before, they now wind differently around the Brillouin zone.
This splitting/reconnecting of Fermi arcs is purely a surface effect (see Appendix~\ref{ssec:supplemental_8_b} for details). 

Breaking time-reversal symmetry with Eq.~\eqref{eq:kappa} also causes the Weyl nodes to wander around the Brillouin zone.
Figure~\ref{fig:weyl_points_8_3b} (a) shows the evolution of the Weyl nodes for $0\leq \kappa \leq 0.25$.
At $\kappa = 3\sqrt{3}-5 \approx 0.2$, two Weyl nodes of opposite chirality meet and mutually annihilate at a high-symmetry point in the Brillouin zone.
In Fig.~\ref{fig:weyl_points_8_3b} (b) is pictured the corresponding evolution of the Fermi arcs in the 001-surface Brillouin zone.
As $\kappa$ is increased from $0$ to $0.2$, the Fermi arcs become more warped as two Weyl nodes of opposite chirality move towards each other.
As $\kappa$ is increased further, the two Fermi arcs touch at a high-symmetry point and become one.
For still larger values of $\kappa$, even more Weyl nodes begin to appear in charge-neutral pairs while others pairs mutually annihilate.

\subsection{(8,3)c}
\label{ssec:8c}

\paragraph*{Lattice structure.--}
The lattice (8,3)c can be viewed most simply as parallel zig-zag chains along the $\hat z$ direction that are coupled by vertices lying in the $x$-$y$ plane. 
It is a hexagonal lattice with eight sites per unit cell at positions
\begin{align}
  \br_1 &= \left( -\frac{1}{5}, \frac{4}{5\sqrt{3}}, \frac{1}{10} \right)\,,
  &\br_2 &= \left( 0, \frac{7}{5\sqrt{3}}, \frac{1}{10} \right)\,, \nonumber\\
  \br_3 &= \left( \frac{1}{5}, \frac{4}{5\sqrt{3}}, \frac{1}{10} \right)\,,
  &\br_4 &= \left( \frac{1}{2}, \frac{1}{2\sqrt{3}}, \frac{3}{10} \right)\,, \nonumber\\
  \br_5 &= \left( 0, \frac{1}{\sqrt{3}}, \frac{1}{10} \right)\,,
  &\br_6 &= \left( \frac{3}{10}, \frac{7}{10\sqrt{3}}, \frac{3}{10} \right)\,, \nonumber\\
  \br_7 &= \left( \frac{1}{2}, \frac{1}{10\sqrt{3}}, \frac{3}{10} \right)\,,
  &\br_8 &= \left( \frac{7}{10}, \frac{7}{10\sqrt{3}}, \frac{3}{10} \right)\,,
\end{align}
and lattice vectors
\begin{align}
  \ba_1 &= \left( 1, 0, 0 \right),&
  \ba_2 &= \left( -\frac{1}{2}, \frac{\sqrt{3}}{2}, 0 \right),&
  \ba_3 &= \left( 0, 0, \frac{2}{5} \right).
\end{align}
The corresponding reciprocal lattice vectors are given by 
\begin{align}
  \bq_1 &= \left( 2\pi, \frac{2\pi}{\sqrt{3}}, 0 \right),&
  \bq_2 &= \left( 0, \frac{4\pi}{\sqrt{3}}, 0 \right),&\nonumber \\
  \bq_3 &= \left( 0, 0, 5\pi \right).
  \label{eq:recip_vec_8_3c}
\end{align}
The unit cell and the  lattice translation vectors are visualized in Fig.~\ref{fig:phase_diagram_8_3c}(a). 
When choosing the assignment of bond types on this lattice, we notice that for each of the chains  there are two possible choices that are in general inequivalent.  
We chose an assignment of bonds such that threefold rotation around the $\hat z$ axis combined with a cyclic permutation of $x$,$y$, and $z$ bonds is a symmetry of the Hamiltonian. 
This is the most natural choice as it ensures that the  phase diagram remains symmetric under interchange of the $J_x$, $J_y$, and $J_z$ couplings.

\begin{figure}
  \includegraphics[width=\linewidth]{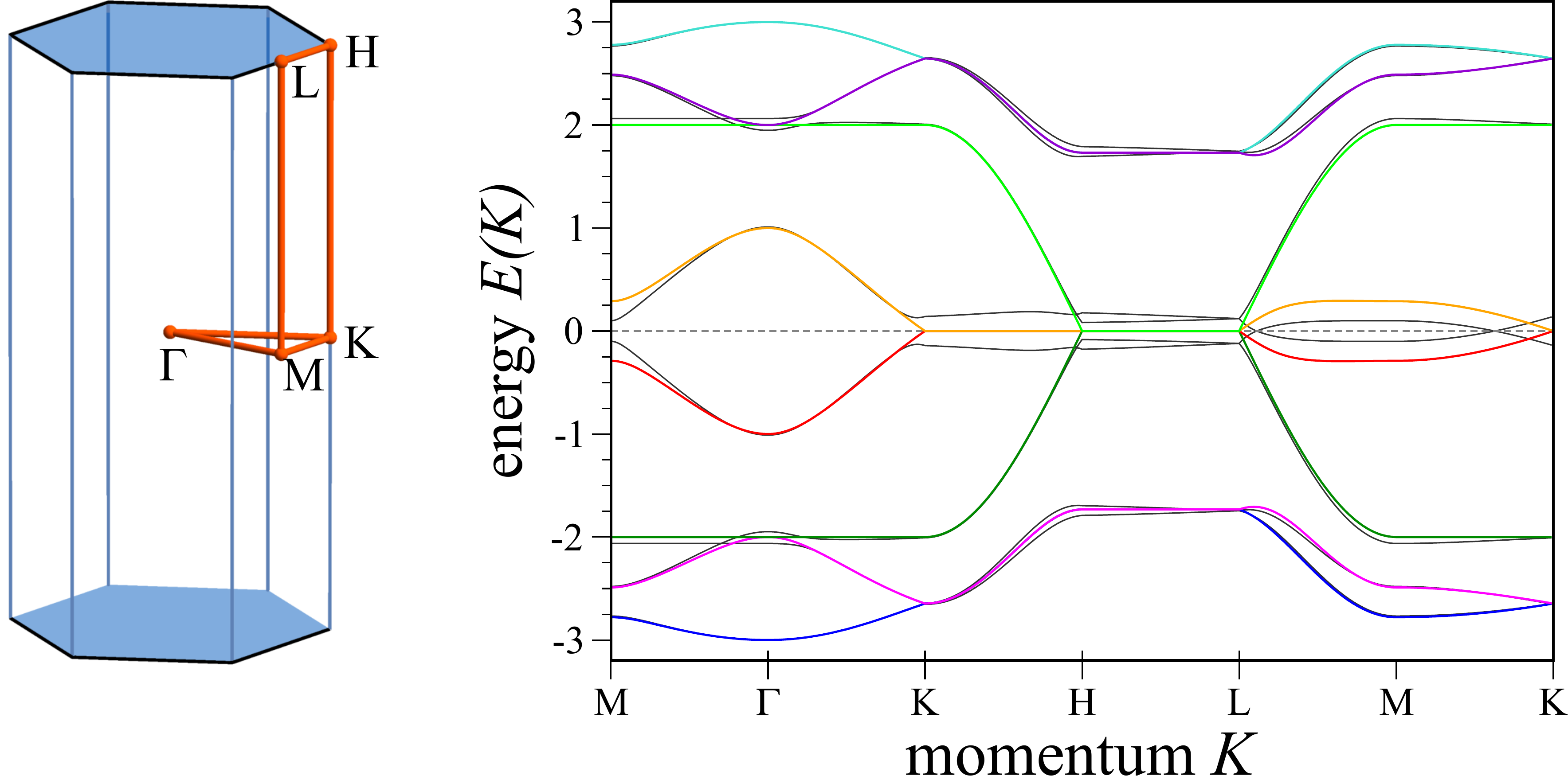}
  \caption{(Color online) The left-hand-side shows the Brillouin zone with high-symmetry points, and the gapless modes indicated in blue for (8,3)c.  The right-hand-side shows the energy dispersion along high-symmetry lines for isotropic couplings (colored) and anisotropic couplings $J_x=3/8$ and $J_z=J_y=(1-J_x)/2=5/16$ (gray). The zero-energy surface immediately gaps completely when departing from the isotropic point. }
  \label{fig:energydispersion8c}
\end{figure}
%

\paragraph*{Gauge structure.--}
For this lattice there are four linearly independent loop operators of length 8.
Following the guidance of Lieb's theorem \cite{Lieb}, one would like to assign $\pi$ flux through all of these plaquettes.
However, this is not possible as the minimal volume constraints on the loop operators indicate that only two out of three eight-length loops in such a volume can have $\pi$ flux (as explained in more detail in Appendix \ref{ssec:supplemental_8_c}).
As such, the minimal volume constraints induce geometric frustration in assigning these $\pi$ fluxes, which gives rise to a macroscopic number of possible \Z\ gauge configurations.
While numerical tests indeed support that the ground state of the \Z\ gauge theory becomes macroscopically degenerate for increasing system sizes, we dismiss this scenario in the following discussion of the Majorana physics.
Instead, we consider the flux sector where all loop operators have eigenvalue $+1$ instead.
This turns out to be the only flux configuration that obeys all the lattice symmetries, i.e., threefold rotation and inversion, but is not geometrically frustrated.
The reader should note that this flux configuration is \textit{not} the ground state sector.
Note, however, that one can always stabilize this flux sector as the ground state by adding terms that penalize $\pi$-flux through plaquettes, similarly to what was done in Ref.~\cite{LaiMotrunich} for the Kitaev model on the two-dimensional square-octagon lattice. 

\paragraph*{Projective symmetries.--}
As the translation vectors of this lattice are identical to those of its two sublattices, sublattice symmetry and time-reversal symmetry are implemented trivially, \ie, with $\kk=0$.
In addition, this lattice possesses inversion symmetry with vanishing $\kt$.
Thus, the relevant energy relations are given by
\begin{align}
  \epsilon(\bk) &= -\epsilon(-\bk)& \mbox{ and }&&
  \epsilon(\bk) &= \epsilon(-\bk),
\end{align}
where the first relation comes from particle-hole symmetry and the second from either time-reversal or inversion symmetry. 

\paragraph*{Majorana metal.--}
The system exhibits a gapless phase around the isotropic point, where the gapless modes form closed {\em nodal lines} in accordance with our projective symmetry analysis. 
At the isotropic point, these lines are located at  $( \pm\frac{4\pi}{3}, 0, k_z )$, as pictured in Fig.~\ref{fig:phase_diagram_8_3c}(c).
In addition, at the isotropic point one also finds a gapless surface at the top and bottom boundaries of the Brillouin zone, $\bq = (k_x,k_y,\pm 5\pi/2)$. 
In contrast to the nodal line, the surface is not a stable zero-mode manifold, as can readily be inferred from time-reversal invariance. 
Any infinitesimal changes in the coupling constants immediately gap the surface, while merely deforming the nodal line. 
This becomes clearly visible in the energy dispersion plot along the high-symmetry lines, shown in Fig.~\ref{fig:energydispersion8c}. 
The colored bands denote the energy dispersion at the isotropic point. 
The zero-energy modes at the top and bottom surfaces of the Brillouin zone gap out immediately when departing from the isotropic point, which is exemplified by the energy dispersion at $J_x=0.375$ and $J_z=J_y=(1-J_x)/2$ marked in gray. 

As discussed in Sec.~\ref{sec:symmetries}, the nodal line is protected by time-reversal symmetry. 
Thus, breaking time-reversal symmetry even infinitesimally causes the Fermi line to gap out almost entirely, leaving just six Weyl nodes which are fixed to zero energy by inversion symmetry.
A similar behavior was studied in Ref.~\cite{wsl2014} for the lattice (10,3)b, discussed in Sec.~\ref{ssec:10b}. 
The appearance of Weyl points is far from being a coincidence.
In fact, it can be argued that time-reversal symmetry \emph{cannot} gap the nodal line completely and Weyl nodes have to occur. 
A simple way to see this is by regarding the 3D model as a 2D model with an additional (dimensional) parameter, \ie, we will view two of the momentum directions as physical and the third, for example, $k_z$ in this case, as a parameter that we can tune. 
Choosing $k_z$ such that the plane of physical momenta does not cut through the nodal line yields an effective two-dimensional system that is a trivial insulator \cite{FootnoteChernNumberArgument}. 
If the plane of physical momenta instead \textit{does} cut the line, the effective system will be gapless and exhibit two-dimensional Dirac points, in complete analogy to the Kitaev honeycomb model in the gapless phase. 
Breaking time-reversal symmetry has different effects for these two cases. 
While it leaves the trivial insulator qualitatively unchanged, it induces a non-trivial gap for the Dirac points and turns the effective system into a \emph{topological Chern insulator}. 
In particular, the latter has non-vanishing Berry flux through the two-dimensional Brillouin zone. 
Thus, we found that changing the momentum parameter in absence of time-reversal symmetry, tunes the system between a trivial insulator and a Chern insulator. 
From general arguments, we know that at the boundary between these two types of insulators, the system has to be gapless.
What is more, the gapless mode has to be a source/sink of Berry flux in order to create the non-vanishing Berry flux in the topological part -- \ie a Weyl point. 
Note that due to the periodicity of the Brillouin zone, these Weyl points always occur in pairs of opposite chirality. 

\begin{figure}
  \includegraphics[width=\linewidth]{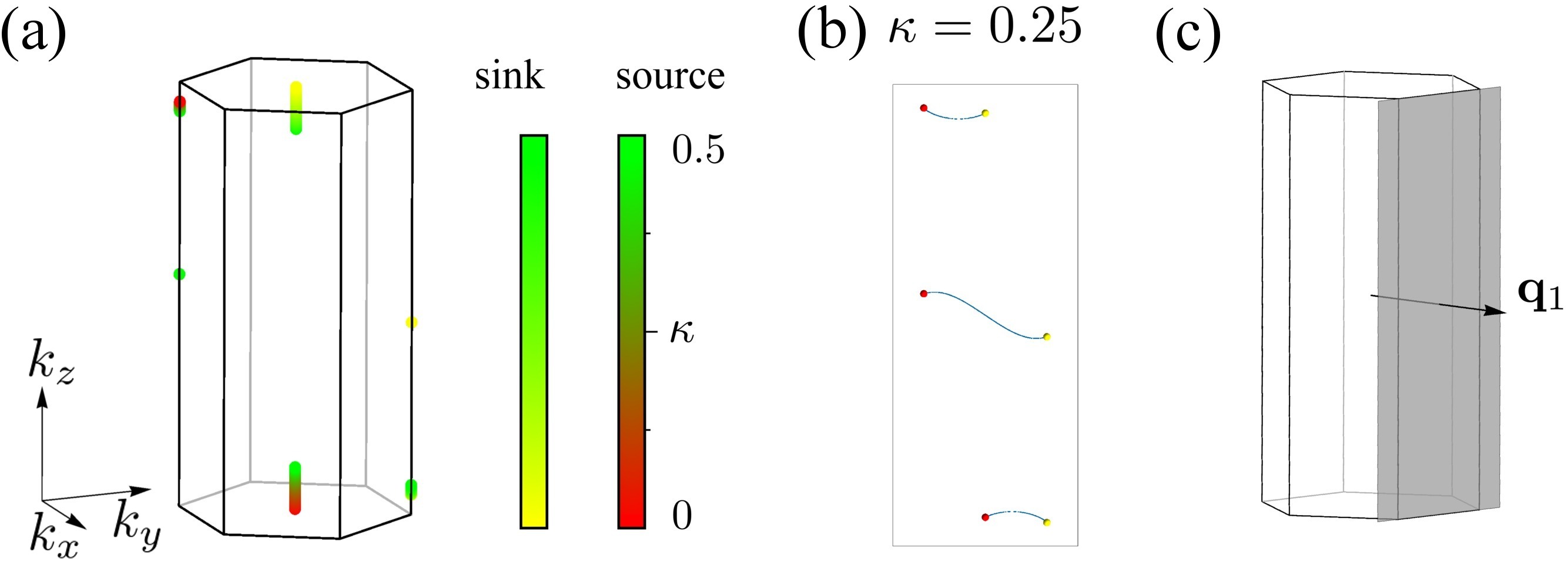}
  \caption{(Color online) (a) Evolution of the Weyl nodes for (8,3)c in presence of a magnetic field for varying $\kappa$ from $0,\ldots,0.5$. (b) Corresponding Fermi arc plot for $\kappa=0.25$. (c) Visualization of the surface Brillouin zone for open boundary conditions in the 100-direction.}
  \label{fig:weyl_points_8_3c}
\end{figure}

As noted above, breaking time-reversal symmetry infinitesimally gaps the Fermi line with the exception of six zero energy Weyl nodes.
Pictured in Fig.~\ref{fig:weyl_points_8_3c} (a) is the evolution of these Weyl nodes for $0 < \kappa \leq 0.5$.
For this range of $\kappa$, the Weyl points move very little in the Brillouin zone along high-symmetry lines.
However, for larger values of $\kappa$, many Weyl points appear in charge-neutral pairs while other pairs mutually annihilate.
Figure~\ref{fig:weyl_points_8_3c} (b) shows the corresponding Fermi arcs in the 100-surface Brillouin zone for $\kappa=0.25$.
Figure~\ref{fig:weyl_points_8_3c} (c) illustrates the projection of the Weyl nodes onto the 100-surface Brillouin zone.

%
\begin{figure*}
  \includegraphics[width=\linewidth]{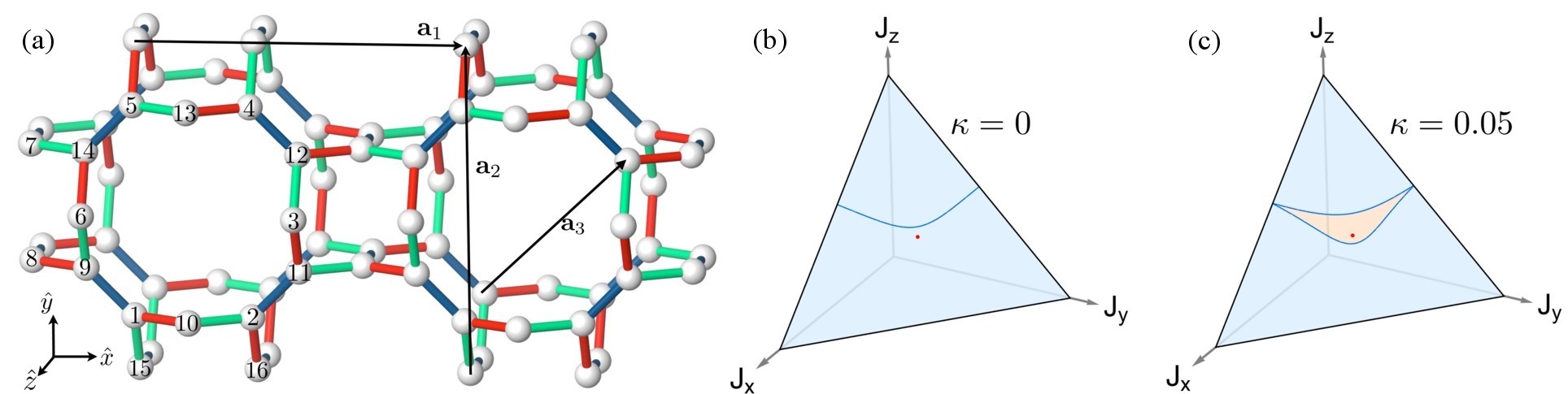}
  \caption{(Color online) (a) Unit cell and translation vectors for the (8,3)n lattice. (b) The Kitaev model on the (8,3)n lattice has no gapless phase. The blue line indicates the phase transition between the two distinct gapped phases, and the red dot marks the isotropic point. (c) Phase diagram for $\kappa=0.05$. For finite $\kappa$, the Dirac nodes split into pairs of oppositely charged Weyl nodes, and the phase transition line evolves to an entire phase of a gapless Weyl spin liquid, marked in light orange. }
  \label{fig:phase_diagram_8_3n}
\end{figure*}

\subsection{(8,3)n}
\label{ssec:8n}

Finally, we turn to the (8,3)n lattice.
This lattice can be viewed as a three-dimensional generalization of the square-octagon lattice, 
where layers of square-octagon lattices are coupled via mid-bond sites. 

\paragraph*{Lattice structure.--}
The (8,3)n lattice is a tetragonal lattice with 16 sites per unit cell.
In order to simplify the notation, we denote all vectors in terms of
\begin{align}
         \ba &= \left( 1, 0, 0 \right),&
  \mathbf{b} &= \left( 0, 1, 0 \right),&
  \mathbf{c} &= \left( 0, 0, \frac{4}{2\sqrt{3} + \sqrt{2}} \right).
\end{align}
The site positions in the unit cell can be written as
\begin{align}
  \br_1    &= x\cdot\ba + \left(\frac{1}{2}-x\right)\cdot\mathbf{b} + \frac{1}{4}\cdot\mathbf{c} \,, \nonumber\\
  \br_2    &= \left(1-x\right)\cdot\ba + \left(\frac{1}{2}-x\right)\cdot\mathbf{b} + \frac{1}{4}\cdot\mathbf{c} \,, \nonumber\\
  \br_3    &= \left(\frac{1}{2}+x\right)\cdot\ba + \frac{1}{2}\mathbf{b} + \left(\frac{1}{2}-z\right)\cdot\mathbf{c} \,, \nonumber\\
  \br_4    &= \left(1-x\right)\cdot\ba + \left(\frac{1}{2}+x\right)\cdot\mathbf{b} + \frac{1}{4}\cdot\mathbf{c} \,, \nonumber\\
  \br_5    &= x\cdot\ba + \left(\frac{1}{2}+x\right)\cdot\mathbf{b} + \frac{1}{4}\cdot\mathbf{c} \,, \nonumber\\
  \br_6    &= \left(\frac{1}{2}-x\right)\cdot\ba + \frac{1}{2}\cdot\mathbf{b} + \left(\frac{1}{2}-z\right)\cdot\mathbf{c} \,, \nonumber\\
  \br_7    &= \left(1-x\right)\cdot\mathbf{b} + z\cdot\mathbf{c} \,, \nonumber\\
  \br_8    &= x\cdot\mathbf{b} + z\cdot\mathbf{c} \,, \nonumber\\
  \br_9    &= \left(\frac{1}{2}-x\right)\cdot\ba + x\cdot\mathbf{b} + \frac{1}{4}\cdot\mathbf{c} \,, \nonumber\\
  \br_{10} &= \frac{1}{2}\cdot\ba + \left(\frac{1}{2}-x\right)\cdot\mathbf{b} + \left(\frac{1}{2}-z\right)\cdot\mathbf{c} \,, \nonumber\\
  \br_{11} &= \left(\frac{1}{2}+x\right)\cdot\ba + x\cdot\mathbf{b} + \frac{1}{4}\cdot\mathbf{c} \,, \nonumber\\
  \br_{12} &= \left(\frac{1}{2}+x\right)\cdot\ba + \left(1-x\right)\cdot\mathbf{b} + \frac{1}{4}\cdot\mathbf{c} \,, \nonumber\\
  \br_{13} &= \frac{1}{2}\cdot\ba + \left(\frac{1}{2}+x\right)\cdot\mathbf{b} + \left(\frac{1}{2}-z\right)\cdot\mathbf{c} \,, \nonumber\\
  \br_{14} &= \left(\frac{1}{2}-x\right)\cdot\ba + \left(1-x\right)\cdot\mathbf{b} + \frac{1}{4}\cdot\mathbf{c} \,, \nonumber\\
  \br_{15} &= x\cdot\ba + z\cdot\mathbf{c} \,, \nonumber\\
  \br_{16} &= \left(1-x\right)\cdot\ba + z\cdot\mathbf{c} \,,
\end{align}
with $x = \frac{\sqrt{3} + \sqrt{2}}{2(2\sqrt{3} + \sqrt{2})}$ and $z = \frac{1}{8}$.
The lattice vectors are given by
\begin{align}
  \ba_1 = \ba,&\qquad
  \ba_2 = \mathbf{b},&
  \ba_3 = \frac{1}{2}(\ba + \mathbf{b} + \mathbf{c}),
\end{align}
with reciprocal lattice vectors
\begin{align}
  \bq_1 &= \left( 2\pi, 0, -\sqrt{\frac{7}{2}+\sqrt{6}}\pi \right),&\nonumber \\
  \bq_2 &= \left( 0, 2\pi, -\sqrt{\frac{7}{2}+\sqrt{6}}\pi \right),&\nonumber \\
  \bq_3 &= \left( 0, 0, 2\sqrt{\frac{7}{2}+\sqrt{6}}\pi \right).
  \label{eq:recip_vec_8_3n}
\end{align}
The unit cell and the lattice translation vectors are shown in Fig.~\ref{fig:phase_diagram_8_3n}(a).
The assignment of bonds was chosen in order to be compatible with fourfold rotation symmetry and inversion. 
It is the unique such choice up to an overall permutation of the $x$, $y$, and $z$ bonds. 
As can be seen from the figure,  $x$- and $y$-bonds are related by lattice symmetries, but the $z$-bonds map only to themselves. 
As a result, the phase diagram is symmetric in interchanging $J_x$ and $J_y$, only.

\paragraph*{Gauge structure.--}
The (8,3)n lattice has eight linearly independent loop operators per unit cell of lengths 8 and 10, respectively.
Lieb's theorem \cite{Lieb} can be faithfully applied for all but a single loop of length eight to determine the flux configuration of the ground state. 
An example of the relevant mirror planes to establish this result is shown in Fig.~\ref{fig:Lieb} (c). 
Using numerical calculations, we verified that also the remaining loop operator has eigenvalue $-1$ in the ground state.

\paragraph*{Projective symmetries.--}
Sublattice symmetry and, consequently, time-reversal symmetry are implemented trivially for the Majorana fermions, \ie, $\kk=0$. 
However, when implementing inversion symmetry, we need to supplement it with a gauge transformation that (artificially) enlarges the unit cell in the $\mathbf{a_3}$-direction. 
Thus, inversion symmetry involves a translation in momentum space by $\kt=\mathbf q_3/2=(0,0,\sqrt{\frac{7}{2} + \sqrt{6}}\pi)$. 

\begin{figure}
  \includegraphics[width=\linewidth]{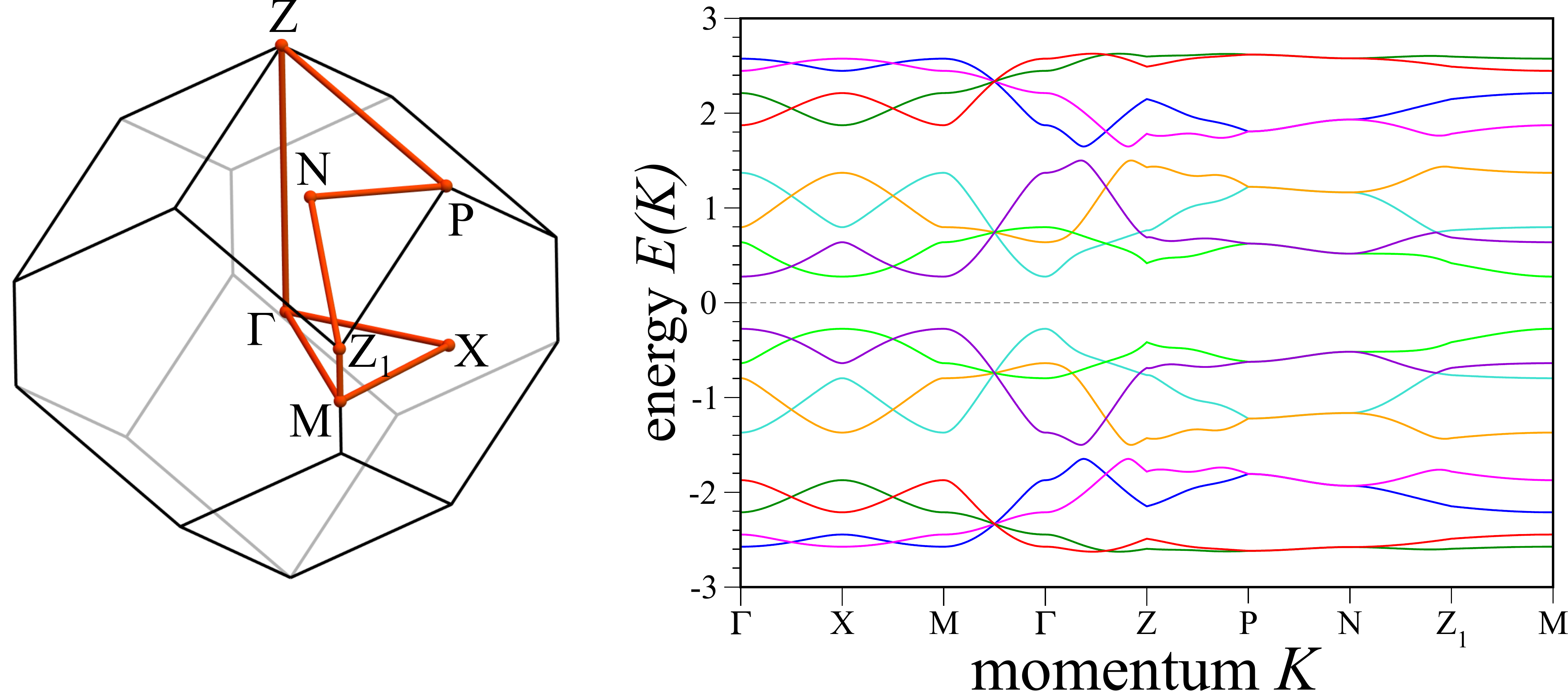}
  \caption{(Color online) Brillouin zone with high-symmetry points and energy dispersion along the corresponding high-symmetry lines for (8,3)n. The spectrum at the isotropic point is fully gapped. }
  \label{fig:energydispersion8n}
\end{figure}
\begin{figure}
  \includegraphics[width=\linewidth]{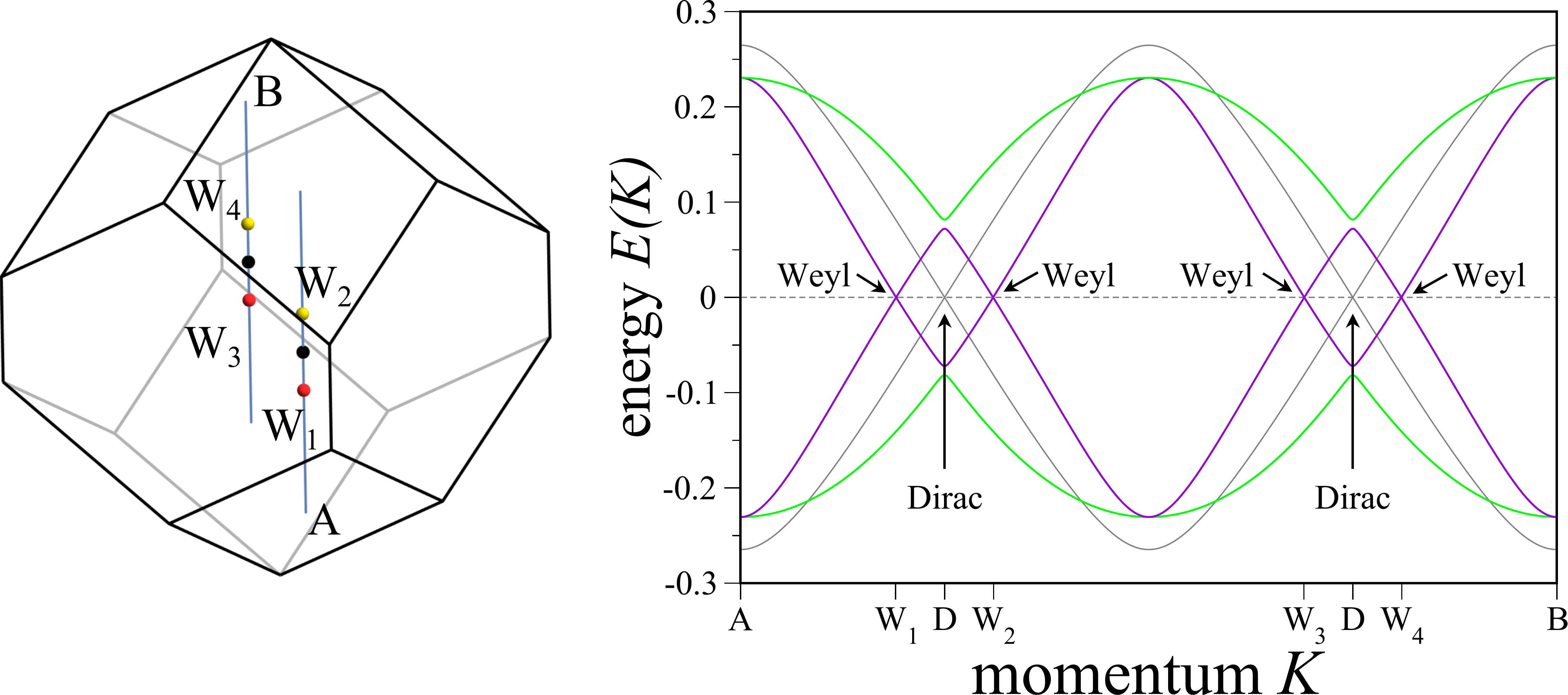}
  \caption{(Color online) Brillouin zone with a path cutting through the Dirac/Weyl nodes 
     			appearing in the energy dispersion for (8,3)n for $\kappa = 0$ (gray lines) and $\kappa = 0.05$ (colored lines), respectively. }
  \label{fig:energydispersion8nII}
\end{figure}
\begin{figure*}[th]
  \includegraphics[width=\linewidth]{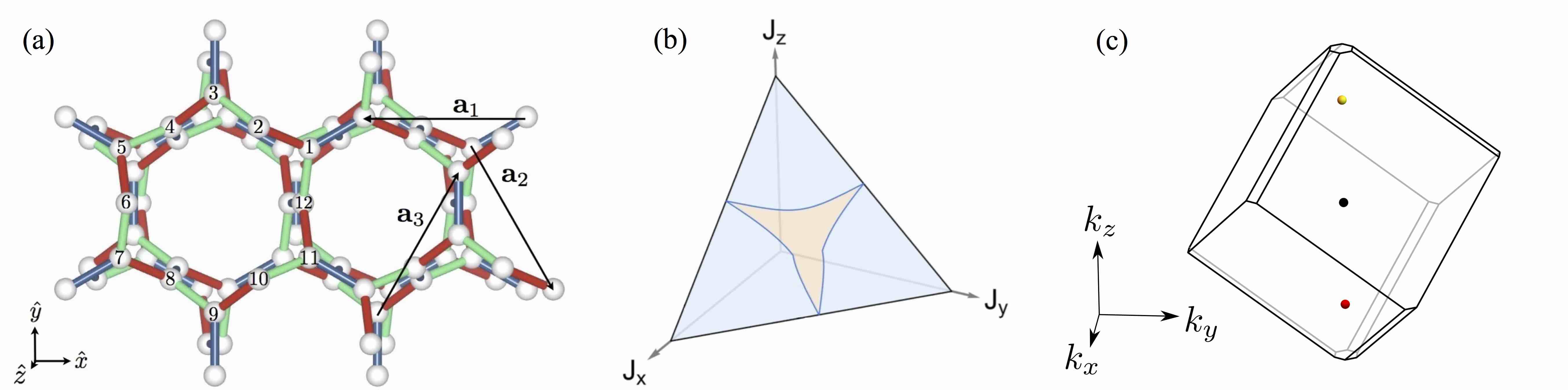}
  \caption{(Color online) (a) Unit cell and translation vectors for the (9,3)a lattice. (b) The phase diagram is symmetric for $J_x\leftrightarrow J_y$. (c) Brillouin zone with position of the gapless Weyl nodes -- red denotes a source, yellow a sink, and black a neutral combination of several Weyl nodes.  }
  \label{fig:comb_9_a}
\end{figure*}
%

\paragraph*{Phase diagram.--}
The (8,3)n lattice is the only one of the lattices considered in this paper that does not exhibit a gapless phase in its phase diagram as shown in Fig.~\ref{fig:phase_diagram_8_3n}(b). 
Instead there are two gapped phases, with Fig.~\ref{fig:energydispersion8n} showing an example of a gapped dispersion relation for the isotropic coupling point $J_x = J_y = J_z$.
The two gapped phases are separated by a line of phase transitions, at which the dispersion exhibits a three-dimensional \emph{Dirac cone} structure of two doubly degenerate bands (see Fig.~\ref{fig:energydispersion8nII}).
One way of thinking of such  three-dimensional Dirac cones is that they are in fact a combination of two (oppositely charged) Weyl cones of singly degenerate bands.
One way of splitting these Weyl cones is by breaking time-reversal symmetry.
Indeed, we find that upon applying a magnetic field along the 111 direction the Dirac cones split into two Weyl nodes each as illustrated in Fig.~\ref{fig:energydispersion8nII} for a small coupling strength $\kappa = 0.05$.
As a result, an extended gapless phase emerges in the phase diagram (around the original line of phase transitions) as illustrated in the phase diagram of Fig.~\ref{fig:phase_diagram_8_3n} (c).
This gapless phase (for small values of $\kappa$) is thus a Weyl spin liquid as discussed already in the context of lattices (8,3)b and (8,3)c.
One should, however, note that no projective symmetry protects this Weyl spin liquid (as it is the case for the other lattices).
In particular, due to the non-trivial implementation of inversion symmetry, the Weyl nodes are not generically fixed to zero energy.
While for small time-reversal breaking strength $\kappa$ they are found to remain strictly at zero energy and move exclusively along high-symmetry lines in the Brillouin zone, this is no longer true for large values of $\kappa$, where the system eventually develops  Majorana Fermi surfaces.
Note that the emerging Fermi surfaces are related by a perfect nesting vector $\kt$ as long as inversion symmetry remains intact. 
 
 \subsection{(9,3)a}
\label{ssec:9a}
We now turn to the (9,3)a lattice, which stands out in our family of tricoordinated, three-dimensional lattices of Table \ref{tab:lattice_overview} as the only lattice with an {\em odd} number of bonds in the elementary loops.
This oddness has important consequences in particular with regard to the projective time-reversal symmetry of the Kitaev model.
While the original spin model is time-reversal symmetric, this is no longer the case for the effective Majorana model, \ie, the effective Majorana model breaks time-reversal symmetry {\em spontaneously}.
This has direct consequences for the emergent Majorana metal as we will discuss in the following.
 
\paragraph*{Lattice structure.--}
The (9,3)a lattice is one of the more complicated lattices with 12 sites per unit cell. 
In order to simplify notation, we denote all vectors in terms of 
\begin{align}
  \mathbf a&=(1,0,0), &\mathbf b&=\left(-\frac 1 2 ,\frac {\sqrt 3} 2,0\right), & \nonumber\\ \mathbf c &=\left(0,0,\frac{\sqrt{6(4+\sqrt{3})}}{1+2\sqrt{3}}\right). 
\end{align}
The complicated value for $\mathbf c$ is needed in order to obtain a lattice with approximately 120 degree bond angles and equal length bonds. 
The site positions in the unit cell  can be written as
\begin{align}
  \mathbf r_1&=\delta_f \cdot  \mathbf a \,,&  
  \mathbf r_2&= 2\delta_h\cdot\mathbf a +\delta_h\cdot\mathbf b +\frac 1 {12} \cdot\mathbf c \,, \nonumber \\
  \mathbf r_3&= \delta_f \cdot  (\mathbf a +\mathbf b) , &
  \mathbf r_4&= \delta_h\cdot\mathbf a+2\delta_h\cdot\mathbf b-\frac 1 {12}\cdot\mathbf c \,, \nonumber \\
  \mathbf r_5&= \delta_f \cdot  \mathbf b \,,&
  \mathbf r_6&= -\delta_h\cdot\mathbf a+\delta_h\cdot\mathbf b+\frac 1 {12}\cdot\mathbf c \,,  \nonumber \\
  \mathbf r_7&=-\delta_f \cdot  \mathbf a \,, &
  \mathbf r_8&=-2\delta_h\cdot\mathbf a -\delta_h\cdot\mathbf b -\frac 1 {12}\cdot \mathbf c \,, \nonumber \\
  \mathbf r_9&=  -\delta_f \cdot  (\mathbf a +\mathbf b) \,,&
  \mathbf r_{10}&= -\delta_h\cdot\mathbf a-2\delta_h\cdot\mathbf b+\frac 1 {12}\cdot\mathbf c \,,\nonumber \\
  \mathbf r_{11}&= - \delta_f \cdot  \mathbf b \,, &
  \mathbf r_{12}&= \delta_h\cdot\mathbf a-\delta_h\cdot\mathbf b-\frac 1 {12}\cdot\mathbf c \,,
\end{align}
with $\delta_f=\frac{\sqrt 3}{1+2\sqrt 3}\approx 0.388$, $\delta_h =\frac{29-3\sqrt{3}}{132}\approx 0.18033$.
The translation vectors are given by 
\begin{align}
  \mathbf a_1&=-\frac 1 3 \mathbf a +\frac 1 3 \mathbf b +\frac 1 3 \mathbf c \,, \nonumber\\
  \mathbf a_2 &=-\frac 1 3 \mathbf a -\frac 2 3 \mathbf b +\frac 1 3 \mathbf c \,, \nonumber\\
  \mathbf a_3&=\frac 2 3 \mathbf a +\frac 1 3 \mathbf b +\frac 1 3 \mathbf c \,.
\end{align}
The unit cell and translation vectors are shown in Fig.~\ref{fig:comb_9_a}(a), as is our choice for assigning the $x$-, $y$-, and $z$-type bonds.  
Up to permutations, the bond assignment shown in Fig.~\ref{fig:comb_9_a}(a) is unique when preserving all the lattice symmetries.
While $x$ and $y$ bonds are related by mirror symmetries,  the $z$ bonds are special. 
This implies that the physics remains unchanged for interchanging $J_x\leftrightarrow J_y$, and the phase diagram  is symmetric in $J_x $ and $J_y$. 

\begin{figure}[b]
  \includegraphics[width=.65\linewidth]{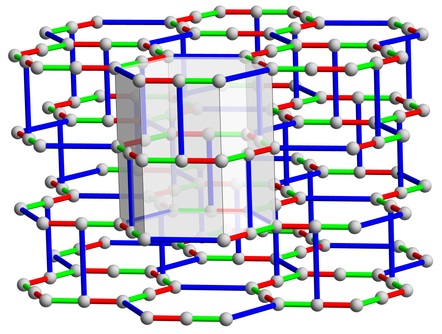}
  \caption{(Color online) A deformed version of the (9,3)a lattice can be obtained by coupling honeycomb layers via mid-bond sites. The eight elementary plaquettes of length nine per unit cell are marked by the gray transparent polygons. A visualization of the plaquettes in the un-deformed (9,3)a lattice can be found in Appendix~\ref{ssec:supplemental_9_a}.  }
  \label{fig:layered_9_a}
\end{figure}
\begin{figure}[b]
  \includegraphics[width=\columnwidth]{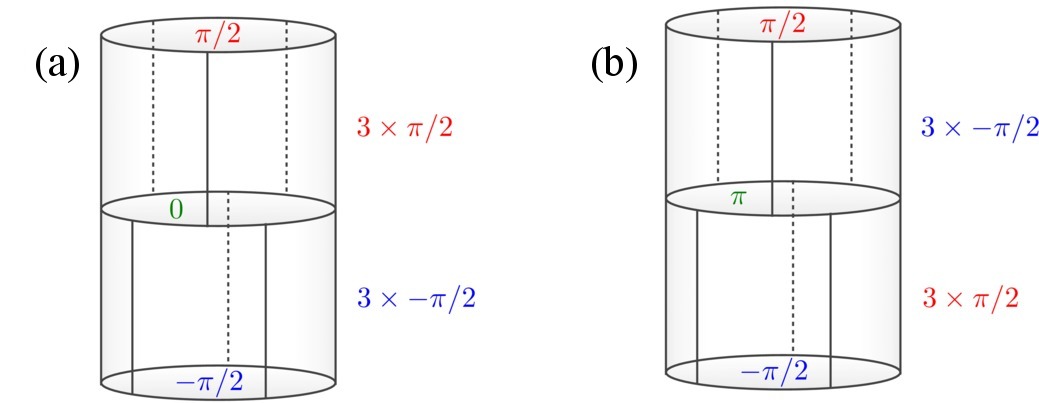}
  \caption{(Color online) Possible flux assignments for the (9,3)a lattice for the eight elementary loops shown in Fig.~\ref{fig:layered_9_a}. }
  \label{fig:CanFigure}
\end{figure}

We note that an equivalent, though deformed version of the (9,3)a lattice can be constructed by joining layers of honeycomb lattices via mid-bond sites, 
as shown in Fig.~\ref{fig:layered_9_a}.
In our subsequent discussion of the gauge structure, we will refer to this deformed lattice structure as it is somewhat easier to visualize than the undeformed one.

\paragraph*{Gauge structure, projective symmetries, and Majorana metal.--}
The (9,3)a lattice has loops with an odd number of bonds, which requires that the corresponding plaquette operators must have eigenvalues $+i$ or $-i$.
Any such flux assignment breaks time-reversal symmetry, as acting with time-reversal flips the sign of all plaquette operators without changing the energy of the eigenstate. 
Note that neither the Hamiltonian nor the plaquette operators break time-reversal symmetry --- both commute with $T$. 
Thus, the effective Majorana model breaks time-reversal symmetry spontaneously and all eigenstates come in time-reversal pairs.
A very similar scenario was put forward in the discussion of Yao and Kivelson~\cite{YaoKivelson} of a chiral spin liquid ground state emerging for a two-dimensional Kitaev model on the 3-12-12 lattice.

\begin{figure*}[ht!]
  \includegraphics[width=\linewidth]{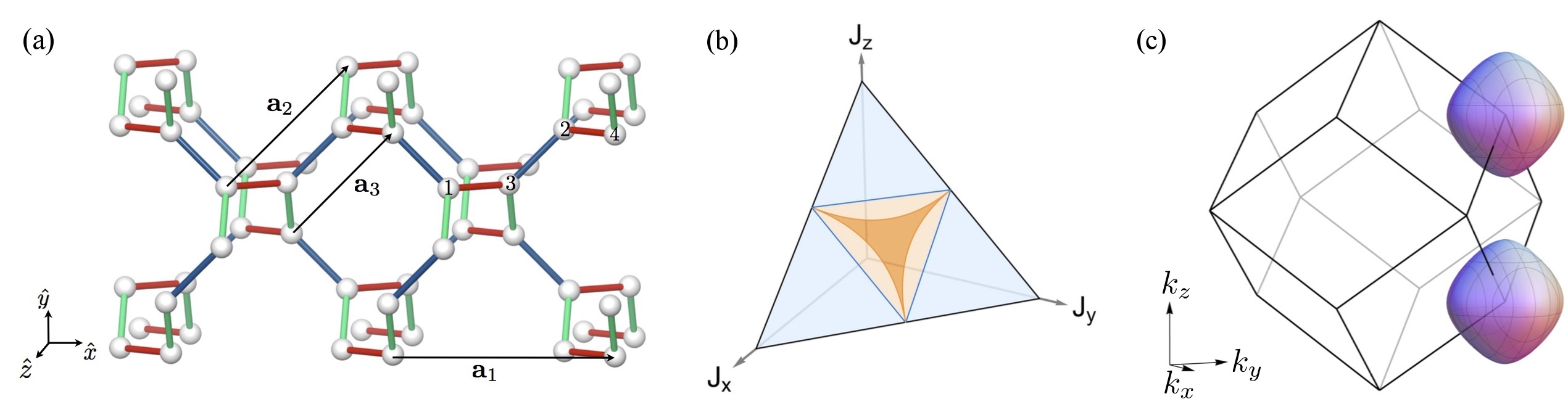}
  \caption{(Color online) (a) Visualization of the Kitaev couplings, unit cell and translation vectors for the (10,3)a lattice. (b) Phase diagram for  (10,3)a -- the gapless region is shaded orange and the gapped blue. The parameter regions shaded darker orange have topological Fermi surfaces, while the lighter orange regions have trivial Fermi surfaces. (c) Visualization of the gapless modes for isotropic couplings.}
  \label{fig:comb_10_a}
\end{figure*}

The (9,3)a lattice has eight 9-loops per unit cell -- in the following denoted by $W_j(\mathbf R)$ -- but only six of them are linearly independent due to the volume constraints shown in Appendix \ref{ssec:supplemental_9_a}.
In contrast to lattices that only have plaquettes with an even number of bonds, we need to specify a direction when assigning eigenvalues to the loop operators. 
We use the convention that the loop is traversed in the mathematically positive direction when viewed from the center of the unit cell, see Fig.~\ref{fig:CanFigure}.
Note that this convention implies that inversion maps a loop operator $W_j(\mathbf R) \rightarrow -W_{j+4}(\mathbf R)$ with the loop subscripts $j=1,\ldots,8$ as defined in Appendix \ref{ssec:supplemental_9_a}.
The threefold rotation permutes the loop operators $W_j(\mathbf R)$ with $j=1,2,3$ ($j=5,6,7$) cyclically, without changing the eigenvalues. 

There are in total two flux configurations, pictorially visualized in Fig.~\ref{fig:CanFigure},  that are compatible with all the lattice symmetries, \ie, threefold rotation around $\hat z$, inversion with respect to the center of the unit cell, as well as lattice translations. 
They differ by the flux ($0$ or $\pi$) through the 12-loop defined by 
\begin{align}
  \label{eq:12-loop}
  W_{12}(\mathbf R)&=\sigma_1^y(\mathbf R) \sigma_2^y (\mathbf R)\sigma_2^x (\mathbf R)\sigma_3^x (\mathbf R)\ldots \sigma_{12}^x (\mathbf R)\sigma_1^x (\mathbf R), 
\end{align}
\ie, the loop is given by the product of all bonds within the unit cell at position $\mathbf R$.
It turns out that the $0$-flux configuration has the lower energy of the two. 
The reader should note, however, that (at least for the system sizes we could test numerically) this symmetric flux sector is \emph{not} the ground state sector, even though the energy difference decreases with system size \cite{Footnote:Scaling}. 
Instead, flux configurations that break the threefold rotational symmetry and/or inversion symmetry appear to have slightly lower energy. 

In the following, we will briefly discuss the properties of the $0$-flux sector.
Note that we can always stabilize this sector as the ground-state sector by assigning an energy to 12-loops that are combinations of two adjacent 9-loops, similar to what was done in Ref.~\cite{Lai2011} for the Kitaev model on the square-octagon lattice (see Appendix \ref{ssec:supplemental_9_a} for details).  
The analysis in this sector turns out to be slightly tedious. 
Even though the flux configuration itself is translation invariant, it requires a \Z\ gauge that enlarges the unit cell by a factor in 2 all three lattice directions, resulting in a 96-site unit cell. 
In this flux sector, the system has an extended gapless phase around the isotropic point,  shown in Fig.~\ref{fig:comb_9_a}(b), with a varying number of gapless Weyl points, depending on the coupling constants.
When increasing one of the coupling constants sufficiently, all the Weyl points annihilate and the system becomes gapped.
At the isotropic point, we find that several of the Weyl points coincide at $\bk=0$, such that the zero-mode is eight-fold degenerate.
However, this multiple zero mode is not stable, and splits into several distinct Weyl points as soon as the coupling constants are altered.
In addition to the eightfold zero mode at $\bk=0$, there are double Weyl nodes at positions $\pm (\mathbf q_1 +\mathbf q_2 +\mathbf q_3)/3$ with charge $\mp 2$, as shown in Fig.~\ref{fig:comb_9_a}(c).
Note that the Brillouin zone  is computed for the enlarged unit cell. 
Breaking time-reversal symmetry does not change this physics qualitatively, but only moves the (double) Weyl nodes in the Brillouin zone. 

We want to emphasize that for general flux configurations that do not break inversion symmetry, the low-energy physics remains qualitatively the same as in the $0$-flux sector discussed above.
In particular, there will be an extended gapless phase around the isotropic point, where the zero-energy modes are Weyl points.
For inversion-symmetry-breaking flux configurations, one generically finds Majorana Fermi surfaces around the isotropic point.

\subsection{(10,3)a}
\label{ssec:10a}

We complete our classification program by discussing the physics of 3D Kitaev models for the three lattices with elementary loop length of 10 (see Table \ref{tab:lattice_overview}).
The Kitaev models for lattices (10,3)a and (10,3)b have already been discussed in the literature, but for the sake of completeness we briefly review these results here. 
For the (10,3)a or hyperoctagon lattice we summarize the results obtained by some of the authors of this paper in Ref.~\onlinecite{hermanns14}, while for the (10,3)b or hyperhoneycomb lattice we report on the results of Ref.~\onlinecite{Mandal09}.
 
\paragraph*{Lattice structure.--}
The (10,3)a lattice can be viewed as another higher dimensional variant of the square-octagon lattice, where the squares and octagons form counter-rotating spirals to form 
a three-dimensional lattice as illustrated in Fig.~\ref{fig:comb_10_a}(a).
Close inspection of this spiral structure reveals that it breaks inversion symmetry and as such the (10,3)a lattice is one of the few chiral lattices in our family of tricoordinated lattices (see Table \ref{tab:lattice_overview}).

More formally, the (10,3)a lattice is a body-centered cubic lattice with four sites per unit cell at positions 
\begin{align}
  \br_1&=\left(\frac 1 8, \frac 1 8 , \frac 1 8 \right) \,, &\br_2&=\left(\frac 5 8, \frac 3 8 , -\frac 1 8 \right) \,,\nonumber\\
  \br_3&=\left(\frac 3 8, \frac 1 8 , -\frac 1 8 \right) \,, &\br_4&=\left(\frac 7 8, \frac 3 8 , \frac 1 8 \right) \,.
\end{align}
The lattice vectors are given by 
\begin{align}
  \ba_1&=\left(1,0,0\right),&
  \ba_2&=\left(\frac{1}{2},\frac{1}{2},-\frac{1}{2}\right),&
  \ba_3&=\left(\frac{1}{2},\frac{1}{2},\frac{1}{2}\right),
\end{align}
and their corresponding reciprocal lattice vectors are
\begin{align}
  \bq_1 &= \left( 2\pi, -2\pi, 0 \right),&
  \bq_2 &= \left( 0, 2\pi, -2\pi \right),&\nonumber \\
  \bq_3 &= \left( 0, 2\pi, 2\pi \right).
  \label{eq:recip_vec_10_3a}
\end{align}
The 3D Kitaev model for this lattice is defined by assigning bond-directional couplings to the bonds as illustrated in Fig.~\ref{fig:comb_10_a}(a). 
Note that all bonds are related to each other by lattice symmetries. 
In particular, threefold rotation combined with a permutation of the $x$, $y$, and $z$ bonds is a symmetry of the Hamiltonian. 
Thus, the phase diagram is symmetric under permutation of  the three different coupling constants.

\paragraph*{Gauge structure.--}
For the (10,3)a lattice there are two linearly independent elementary loop operators per unit cell, both of which have length 10.
In the following, we will consider the flux sector, for which all the plaquette operators have eigenvalue $+1$, as suggested by Lieb's theorem \cite{Lieb} were it to apply. 
We have verified numerically that this is indeed the ground-state flux sector. 
In addition, it is the unique flux sector that obeys all the lattice symmetries; note that choosing all plaquettes to have $\pi$ flux is forbidden due to the volume constraints comprising three adjacent plaquette operators (see Appendix \ref{ssec:supplemental_10_a} for a more detailed discussion on these volume constraints).

\begin{figure}
  \includegraphics[width=\linewidth]{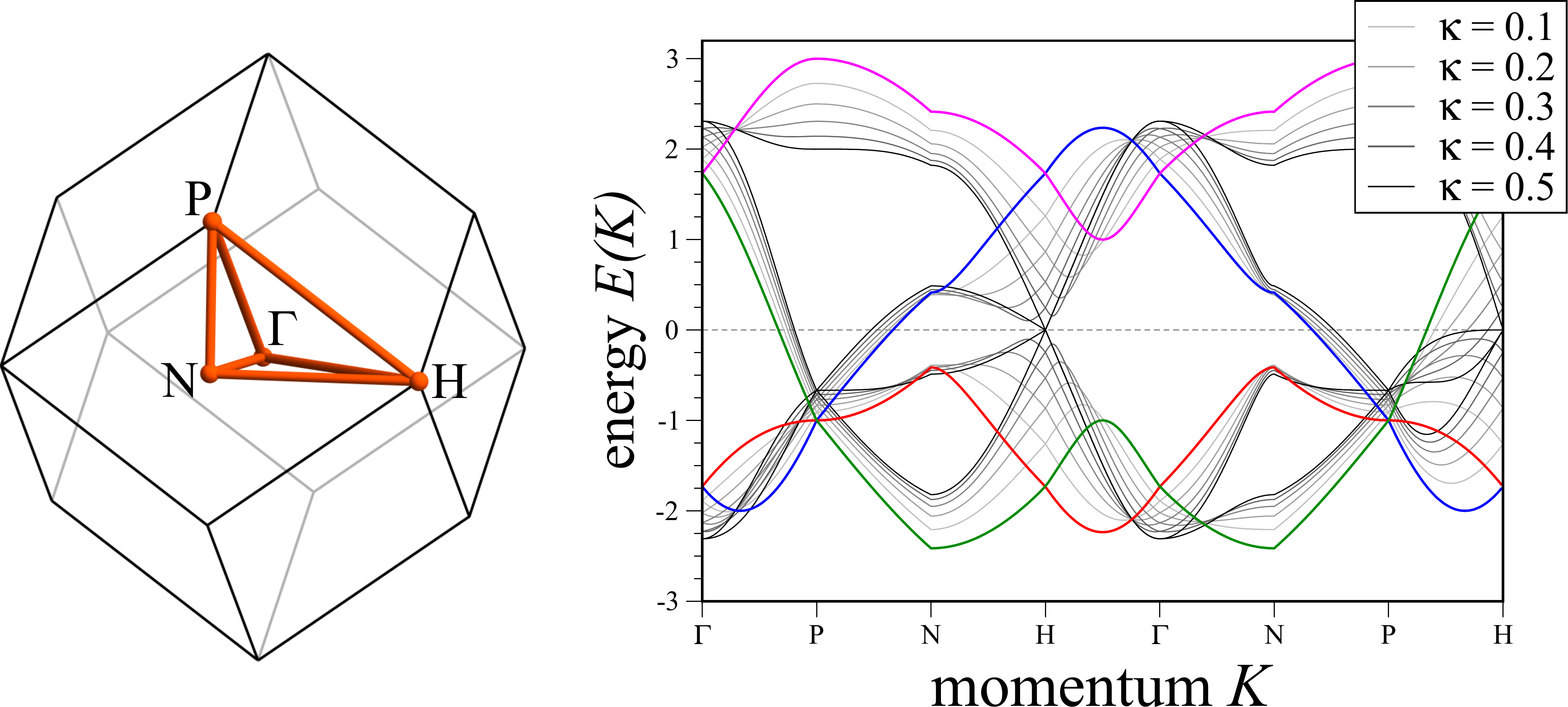}
  \caption{(Color online) Brillouin zone with high-symmetry points (left) and energy dispersion along the corresponding high-symmetry lines for lattice (10,3)a (right).
     		 Gray shaded curves indicate energy dispersion upon time-reversal symmetry breaking with magnetic field strength $\kappa$.  
     		 }
  \label{fig:energydispersion10a}
\end{figure}
\begin{figure}[t]
  \includegraphics[width=\linewidth]{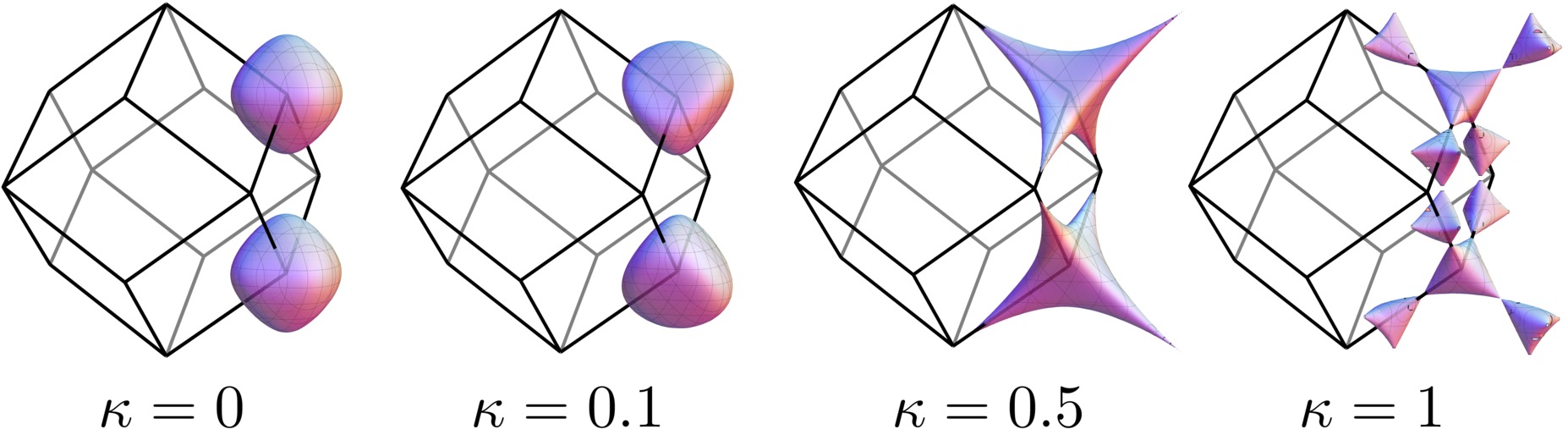}
  \caption{(Color online) Deformation of the Fermi surface of the Majorana metal for the (10,3)a lattice when breaking time-reversal symmetry. 
     			Plots are shown for varying $\kappa$ parametrizing the magnetic field strength in Eq.~\eqref{eq:kappa}. }
  \label{fig:10a_kappa}
\end{figure}
%

\paragraph*{Projective symmetries.--}
The (10,3)a lattice has the property that the translation vectors $\ba_2$ and $\ba_3$ map the two sublattices onto each other.
Therefore, as discussed in Sec.~\ref{sec:symmetries}, sublattice symmetry (and consequently also time-reversal symmetry) involves a non-vanishing translation in momentum space by $\kk =(-\mathbf q_2 +\mathbf q_3)/2 = (0,0,2\pi)$, where $\mathbf q_2$ and $\mathbf q_3$ are the reciprocal lattice vectors defined above.
As the lattice is chiral, the relevant energy relations are given by particle-hole and time-reversal symmetry
\begin{align}
	\epsilon(\bk) &= - \epsilon(- \bk)& \mbox{ and }&& \epsilon(\bk) &= \epsilon(- \bk+\kk) \,. 
\end{align} 
%

\paragraph*{Majorana metal.--}
The phase diagram for the Kitaev model on lattice (10,3)a is shown in Fig.~\ref{fig:comb_10_a}(b); the system exhibits a gapless phase around the isotropic point, where the gapless modes sit on two Majorana Fermi surfaces which are visualized in Fig.~\ref{fig:comb_10_a}(c), see Fig.~\ref{fig:energydispersion10a}.
The surfaces are centered around the corners of the Brillouin zone at $(\pi,\pi,\pi)$ and $(\pi,\pi,-\pi)$.
The darker shaded orange region of the phase diagram denotes the parameter space where these Majorana Fermi surfaces are topological, \ie, they enclose a Weyl node at finite energy, a scenario which we discuss in further detail in Sec.~\ref{ssec:topFermiSurface}. 

The two Majorana Fermi surfaces can be mapped onto each other by the perfect nesting vector $\kk$, as can be seen from Fig.~\ref{fig:comb_10_a}(c).
This has important consequences.
In particular the system is susceptible to a BCS-type spin-Peierls instability~\cite{SpinPeierlsHyperoctagon} driven by interactions between the Majorana fermions, which can be induced by additional spin exchanges such as a Heisenberg term augmenting the pure Kitaev model.
A short discussion on the spin-Peierls instability can be found in Sec.~\ref{sec:spin-peierls}.

Breaking time-reversal symmetry does not change the nature of the Majorana metal, \ie, the Fermi surfaces remain in place when including the $\kappa$ term of Eq.~\eqref{eq:kappa}. However, they do deform in a non-trivial way with increasing $\kappa$ as illustrated in Fig.~\ref{fig:10a_kappa}.

%
\begin{figure}[b]
  \includegraphics[width=\linewidth]{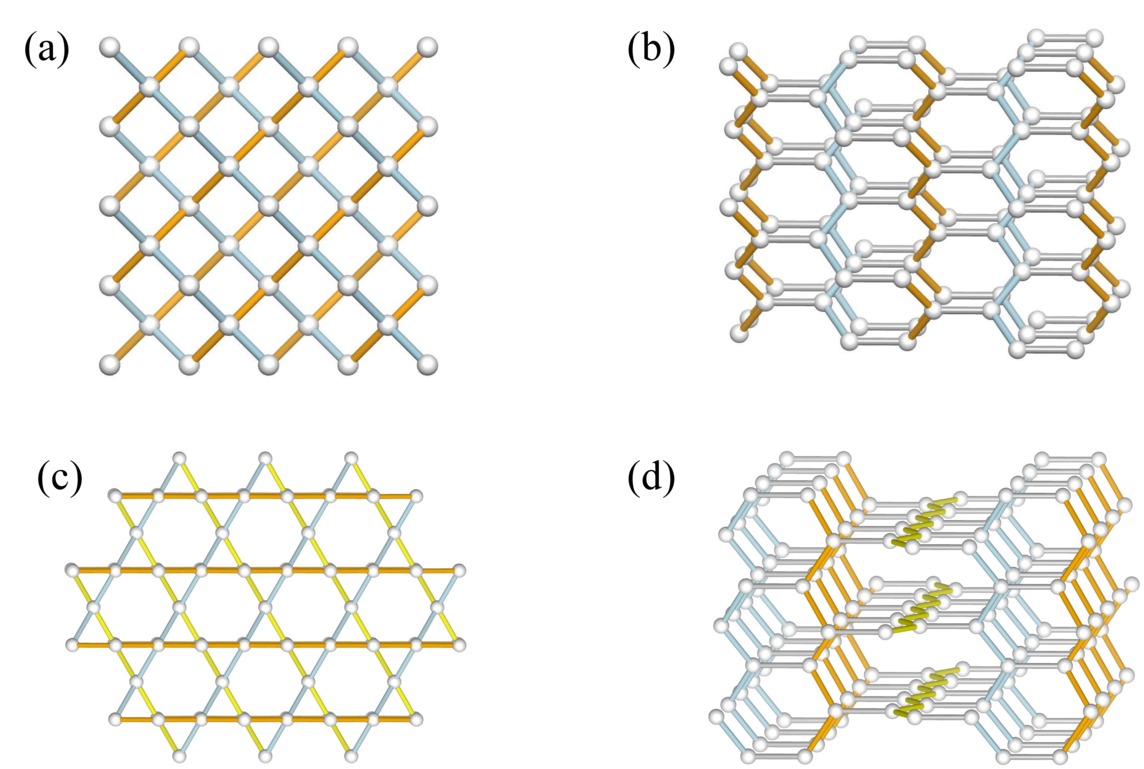}
  \caption{(Color online)  View on (10,3)b along (a) (1,0,0). and (b) (0,1,10).  View on (10,3)c along (c) (1,0,0) and (d) along  (1,10,1) }
  \label{fig:comparison_10_3_b_c}
\end{figure}
\begin{figure*}
  \includegraphics[width=\linewidth]{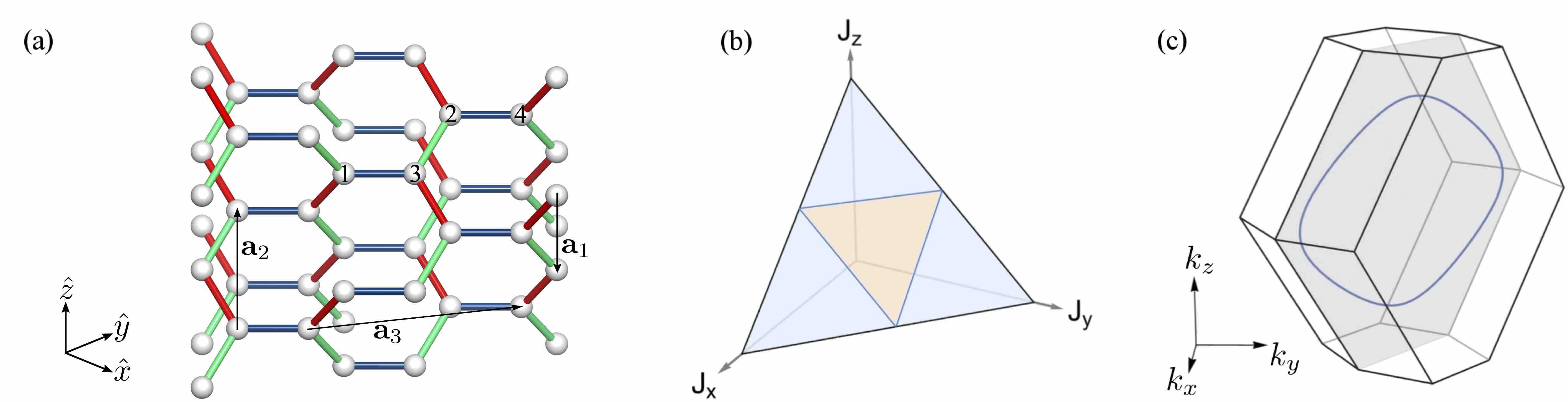}
  \caption{(Color online) (a) Visualization of the Kitaev couplings, unit cell and translation vectors for the (10,3)b lattice. (b) Phase diagram for  (10,3)b. (c) At the isotropic point, the gapless modes form a ring in the $k_x+k_y=0$ plane, indicated in gray. }
  \label{fig:phase_diagram_10_3b}
\end{figure*}

\subsection{(10,3)b}
\label{ssec:10b}

The (10,3)b lattice is probably the best known tricoordinated lattice in three spatial dimensions and is now typically referred to as the hyperhoneycomb lattice \cite{beta}.
The 3D Kitaev model for this lattice has recently been discussed extensively  \cite{Mandal09,hyperhoneycomb1,finiteTGauge2,Nasu14,wsl2014} in the context of the iridate \bLiIrO\ \cite{beta},
for which spin-orbit entangled $j=1/2$ moments form on the iridium sublattice, which is precisely the (10,3)b hyperhoneycomb lattice. 

The most symmetric form of the (10,3)b hyperhoneycomb lattice can best be visualized as parallel $xy$-zigzag chains along two distinct directions (90$^\circ$ rotated with respect to each other \cite{FootNoteSym10b}) that are coupled by $z$-bonds [see Figs.~\ref{fig:comparison_10_3_b_c}(a) and (b)]. 
It is a close cousin of the third lattice with elementary loop length 10, the (10,3)c lattice. 
The latter is made up of three parallel $xy$-zigzag chains (120$^\circ$ rotated with respect to each other) that are coupled by  $z$-bonds [see Figs.~\ref{fig:comparison_10_3_b_c}(c) and (d)].

\paragraph*{Lattice structure.--}
More formally, the (10,3)b lattice is a tetragonal lattice with four sites per unit cell at positions
\begin{align}
  \br_1 &= (0, 0, 0), &\br_2 &= (1, 2, 1), \nonumber\\
  \br_3 &= (1, 1, 0), &\br_4 &= (2, 3, 1).
\end{align}
The lattice vectors are given by 
\begin{align}
  \ba_1 &= \left( -1, 1, -2 \right),&
  \ba_2 &= \left( -1, 1, 2 \right),&
  \ba_3 &= \left( 2, 4, 0 \right),
\end{align}
and their corresponding reciprocal lattice vectors are
\begin{align}
  \bq_1 &= \left( -\frac{2\pi}{3}, \frac{\pi}{3}, -\frac{\pi}{2}  \right),&
  \bq_2 &= \left( -\frac{2\pi}{3}, \frac{\pi}{3}, \frac{\pi}{2}  \right),&\nonumber \\
  \bq_3 &= \left( \frac{\pi}{3}, \frac{\pi}{3}, 0 \right).
  \label{eq:recip_vec_10_3b}
\end{align}
The unit cell with translation vectors, as well as the assignment of bond types that defines the  Kitaev Hamiltonian are illustrated in Fig.~\ref{fig:phase_diagram_10_3b} (a).
Note that only the $x$- and $y$-bonds are related by lattice symmetries, which results in a phase diagram that is symmetric under exchange of couplings $J_x$ and $J_y$.

\begin{figure}
  \includegraphics[width=\linewidth]{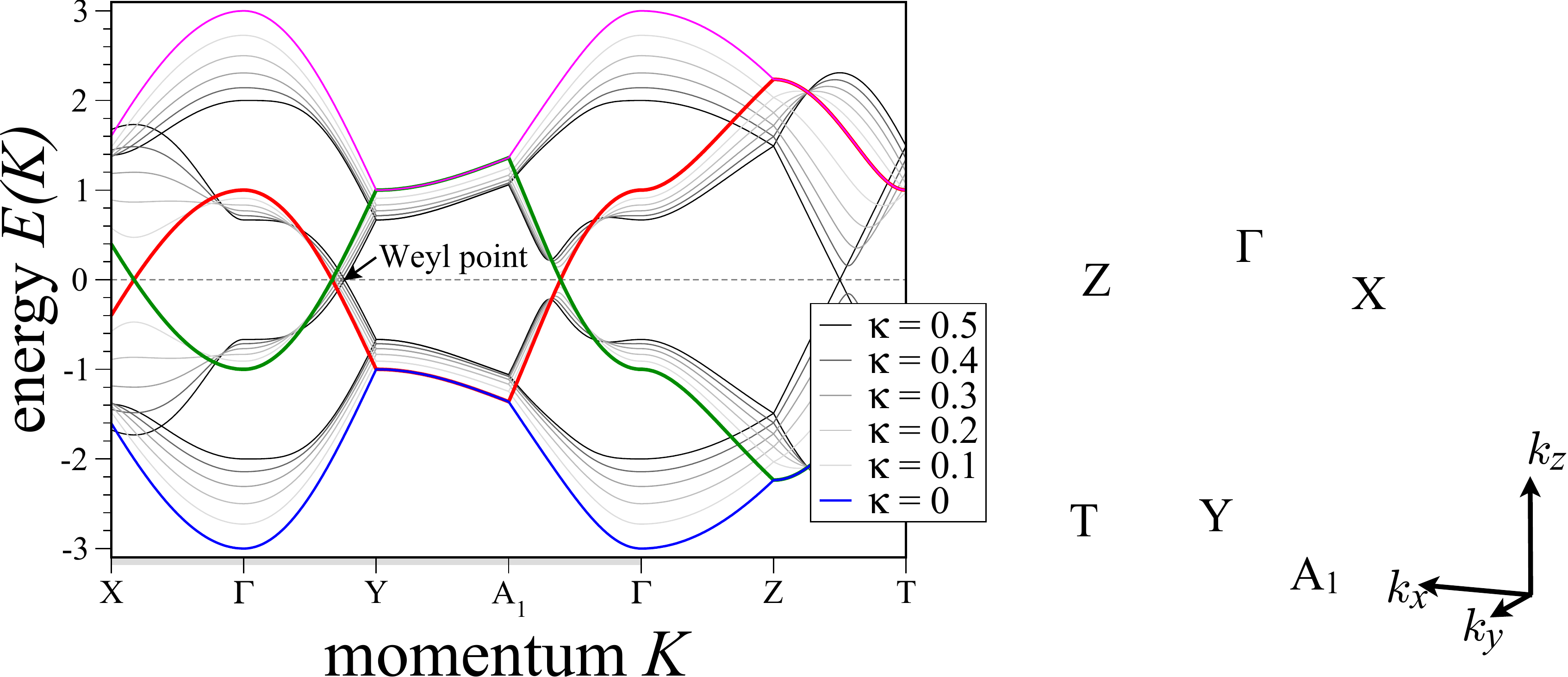}
  \caption{(Color online) 
     			The energy dispersion of the  Kitaev model on the (10,3)b hyperhoneycomb lattice
			for various values of $\kappa$ (parametrizing the effective magnetic field) along certain high-symmetry lines indicated in the Brillouin zone on the right-hand side. 
			The gray hexagon indicates the plane $k_x=-k_y$ on which the line of gapless mode (black line) is located.
     			Figure adapted from Ref.~\onlinecite{wsl2014}. }
  \label{fig:energydispersion10b}
\end{figure}
\begin{figure*}
  \includegraphics[width=\linewidth]{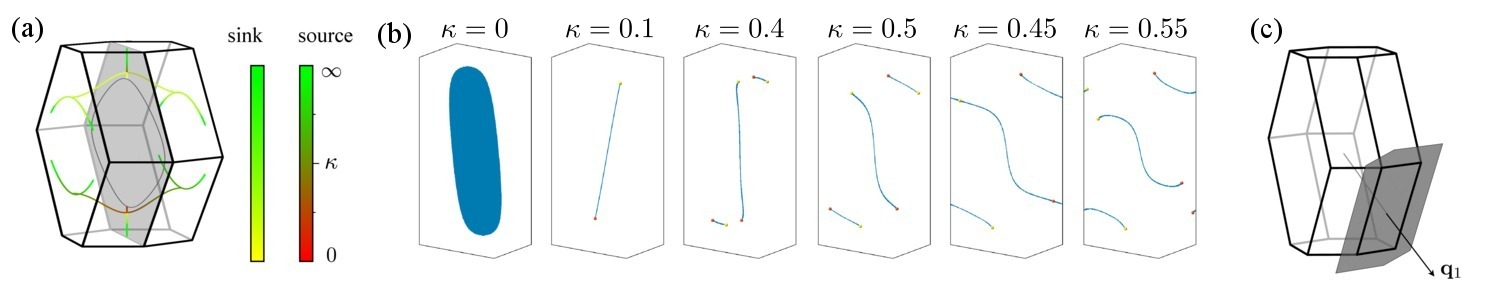}
  \caption{(Color online) (a) Evolution of the Weyl nodes  for (10,3)b in the presence of a magnetic field for varying $\kappa$ from $0,\ldots,\infty$. 
     			(b) Corresponding Fermi arc evolution. 
			(c) Visualization of the surface Brillouin zone for open boundary conditions in the 100-direction. 
			Figure adapted from Ref.~\onlinecite{wsl2014}. }
  \label{fig:10bFermiArcs}
\end{figure*}
%

\paragraph*{Gauge structure.--}
For the (10,3)b lattice there are two linearly independent loop operators per unit cell, both of which have length 10 (see Appendix \ref{ssec:supplemental_10_b}).
In the following, we consider the flux sector for which all loop operators have eigenvalue $+1$.
It has been verified numerically in Ref.~\cite{Mandal09}, and independently by us, that this is indeed the ground state flux sector.
The vison gap, calculated as the energy gap arising from flipping either an $x$- or a $y$-bond, is among the highest found for the lattices considered here with $\Delta \sim 0.13$ in units of the Kitaev coupling at the isotropic coupling point $J_x = J_y = J_z$. 
Flipping an $x$- or a $y$-bond operator implies switching signs of six plaquette operators (see Appendix \ref{ssec:supplemental_10_b} for a visualization).

\paragraph*{Projective symmetries.--}
As the translation symmetry of this lattice is equivalent to that of its two sublattices, sublattice symmetry and time-reversal symmetry are implemented trivially, \ie, with vanishing $\kk$.
 As this lattice also possesses inversion symmetry with vanishing $\kt$, the relevant energy relations become 
\begin{align}
  \epsilon(\bk) &= -\epsilon(-\bk)& \mbox{ and }&&
  \epsilon(\bk) &= \epsilon(-\bk) \,,
\end{align}
the first originating from particle-hole symmetry and the second from time-reversal or inversion symmetry.

\paragraph*{Majorana metal.--}
The phase diagram for the Kitaev model on lattice (10,3)b is shown in Fig.~\ref{fig:phase_diagram_10_3b} (b); the system exhibits a gapless phase around the isotropic point where the gapless modes form a closed line of Dirac nodes, pictured in Fig.~\ref{fig:phase_diagram_10_3b} (c).
The Majorana Fermi line, which lies in the $k_x = -k_y$ plane, is protected by time-reversal symmetry.
Breaking time-reversal symmetry  causes the Fermi line to gap out almost entirely, leaving just two Weyl nodes, which are fixed to zero energy as long as inversion symmetry remains intact \cite{wsl2014} (see Fig.~\ref{fig:energydispersion10b}).  

Note that the behavior of (10,3)b is completely analogous to that of the (8,3)c lattice in Sec.~\ref{ssec:8c}, although the details differ. 
In particular, we can use the same argument for why time-reversal symmetry breaking cannot gap the system completely, and  Weyl points have to occur.
For small values of $\kappa$, the Weyl points move along the $\hat k_z$-axis up until $\kappa=\frac 1 2 \sqrt{\frac 3 5 }$, where four additional Weyl nodes appear. 
The full evolution of the Weyl nodes and their corresponding Fermi arc surface states are visualized in Figs.~\ref{fig:10bFermiArcs} (a) and (b).
While the Weyl points which move along the $k_z$ axis recombine at $\kappa \rightarrow \infty$, the ones on the front/back surface do not. 
Instead, one of the velocities of the Weyl points vanishes and they become part of  gapless nodal lines at $\kappa=\infty$. 
 
\subsection{(10,3)c}
\label{ssec:10c}

\begin{figure*}
  \includegraphics[width=\linewidth]{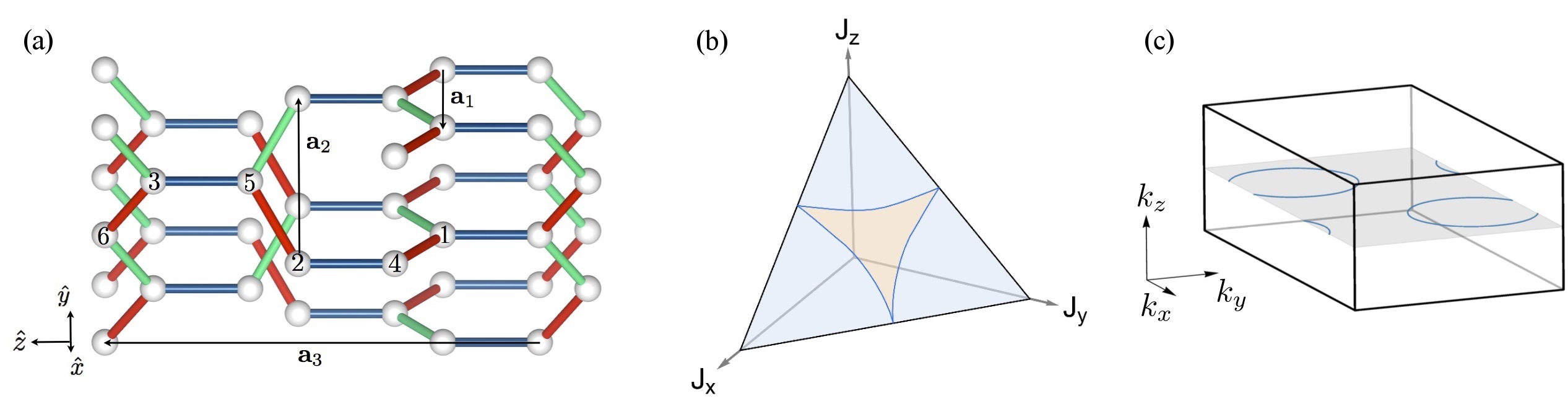}
  \caption{(Color online)  (a) Visualization of the Kitaev couplings, the unit cell and the translation vectors for the lattice (10,3)c. (b) Phase diagram for (10,3)c. (c) Visualization of the gapless nodal line in the $k_z=0$ plane, indicated in gray. }
  \label{fig:phase_diagram_10_3c}
\end{figure*}
%

The (10,3)c lattice is a close cousin of (10,3)b as already mentioned above.
One main distinction between the two lattices is that  (10,3)c is a chiral lattice, while (10,3)b is inversion symmetric.  
The chirality has important consequences for the behavior for broken time-reversal, as we will see in the following. 

\paragraph*{Lattice structure.--}
More formally, the (10,3)c lattice is a trigonal lattice with six sites per unit cell at positions 
\begin{align}
  \br_1 &= \left( \frac{1}{4}, \frac{1}{4\sqrt{3}}, \frac{1}{2\sqrt{3}} \right)\,, &\br_2 &= \left( \frac{3}{4}, \frac{1}{4\sqrt{3}}, \frac{2}{\sqrt{3}} \right)\,, \nonumber\\ 
  \br_3 &= \left( \frac{1}{2}, \frac{1}{\sqrt{3}}, \frac{7}{2\sqrt{3}} \right)\,, &\br_4 &= \left( \frac{3}{4}, \frac{1}{4\sqrt{3}}, \frac{1}{\sqrt{3}} \right)\,, \nonumber\\ 
  \br_5 &= \left( \frac{1}{2}, \frac{1}{\sqrt{3}}, \frac{5}{2\sqrt{3}} \right)\,, &\br_6 &= \left( \frac{1}{4}, \frac{1}{4\sqrt{3}}, \frac{4}{\sqrt{3}} \right)\,. 
\end{align}
The lattice vectors are given by 
\begin{align}
  \ba_1 &= \left( 1, 0, 0 \right),&
  \ba_2 &= \left( -\frac{1}{2}, \frac{\sqrt{3}}{2}, 0 \right),&\nonumber \\
  \ba_3 &= \left( 0, 0, \frac{3\sqrt{3}}{2} \right),
\end{align}
and their corresponding reciprocal lattice vectors are
\begin{align}
  \bq_1 &= \left( 2\pi, \frac{2\pi}{\sqrt{3}}, 0 \right),&
  \bq_2 &= \left( 0, \frac{4\pi}{\sqrt{3}}, 0 \right),&\nonumber \\
  \bq_3 &= \left( 0, 0, \frac{4\pi}{3\sqrt{3}} \right).
  \label{eq:recip_vec_10_3c}
\end{align}

The choice of bond types for implementing the Kitaev Hamiltonian is illustrated in Fig.~\ref{fig:phase_diagram_10_3c}(a).
Note that we chose the $x$- and $y$-bonds on each of the chains such that the lattice is invariant under a 120$^\circ$ screw-rotation.
This ensures that the phase diagram is symmetric under exchanging $J_x\leftrightarrow J_y$.

\paragraph*{Gauge structure.--}
For this lattice there are three loop operators of length 10 and three of length 12 per unit cell.
These six loop operators form three closed volumes which leads to only three linearly independent loop operators per unit cell (see Appendix \ref{ssec:supplemental_10_c}).
In what follows, we consider the flux sector where all loop operators of length 10 have eigenvalue $+1$ and all loop operators of length 12 have eigenvalue $-1$.
This configuration of fluxes respects all lattice symmetries and, although this lattice does not possess the symmetries required for rigorous application of Lieb's theorem \cite{Lieb}, is consistent with the flux assignments one would expect were Lieb's theorem  to hold. 
It should be noted that, although this flux configuration breaks no lattice symmetries, fixing a compatible gauge requires an enlargement of the unit cell in the 010-direction (see Appendix~\ref{ssec:supplemental_10_c}).

\paragraph*{Projective symmetries.--}
As the translation symmetry of this lattice is equivalent to that of its two sublattices, sublattice symmetry and time-reversal symmetry are implemented trivially, \ie, $\kk=0$.
The lattice lacks inversion symmetry, as discussed above, and the relevant energy relations are therefore given by particle-hole and time-reversal symmetry
\begin{align}
  \epsilon(\bk) &= -\epsilon(-\bk)& \mbox{ and }&&
  \epsilon(\bk) &= \epsilon(-\bk) \,.
\end{align}
\begin{figure}
  \includegraphics[width=\linewidth]{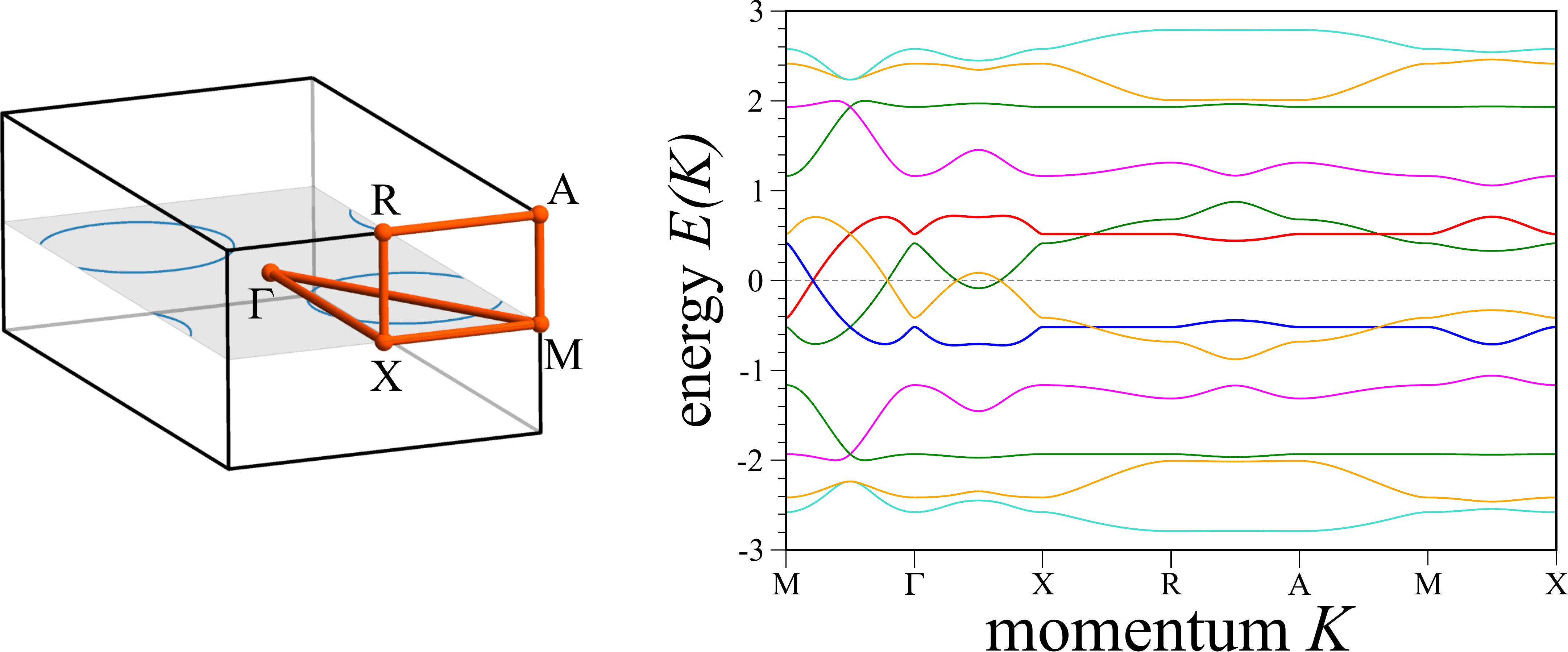}
  \caption{(Color online) Brillouin zone with high-symmetry points and energy dispersion along the corresponding high-symmetry lines for (10,3)c.}
  \label{fig:energydispersion_10_c}
\end{figure}
\begin{figure}
  \includegraphics[width=\linewidth]{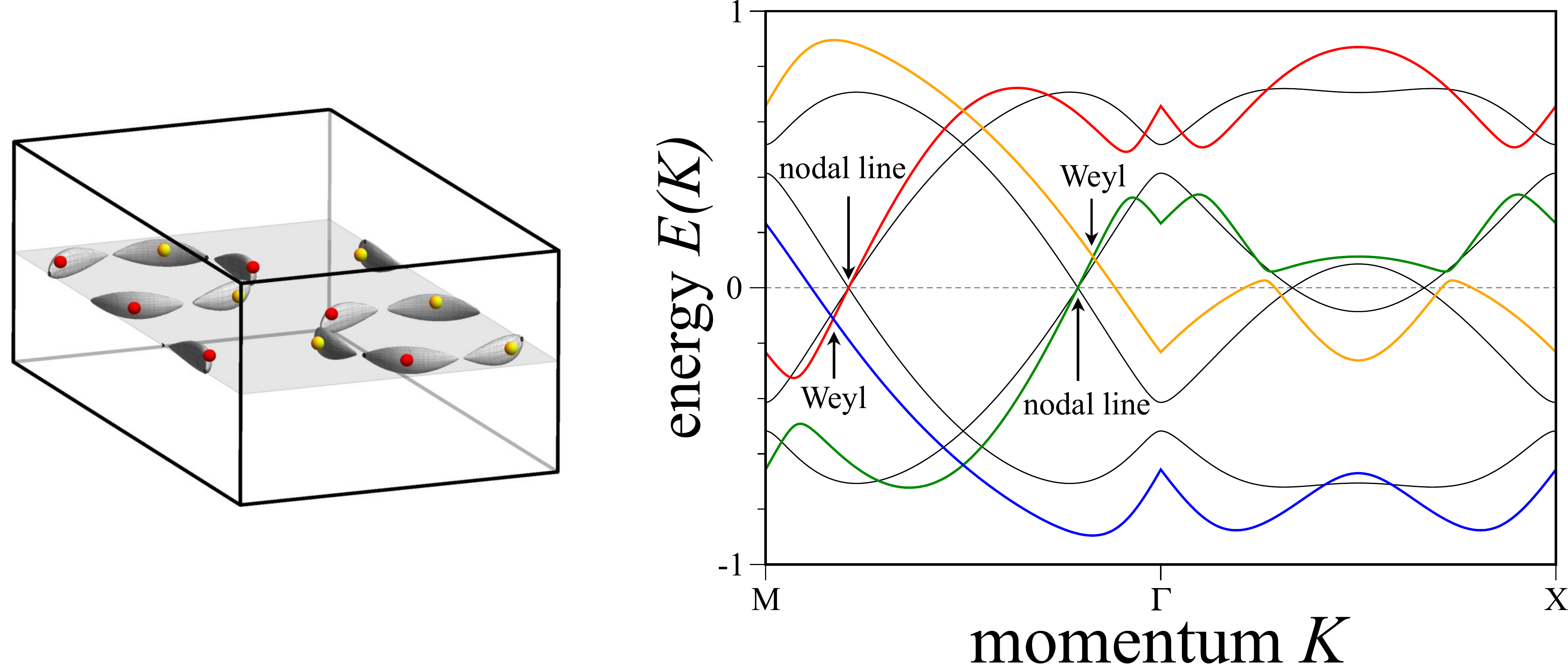}
  \caption{(Color online) Illustration of the 12 Fermi pockets enclosing a Weyl node (indicated by the yellow and red dots) and the energy dispersion 
     					upon breaking of time-reversal symmetry for (10,3)c.}
  \label{fig:energydispersion_10_c_kappa}
\end{figure}
%

\paragraph*{Majorana metal.--}
The phase diagram for the Kitaev model for lattice (10,3)c is shown in Fig. \ref{fig:phase_diagram_10_3c} (b). 
The model exhibits a gapless phase around the isotropic point where the gapless modes form two closed {\em nodal lines} with a linear Dirac-type dispersion in the two directions orthogonal to the nodal line (see the energy dispersion in Fig.~\ref{fig:energydispersion_10_c}).
These nodal lines, lying in the $k_z = 0$ plane, are fixed to zero energy by time-reversal symmetry.

Breaking time-reversal symmetry lifts the dispersion along the original nodal line from zero energy.
In particular, 12 Weyl nodes form in the dispersion, whose locations are related by the 120$^\circ$ screw rotation.
As the (10,3)c lattice lacks inversion symmetry, the Weyl nodes are not fixed to zero energy and are found to indeed move to energies above and below zero energy for arbitrarily small values of $\kappa$.
As a result, the remaining nodal structure upon breaking of time-reversal symmetry is given by 12 Fermi pockets, \ie, small Fermi surfaces that each enclose one of the 12 Weyl nodes.
The formation of these Fermi surfaces can nicely be tracked in the energy dispersion along high-symmetry lines shown in Fig.~\ref{fig:energydispersion_10_c_kappa}, where the gray bands show the energy dispersion for the time-reversal invariant system (exhibiting nodal lines) and the colored lines correspond to the energy dispersion for $\kappa=0.2$ (in units of the Kitaev coupling).
It can clearly be seen that there are two band crossings between points $M$ and $\Gamma$; the one nearer to $M$ is located below zero energy and the other above zero energy. 
Note that because the Fermi surfaces enclose a Weyl node, they inherit topologically non-trivial features as discussed in Section \ref{ssec:topFermiSurface} in more detail.

Finally, we point out that the (10,3)c lattice is special in that it is the only lattice for which breaking of time-reversal symmetry {\em increases} the nodal manifold of the Majorana metal (from a line to a surface) and, thus, also the associated density of states.


\section{Weyl physics}
\label{sec:Weyl}

A generic feature found in the dispersion relation of {\em all} 3D Kitaev models studied in this paper is the occurrence of Weyl nodes -- either right at the Fermi energy or above/below it. 
If the Weyl nodes sit right at the Fermi energy (\ie, zero energy), we encounter a spin liquid analog of the electronic Weyl semimetal \cite{WeylSM}, a state which we have dubbed a Weyl spin liquid in previous work \cite{wsl2014}.
If the Weyl nodes sit above/below the Fermi energy, the system exhibits topological Fermi surfaces (each enclosing at least one Weyl node), the spin analog of the so-called Weyl metal \cite{WeylM}.
Kitaev models in the first category are those defined for lattices (8,3)b, (8,3)c, (8,3)n, (9,3)a, and (10,3)b, while the Kitaev models for lattices (8,3)a, (10,3)a, and (10,3)c are in the second category 9see also Table \ref{tab:majorana_metals}).
We discuss these two scenarios in further detail in the following two subsections.

\subsection{Weyl spin liquids}
\label{ssec:weylspinliquids}

We will first concentrate on the case where the Weyl nodes for one of our 3D Kitaev models sit precisely at the Fermi energy.
The Majorana energy dispersion relation for these systems is then in precise analogy to those of electronic Weyl semimetals \cite{WeylSM}.
These electronic Weyl semimetals have garnered considerable attention for their recent observation in TaAs \cite{TaAs} and photonic materials \cite{WeylSM-Photonic}.
An intense experimental effort is currently underway to observe the unusual response of these electronic Weyl semimetals to electromagnetic fields such as the chiral anomaly \cite{chiralAnomaly,chiralAnomalyExp} and unusual negative magnetotransport \cite{Magnetoresistance}.

For electronic systems, it has been realized early on that Weyl physics can emerge by either breaking time-reversal or inversion symmetry \cite{WeylSM}.
For the spin systems at hand it turns out that different symmetry scenarios are at play that give rise to Weyl physics.
Primarily, we distinguish three different scenarios with respect to the role that time-reversal symmetry plays.
In analogy to the electronic systems, we can find Weyl physics when breaking time-reversal symmetry {\em explicitly}.
This is the case, for instance, for the 3D Kitaev models for lattices (10,3)b and (10,3)c exhibiting nodal lines in the presence of time-reversal symmetry and a number of Weyl nodes when breaking time-reversal symmetry.
This also includes the case of the Kitaev model for the lattice (8,3)n, for which the emergence of Weyl nodes arises from the splitting of a Dirac cone upon time-reversal symmetry breaking.
This scenario should be carefully distinguished from the physics that plays out for the Kitaev model on the lattice (9,3)a where time-reversal symmetry is broken {\em spontaneously} and the system is then found to exhibit Weyl physics.
The third and most unusual scenario is the one found for the Kitaev model for lattice (8,3)b.
Here, {\em neither} time-reversal nor inversion symmetry are broken, but the system nevertheless exhibits Weyl physics, a symmetry scenario that is not possible for electronic systems.
The reason that we can observe Weyl physics in the spin system without breaking time-reversal nor inversion symmetry is that the {\em projective} time-reversal and inversion symmetry for the underlying Majorana fermions are implemented in a non-trivial way.
In particular, these projective symmetries incorporate momentum shifts by $\kk$ and a pair of Weyl nodes at $\pm \bk$ always has a pair of time-reversal partners at $\mp (\bk-\kk)$
(as long as time-reversal symmetry is intact).
Second, we note that due to particle-hole symmetry (inherent on the level of the Majorana fermions), each Weyl point at position $\mathbf k$ (and energy $\epsilon$) will have a particle-hole partner of opposite chirality at position $-\mathbf k$ (and energy $-\epsilon$).
In contrast to the Weyl points in an electronic Weyl semimetal, one cannot regard these as two independent Weyl points.
Instead, it is more instructive to think of them as a single Weyl point of a \emph{complex} fermion (rather than two Weyl points of Majorana fermions).
This is a direct consequence of the fact that the system at hand does not possess $U(1)$ symmetry, but only \Z\  symmetry.
Thus, if neither time-reversal nor inversion symmetry are broken, we will encounter multiples of {\em four} Weyl nodes for the underlying Majorana system as it is indeed the case for the Kitaev model of lattice (8,3)b.
Note that while each pair of Weyl nodes has a corresponding chiral surface state, their effects cancel out each other exactly and the resulting spin liquid is not chiral.
Upon breaking of time-reversal symmetry (e.g. by applying a magnetic field) the perfect nesting between the two pairs of Weyl nodes as well as the exact cancellation of their respective Fermi arcs are destroyed.

The lack of $U(1)$ symmetry in Weyl spin liquids has important consequences for various physical observables.
Most importantly, Weyl spin liquids will not exhibit the usual chiral anomaly as charge pumping between Weyl nodes is intimately connected to charge conservation and thus $U(1)$ symmetry.
Instead, Weyl spin liquids will exhibit a more subtle incarnation of the chiral anomaly, to be discussed elsewhere. 

On a more formal level, the different roles of symmetries in the Weyl physics of electronic and spin systems is reflected in the classification of their underlying free-fermion Hamiltonians in the 10-fold way symmetry classification scheme of Altland and Zirnbauer \cite{AltlandZirnbauer}.
For electronic systems, Weyl semimetals are found in symmetry classes A or AII, corresponding to the breaking of either time-reversal or inversion symmetry, respectively.
In contrast, for the Kitaev models at hand, we find Weyl physics in symmetry classes D or BDI depending on whether time-reversal symmetry is broken (explicitly or spontaneously) or not.
This distinction of Weyl physics in electronic and spin systems will likely impact their sensitivity to disorder, a direction that we will pursue in the future.


\subsection{Topological Fermi surfaces}
\label{ssec:topFermiSurface}
We now want to turn to Majorana metals where the Weyl points occur at {\em finite} energy, which necessarily implies that they are encapsulated by a  Fermi surface. 
Such Fermi surfaces are called topological Fermi surfaces in the following, as they are found to inherit some of the topological features of the enclosed Weyl points. 
The physical properties discussed here are relevant for the lattices (10,3)a and (8,3)a, as well as (10,3)c for broken time-reversal symmetry. 

\begin{figure}
  \includegraphics[width=\linewidth]{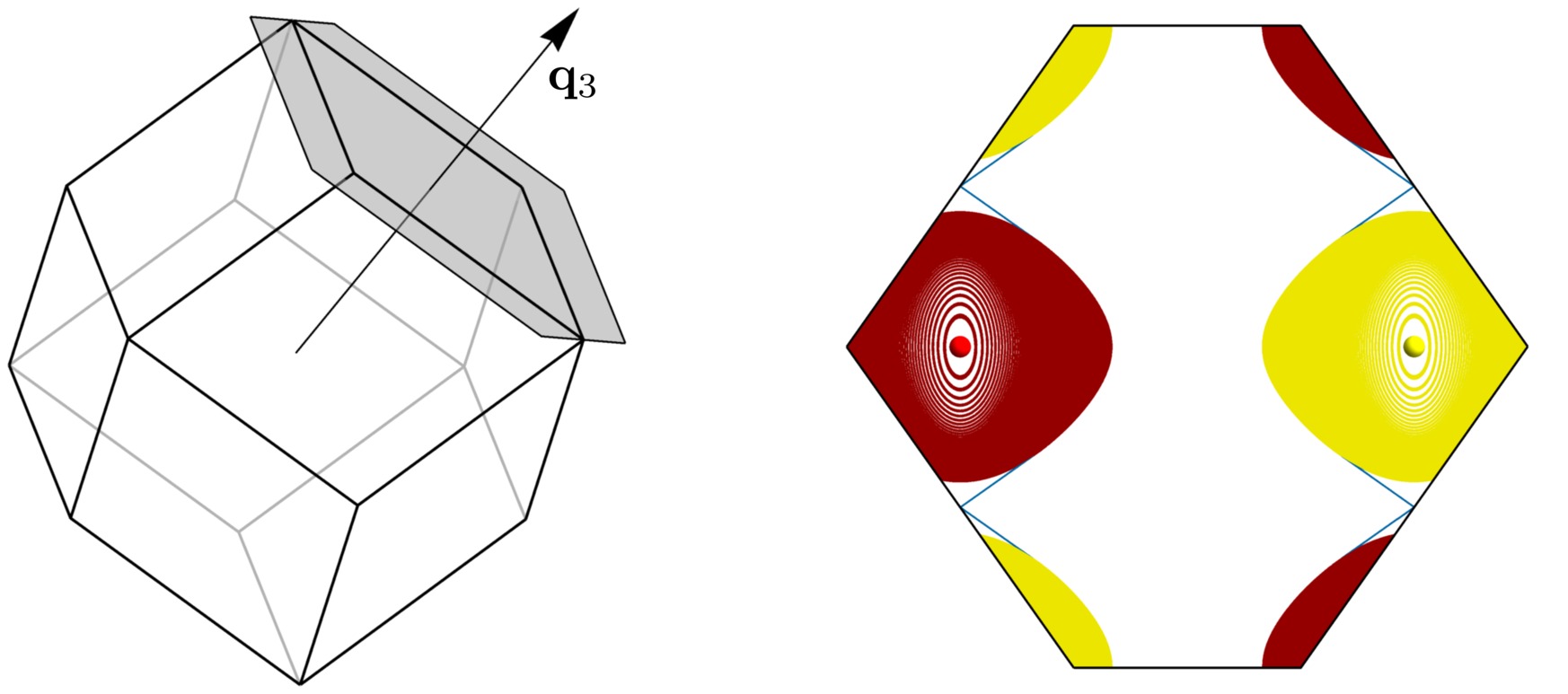}
  \caption{(Color online) 
    		\textit{(Left)} Visualization of the surface Brillouin zone for the 001 direction of lattice (10,3)a. 
    		\textit{(Right)} Gapless surface modes illustrated in the surface Brillouin zone. 
				    Aside from two puddles arising from the projection of the Majorana Fermi surfaces,
				    two Fermi arcs (indicated by the blue line) form between the projected Weyl nodes (indicated by the red and yellow dots).}
  \label{fig:10_3_a_fermi_arc}
\end{figure}

Let us first note that the Chern number of a closed 2D surface is still well-defined, as long as it does not cut through any of the Fermi surfaces.
Further the value of the associated Chern number cannot depend on whether or not the enclosed Weyl point sits at zero energy.
Thus, even in the presence of Fermi surfaces, we find that the effective two-dimensional Hamiltonian defined on an arbitrary closed surface that surrounds a  topological Majorana Fermi surface is that of a topological Chern insulator. 
Using the same arguments as above, we find that in the presence of boundaries the Majorana Fermi surfaces cannot lie isolated in the surface Brillouin zone, but must again be connected by (chiral) Fermi arcs, see Fig.~\ref{fig:10_3_a_fermi_arc} for an example. 

\begin{figure}
  \includegraphics[width=\linewidth]{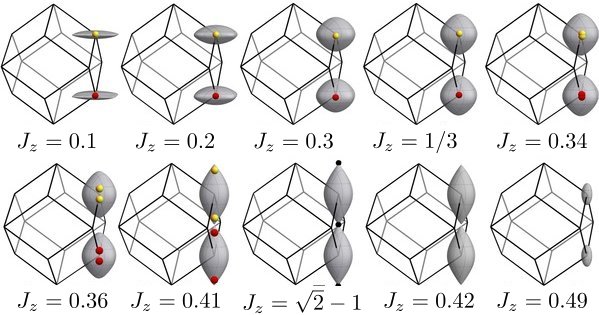}
  \caption{(Color online)  Deformation of the Majorana Fermi surface of the (10,3)a lattice
    		when changing the $J_z$ coupling constant from 0 to $J_z=1/2$ where the system becomes gapped. 
    		Red (yellow) dots denote the position of Weyl points of positive (negative) chirality. 
		The black dots are pairs of Weyl points of opposite chirality.  For $J_z>1/3$ the Weyl points start to split and annihilate at  $J_z=\sqrt{2}-1$. }
  \label{fig:10_3_a_jz}
\end{figure}

\begin{figure}[t]
  \includegraphics[width=\linewidth]{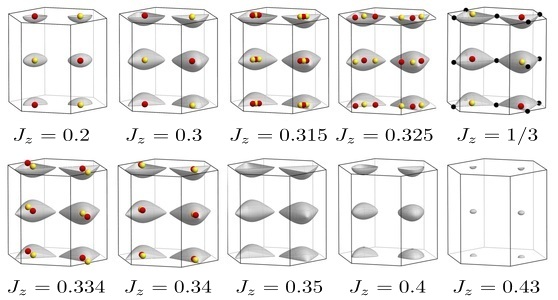}
  \caption{(Color online)  Deformation of the (topological) Majorana Fermi surfaces of lattice (8,3)a when changing  $J_z$. 
    		Red (yellow) dots denote the position of Weyl points of positive (negative) chirality. 
		The black dots are pairs of Weyl points of opposite chirality.  
		Around $J_z\approx 0.31$ each Weyl point splits into three. 
		At the isotropic point, two pairs of Weyl points with opposite charges are annihilated while one is created at the touching points of the Fermi surfaces. 
		This results in topologically trivial Fermi surfaces for $J_z >\frac 1 3 $.  }
  \label{fig:8_3_a_jz}
\end{figure}

The fact that Weyl nodes cannot be gapped out individually readily implies that the same is true for topological Fermi surfaces.
In fact, in order to gap them, one usually has to first annihilate their enclosed Weyl nodes pairwise.
For the lattices (10,3)a and (8,3)a, such pair annihilation can be observed when varying the relative strength of the three Kitaev couplings.
In particular, we find that pair annihilation occurs at touching points of two Fermi surfaces.
These touching points are, in fact, a pair of Weyl nodes of opposite chirality.
Figures \ref{fig:10_3_a_jz} and \ref{fig:8_3_a_jz} show the evolution of the Fermi surfaces along the line $J_x=J_y$ with the position of the enclosed Weyl nodes denoted by the red (positive) and yellow (negative) sphere. 
For the lattice (10,3)a, the  Weyl nodes have charge $\pm 2$ and are located at $(\pi,\pi, \pm \pi)$ for $J_z<1/3$.
At $J_z=1/3$ they split and pair-annihilate with the Weyl nodes of opposite charge at $J_z=\sqrt{2}-1$.
The resulting trivial Fermi surface persists until $J_z=1/2$, where the system becomes gapped. 
The behavior for the Kiteav model on lattice (8,3)a is slightly different.
The Fermi surfaces are topological for $J_z<1/3$.
The touching point at $J_z=1/3$ marks the pair creation/annihilation of opposite-charge Weyl nodes that split for $J_z>1/3$ and annihilate with the original Weyl points at $J_z\approx 0.34$, see the sequence of Fig.~\ref{fig:8_3_a_jz}.

%


\section{Spin-Peierls instabilities}
\label{sec:spin-peierls}
While our entire discussion so far has concentrated on the pure Kitaev model, it is of course interesting to also discuss the effect of additional interactions (e.g., Heisenberg exchange) on the nature of the gapless spin-liquid ground state. 
Such additional terms have in general two effects: the vison excitations of the \Z\ gauge field gain dynamics and obtain a dispersion, while the Majorana fermions generically become interacting. 
The former effect may be ignored for sufficiently small perturbations, as the vison excitations remain gapped.
The effect of interactions between Majorana fermions depends crucially on the nature of the gapless modes. 
For quantum spin liquids with a Fermi line or Weyl points, a  scaling analysis as, e.g., done in Ref.~\cite{hyperhoneycomb1} shows that interaction terms are irrelevant at the Kitaev point and that the spin liquid is therefore stable against small perturbations. 
For quantum spin liquids with a Majorana Fermi surface, interactions turn out to be marginal.
A careful analysis for the (10,3)a hyperoctagon lattice in Ref.~\cite{SpinPeierlsHyperoctagon} has shown that  time-reversal symmetric interactions generically destabilize the Majorana Fermi surfaces even for infinitesimal coupling strength, and the surfaces gap out except for an odd number of nodal lines. 
In the following, we will briefly review the underlying mechanism and the main features of this instability, which will be referred to as a spin-Peierls BCS instability for reasons that will become clear in the following.
For further details we refer the reader to Ref.~\cite{SpinPeierlsHyperoctagon}.

\begin{figure}
  \includegraphics[width=\linewidth]{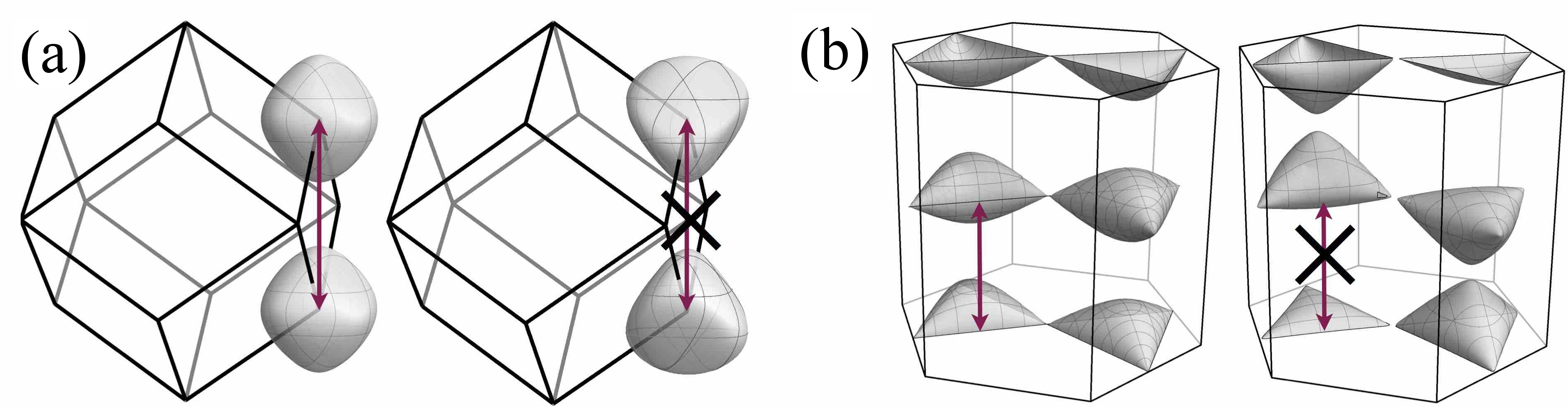}
  \caption{(Color online) Effect of time-reversal symmetry breaking on (a) the (10,3)a and (b) the (8,3)a lattice, with the time-reversal invariant system on the left and the time-reversal symmetry breaking system ($\kappa=J_K/10$) on the right, respectively. Breaking time-reversal symmetry destroys the perfect nesting condition with wave vector $\kk$, marked by the arrow.  }
  \label{fig:SpinPeierls}
\end{figure}

As we have seen in Section~\ref{sec:symmetries}, a Majorana Fermi surface can only be stable if time-reversal symmetry is implemented non-trivially, \ie, with $\epsilon(\bk)=\epsilon(-\bk +\kk)$. 
Combining this relation with particle-hole symmetry, we find that the spectrum necessarily exhibits perfect nesting of the Majorana Fermi surfaces, $\epsilon(\bk)=-\epsilon(\bk +\kk)$.
This perfect nesting for lattices (10,3)a and (8,3)a is visualized on the left-hand-sides of Figs.~\ref{fig:SpinPeierls} (a) and (b), respectively, with the perfect nesting vector indicated by the arrow.
Alternatively, we could have chosen to express the system in terms of complex fermions 
\begin{align}
  f_j(\bk)= \gamma_j(\bk)/\sqrt{2}&\qquad\mbox{ with }& j=1,\ldots,n,
\end{align} 
where $\gamma_j(\bk)$ denote the eigenmodes of the Majorana Hamiltoninan with $\epsilon_j(\bk)>\epsilon_i(\bk)$ for $j<i$ and $\gamma_j(\bk)^\dagger =\gamma_{2n+1-j}(-\bk)$ due to particle-hole symmetry. 
The 2$n$ Majorana Fermi surfaces, thus, combine into $n$ complex Fermi surfaces, and the perfect nesting condition becomes the usual BCS pairing condition $\epsilon(\kk/2 + \bk)=\epsilon(\kk/2-\bk)$, centered around $\kk/2$. 
Note that there is no $U(1)$ symmetry in the system. 
Instead, a non-vanishing pair correlator $\langle f_\alpha ^\dagger (\kk/2 + \bk)f_\beta^\dagger(\kk/2 - \bk)\rangle$ breaks translation symmetry spontaneously.
The resulting dimerization of the system is reflected, e.g., in the spin-spin correlations that acquire a staggered component. 
Due to the spontaneous breaking of translation symmetry, this  BCS-type instability  shows similar features to the usual spin-Peierls instability, except that the dimerization always sets in for infinitesimal interaction strength. 
As was shown in Ref.~\cite{SpinPeierlsHyperoctagon}, any additional time-reversal symmetric interaction will, independent of the microscopic details,  give rise to this kind of instability. 
For the Kitaev model on lattice (10,3)a, time-reversal symmetry ensures that the Fermi surface cannot be gapped out completely,  and  an odd number of nodal lines remains.
For the Kitaev model on lattice (8,3)a, one finds four Majorana surfaces instead of (the minimal number of) two.
This additional freedom, in principle, allows for interactions to gap out the system completely. 

One way to stabilize the Majorana Fermi surfaces is by breaking time-reversal symmetry.
This leads to a deformation of the Majorana Fermi surfaces (as illustrated, e.g., in Fig.~\ref{fig:10a_kappa}) that destroys the perfect nesting condition, \ie, the original nesting vector $\kk$ does not map the two Majorana Fermi surfaces onto each other any longer.
For the lattice (10,3)a, breaking time-reversal symmetry causes elongating/flattening of the surface along the four 111 directions as shown in Fig.~\ref{fig:SpinPeierls} (a); for (8,3)a the surfaces are elongated/shortened along the 001 direction as shown in Fig.~\ref{fig:SpinPeierls} (b).
In both cases, translation along $\kk$ does not map the surfaces exactly onto each other.
The resulting mismatch in energy cuts off the BCS instability at low enough temperatures and restores the  Fermi surface.


\section{Outlook}
\label{sec:outlook}

\begin{figure*}[ht!]
  \includegraphics[width=\linewidth]{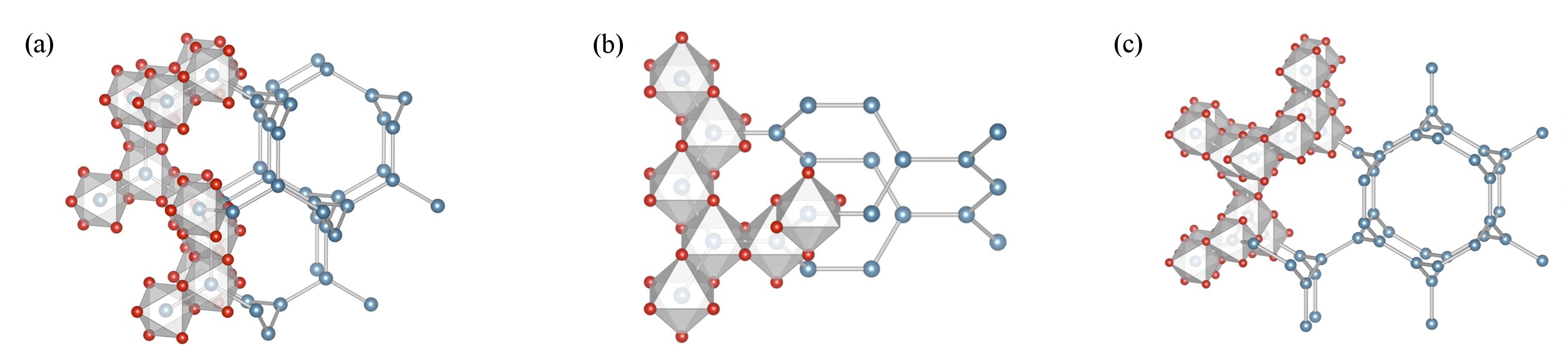}
  \caption{(Color online) The lattices (a) (10,3)a, (b) (10,3)b, and (c) (8,3)b can be embedded in a network of edge-sharing octahedra -- the elementary building block of all candidate materials realizing bond-directional Kitaev-type exchange \cite{Khaliullin}. Lattice (10,3)b is realized in the iridate \bLiIrO, while the other two lattice structures still await a material realization in the form of a spin-orbit entangled Mott insulator.}
  \label{fig:materials}
\end{figure*}

The physics of  fractionalization, accompanied by the formation of spin-liquid ground states, is beautifully embodied in the Kitaev model.
The model is unique in that it allows to precisely track on an analytical level the splitting of the original spin degrees of freedom into Majorana fermions and a \Z\ gauge field.
This study has explored this phenomenon for three-dimensional Kitaev models with a focus on the collective physics of the itinerant Majorana fermions, the formation of Majorana metals whose nature intimately depends on the topology of the underlying lattice.
We have provided a comprehensive classification of these Majorana metals for the most elementary tricoordinated lattices in three spatial dimensions (summarized in Tables \ref{tab:lattice_overview} and  \ref{tab:majorana_metals}), which is rooted in an elementary symmetry analysis of the projective time-reversal and inversion symmetries for these lattices.
Focusing primarily on the Majorana physics, our study already attests to the rich physics of three-dimensional Kitaev models, while also pointing to a number of future research directions which we briefly comment on in the following.

On a conceptual level, it will be interesting to complement the current analysis with a more rigorous study of the physics of the \Z\ gauge field for the family of three-dimensional Kitaev models at hand.
Even for the pure Kitaev model, for which the gauge field remains static, the ground state of the gauge field might be somewhat non-trivial for some of the lattices under scrutiny in this study.
In particular, for lattice (9,3)a we have found possible evidence that some of the low-energy states or even the  ground state of the  \Z\ gauge field might break some of the point-group symmetries of the lattice.
To further elucidate this possibility, a more stringent numerical approach is needed such as the Monte Carlo sampling approach recently developed by the Motome group \cite{finiteTGauge1}.
Such an unbiased, ergodic sampling approach would also be helpful in understanding the low-energy physics of the \Z\ gauge field for lattice (8,3)c, for which we find evidence of geometric frustration in the assignment of \Z\ fluxes, possibly leading to an extensive degeneracy of gauge field configurations in the ground state.
It will be interesting to further explore how this possibly somewhat unusual physics of the \Z\ gauge field couples back to the formation of a collective state of the Majorana fermions.

One key distinction between Kitaev models in two and three spatial dimensions is their finite-temperature behavior.
For the three-dimensional Kitaev models we expect to observe a finite-temperature phase transition at which the \Z\ gauge field orders \cite{SenthilFisher,finiteTGauge2}.
Indeed such a finite-temperature phase transition has recently been observed in Monte Carlo simulations of the (10,3)b hyperhoneycomb lattice \cite{finiteTGauge1}.
Generically, this transition is expected to be a continuous phase transition in the inverted 3D Ising universality class.
However, this must not be the case for lattice geometries where the zero-temperature physics indicates a breaking of time-reversal and/or point-group symmetries in addition to the \Z\ symmetry associated with the gauge theory. 
It will be interesting to numerically explore whether these systems exhibit a {\em single} continuous finite-temperature phase transition, at which simultaneously multiple symmetries are broken spontaneously.
If so, they will establish remarkable instances of lattice models whose critical behavior evades a description in terms of the Landau-Ginzburg-Wilson (LGW) paradigm.
In fact, such unconventional finite-temperature transitions in Kitaev models might provide the most ``natural'' instances of non-LGW  criticality in three-dimensional systems, which so far has been explored in the context of classical dimer \cite{nonLGW-classical1} and loop \cite{nonLGW-classical2} models -- somewhat more artificially constructed systems that have no direct realization in a microscopic setting.

Complementary to this study of the gapless spin liquids arising for roughly equal coupling strength $J_x \sim J_y \sim J_z$ in three-dimensional Kitaev models, it should be compelling to systematically investigate the gapped spin liquids which arise when one of the three couplings dominates over the other two.
In particular, it will be interesting to go beyond initial studies of the (10,3)b hyperhoneycomb lattice \cite{Nasu14} and see whether some of these gapped phases give rise to non-trivial loop statistics \cite{Levin14}.

Another important avenue will be to study the effect of disorder in these three-dimensional Kitaev models.
Such an analysis is of particular interest for the Weyl spin liquids found in a number of lattice geometries (see Table \ref{tab:majorana_metals} and the discussion in Sec.~\ref{sec:Weyl}). 
What sets these Weyl spin liquids apart from their electronic counterparts is their respective symmetry classification in terms of the ten-fold way classification scheme of free-fermion systems \cite{AltlandZirnbauer}.
Weyl semimetals in electronic systems can arise from breaking either time-reversal or inversion symmetry, putting these systems into symmetry classes A or AII, respectively.
In contrast, Weyl physics in Majorana systems (which per se exhibit particle-hole symmetry) can arise from breaking time-reversal symmetry (akin to the electronic system) but also without breaking of time-reversal symmetry {\it nor} inversion symmetry, putting the Majorana systems into symmetry classes D or BDI, respectively. 
It will be interesting to identify to what extent  disorder physics in Weyl semimetals is sensitive to the underlying symmetry class.

One of the most salient future research directions will be to go beyond the pure Kitaev model and to study the physics arising from additional spin exchanges such as a Heisenberg exchange.
These magnetic interactions have two effects.
They render dynamics to the \Z\ gauge field, allowing the elementary vison excitations to disperse through the lattice.
At the same time these additional spin exchanges also induce interactions between the Majorana fermions, possibly destabilizing the Majorana metal of the pure Kitaev model as discussed in Sec. \ref{sec:spin-peierls} on the spin-Peierls instability of Majorana metals with a Majorana Fermi surface \cite{SpinPeierlsHyperoctagon}.

Finally, we hope that our study highlighting the rich physics arising from fractionalization in three-dimensional Kitaev models will provide further stimulus to the ongoing search for materials realizing spin-orbit entangled $j=1/2$ Mott insulators with strong bond-directional Kitaev-type exchanges. It would be most enthralling if candidate $j=1/2$ Mott materials for tricoordinated lattice structures beyond the (10,3)b hyperhoneycomb lattice of \bLiIrO\ (and its higher harmonic in \cLiIrO) could be realized. While certainly not the focus of the current study, we might offer some minimal guidance for this experimental search in noting that one prerequisite facilitating the emergence of strong bond-directional Kitaev exchanges is the occurrence of double exchange paths  between the spin-orbit entangled moments on the tricoordinated lattice structure \cite{Khaliullin}. One way to realize such a scenario is found in the iridates  $A_2$IrO$_3$ (with $A=$Na, Li) where $5d^5$ Ir$^{4+}$ ions are embedded in IrO$_6$ oxygen cages that form an {\em edge-sharing} network. Such an edge-sharing network structure can also be realized for the lattices (10,3)a and (8,3)b visualized in Fig.~\ref{fig:materials} in addition to the experimentally observed realization of the lattice (10,3)b in \bLiIrO. It would be thus highly tantalizing if spin-orbit entangled $j=1/2$ Mott insulators could be realized for one of these lattice structures.

%
%
%
%


\acknowledgments
We thank L. Balents, Y. Motome, J. Nasu, A. Rosch, and A. Vishwanath for insightful discussions. 
M.H. acknowledges partial support through the Emmy-Noether program of the DFG.
The numerical simulations were performed on the CHEOPS cluster at RRZK Cologne.
All authors acknowledge hospitality of the KITP during the final stages of this work.
This research was supported in part by the National Science Foundation under Grant 
No. NSF PHY11-25915.



\appendix

\section{Three-dimensional Kitaev lattices}
\label{sec:appendix3DKitaev}

In this first appendix we expand the discussion of the lattice structure for each  of the tricoordinated lattices of Table~\ref{tab:lattice_overview}.
In particular, we show visualizations of each lattice structure along various high-symmetry projections, with the underlying VESTA \cite{VESTA} 
visualization files provided in the Supplemental Material of this  paper. The latter also define the crystallographic axes referred to in the captions. 
In addition, we give detailed information on the space group of each lattice and the Wyckoff positions for their unit cell. 

\subsection{(8,3)a}
Lattice (8,3)a is described by the hexagonal space group $P6_2 22$ (No. 180) with $c/a=(3\sqrt{2})/5$.
The Wyckoff positions for the unit cell are $6(i)$ with $x=\frac{2}{5}$.

This lattice is bipartite with $\kk = \bq_3/2$, where $\bq_3$ is a reciprocal lattice vector defined in Eq.~\eqref{eq:recip_vec_8_3a}, due to $\bq_3$ connecting different sublattices, and lacks inversion symmetry.
There are two distinct sets of $x$-, $y$- and $z$-bonds which cannot be related by symmetries, those which make up the co-rotating spirals [see Fig.~\ref{fig:comparison_8_3_a_b}(a)] and those that connect two nearest-neighbor spirals.
All bonds of a given set are related by a combination of $C_2$ symmetry and a three-fold screw rotation.
The symmetry between $x$-, $y$- and $z$-bonds is reflected in a phase diagram symmetric in all couplings.

\begin{figure}[h!]
  \includegraphics[width=\linewidth]{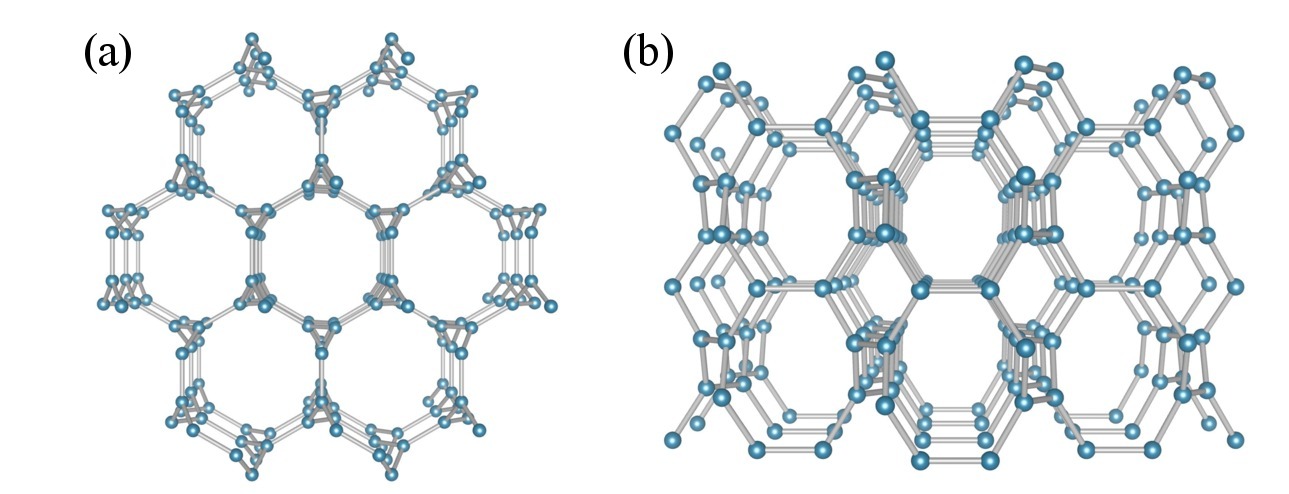}
  \caption{(Color online) The (8,3)a lattice viewed along the (a) crystallographic $\hat c$ axis  and (b) crystallographic $\hat a$ axis .}
  \label{fig:sym_8_3_a}
\end{figure}

The lattice (8,3)a is depicted in certain high-symmetry projections in Fig.~\ref{fig:sym_8_3_a}:
in (a) the lattice is viewed along a three-fold screw axis
and (b) shows the lattice along a two-fold rotation axis.

The high-symmetry points shown in the dispersion plot of the main text are defined as
\begin{align}
  \Gamma  &= \Big( 0, 0, 0 \Big)\,,
  &M      &= -\left( \pi, \frac{\pi}{\sqrt{3}}, 0 \right)\,, \nonumber\\
  K       &= -\left( \frac{2\pi}{3}, \frac{2\pi}{\sqrt{3}}, 0 \right)\,,
  &A      &= -\left( 0, 0, \frac{5\pi}{3\sqrt{2}} \right)\,, \nonumber\\
  H       &= -\left( \frac{2\pi}{3}, \frac{2\pi}{\sqrt{3}}, \frac{5\pi}{3\sqrt{2}} \right)\,,
  &L      &= -\left( \pi, \frac{\pi}{\sqrt{3}}, \frac{5\pi}{3\sqrt{2}} \right)\,.
\end{align}
%

\subsection{(8,3)b}
Lattice (8,3)b is described by the trigonal space group $R\bar{3}m$ (No. 166) with $c/a = \sqrt{6}/5$.
The Wyckoff positions for the (hexagonal) unit cell are $18(f)$ with $x=\frac{2}{5}$.

This lattice is bipartite with $\kk = (\bq_1 + \bq_3)/2$, where $\bq_1$ and $\bq_3$ are reciprocal lattice vectors defined in Eq.~\eqref{eq:recip_vec_8_3b}, due to $\bq_1$ and $\bq_3$ connecting different sublattices, and possesses inversion symmetry with vanishing $\kt$.
There are two distinct sets of $x$-, $y$- and $z$-bonds which cannot be related by symmetries, those which make up the counter-rotating spirals [see Fig.~\ref{fig:comparison_8_3_a_b}(b)] and those which connect them.
All bonds of a given set are related by a combination of $C_3$ and inversion symmetries.
The symmetry between $x$-, $y$- and $z$-bonds is reflected in a phase diagram symmetric in all couplings.

\begin{figure}
  \includegraphics[width=\linewidth]{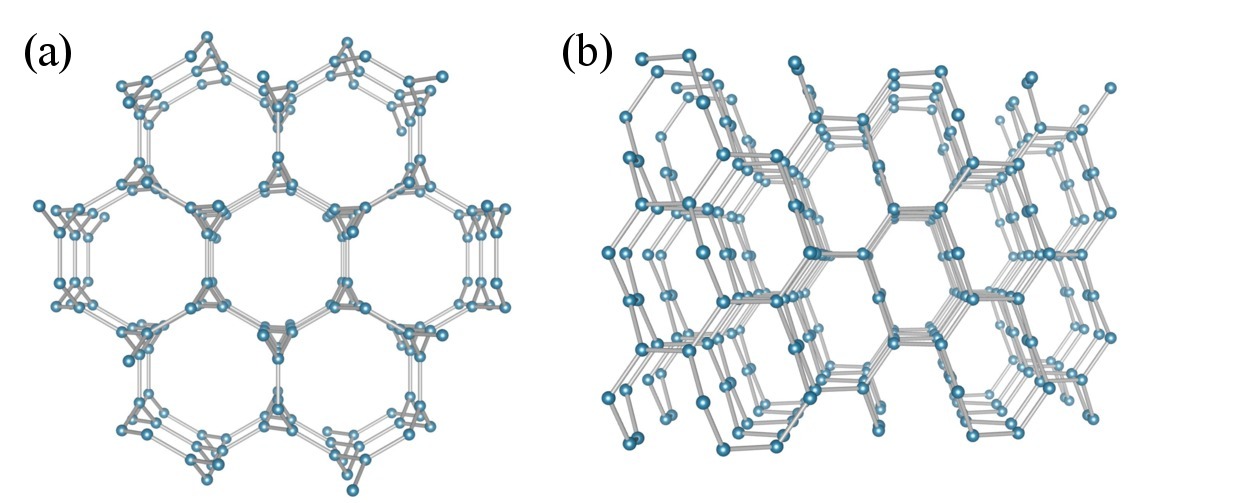}
  \caption{(Color online) The (8,3)b lattice viewed along the (a) crystallographic $\hat c$ axis   and (b) crystallographic $\hat a$ axis.}
  \label{fig:sym_8_3_b}
\end{figure}

The lattice (8,3)b is depicted in certain high-symmetry projections in Fig.~\ref{fig:sym_8_3_b}:
in (a) the lattice is viewed along a three-fold rotation axis
and (b) shows the lattice along a two-fold rotation axis.
%

\subsection{(8,3)c}

Lattice (8,3)c is described by the hexagonal space group $P6_3/mmc$ (No. 194) with $c/a = 2/5$.
The Wyckoff positions for the unit cell are $2(c)$ and $6(h)$ with $x=\frac{7}{15}$.

This lattice is bipartite with vanishing $\kk$ and possesses inversion symmetry with vanishing $\kt$.
There are two distinct sets of $x$-, $y$- and $z$-bonds; those forming the zig-zag chains along $\hat z$, and those lying in the $xy$-plane. 
The sets of all $x$-, $y$- and $z$-bonds are related to each other by a six-fold screw rotation, which is reflected in a phase diagram symmetric in all couplings.

\begin{figure}[h!]
  \includegraphics[width=\linewidth]{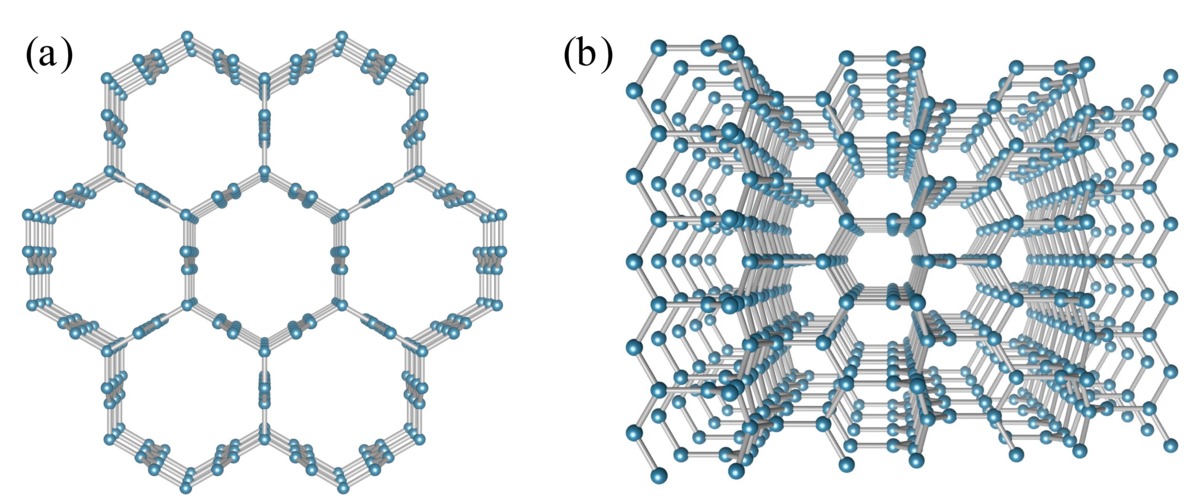}
  \caption{(Color online) The (8,3)c lattice viewed along the (a) crystallographic $\hat c$ axis and (b) crystallographic $\hat a$ axis.}
  \label{fig:sym_8_3_c}
\end{figure}

Figure~\ref{fig:sym_8_3_c}(a) depicts the lattice along this six-fold screw rotation axis (alternatively three-fold rotation axis), while (b) shows the lattice along the crystallographic $\hat a$ axis.

The high-symmetry points shown in the dispersion plot of the main text are defined as
\begin{align}
  \Gamma  &= \Big( 0, 0, 0 \Big)\,,
  &M      &= \left( \pi, \frac{\pi}{\sqrt{3}}, 0 \right)\,, \nonumber\\
  K       &= \left( \frac{2\pi}{3}, \frac{2\pi}{\sqrt{3}}, 0 \right)\,,
  &H      &= \left( \frac{2\pi}{3}, \frac{2\pi}{\sqrt{3}}, \frac{5\pi}{2} \right)\,, \nonumber\\
  L       &= \left( \pi, \frac{\pi}{\sqrt{3}}, \frac{5\pi}{2} \right)\,.
\end{align}
%

\subsection{(8,3)n}

Lattice (8,3)n is described by the tetragonal space group $I4/mmm$ (No. 139) with $c/a = 4/(2\sqrt{3} + \sqrt{2})$.
The Wyckoff positions for the unit cell are $16(k)$ $(x, \frac{1}{2}+x, \frac{1}{4})$ with $x=\frac{\sqrt{3} + \sqrt{2}}{2(2\sqrt{3} + \sqrt{2})}$, and $16(n)$ $(0xz)$ with $x=\frac{\sqrt{3} + \sqrt{2}}{2(2\sqrt{3} + \sqrt{2})}$ and $z=\frac{1}{8}$.

This lattice is bipartite with vanishing $\kk$ and possesses inversion symmetry with $\kt = \bq_3/2$, where $\bq_3$ is a reciprocal lattice vector defined in Eq.~\eqref{eq:recip_vec_8_3n}.
All $x$- and $y$-bonds are related by a combination of inversion, four-fold rotation, and mirror symmetries.
The $z$-bonds come in two distinct sets: those that lie in the $xy$ plane and connect nearest-neighbor "squares", and those that are along $\hat z$ and connect neighboring "square-octagon planes."
Within each set, the bonds are related to each other by four-fold rotation symmetry. 
However, $z$-bonds  are not related to any other bond type by lattice symmetries.
The symmetry between $x$- and $y$- bonds is reflected in a phase diagram symmetric under $J_x \leftrightarrow J_y$.

\begin{figure}[h!]
  \includegraphics[width=\linewidth]{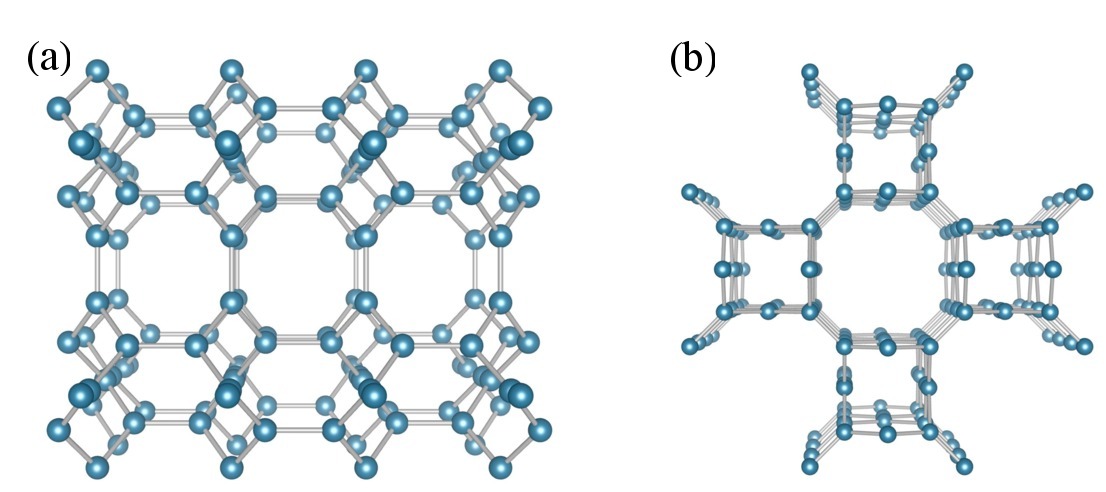}
  \caption{(Color online) The (8,3)n lattice viewed along  (a)  $\hat a+\hat b$ axis  and (b)  $\hat c$ axis.}
  \label{fig:sym_8_3_n}
\end{figure}

The lattice (8,3)n is depicted in certain high-symmetry projections in Fig.~\ref{fig:sym_8_3_n} --
in (a) the lattice is viewed along a two-fold rotation axis
and (b) shows the lattice along a four-fold rotation axis.

The high-symmetry points shown in the dispersion plot of the main text are defined as
\begin{align}
  \Gamma  &= \Big( 0, 0, 0 \Big)\,, \nonumber\\
  X       &= \Big( \pi, -\pi, 0 \Big)\,, \nonumber\\
  M       &= \Big( 0, -2\pi, 0 \Big)\,, \nonumber\\
  Z       &= \left( 0, 0, \frac{1}{20}(-3\sqrt{2} + 26\sqrt{3}) \pi \right)\,, \nonumber\\
  P       &= \left( \pi, -\pi, \frac{1}{4}(\sqrt{2} + 2\sqrt{3}) \pi \right)\,, \nonumber\\
  N       &= \left( 0, -\pi, \frac{1}{4}(\sqrt{2} + 2\sqrt{3}) \pi \right)\,, \nonumber\\
  Z_1     &= \left( 0, -2\pi, \frac{1}{20}(13\sqrt{2} - 6\sqrt{3}) \pi \right)\,.
\end{align}
%

\subsection{(9,3)a}

Lattice (9,3)a is described by the trigonal space group $R\bar{3}m$ (No. 166) with $c/a = \frac{\sqrt{6(4 + \sqrt{3})}}{1 + 2\sqrt{3}}$.
The Wyckoff positions for the (hexagonal) unit cell are $18(f)$ $(xyz)$ with $x=\frac{\sqrt{3}}{1 + 2\sqrt{3}}$, $y = z = 0$, and $18(h)$ $(x0z)$ with $x=\frac{1+\sqrt{3}}{4(1 + 2\sqrt{3})}$ and $z=\frac{3}{4}$.

This lattice is the only \textit{non}-bipartite lattice considered in this paper.
It possesses inversion symmetry with vanishing $\kt$.
All $x$- and $y$-bonds are related by a combination of $C_3$ and mirror symmetries.
There are two distinct sets of $z$-bonds which are not related by any symmetries, however, all bonds of a given set may be mapped to each other via $C_3$ symmetry.
The symmetry between $x$- and $y$-bonds is reflected in a phase diagram symmetric under $J_x \leftrightarrow J_y$.

\begin{figure}[h!]
  \includegraphics[width=\linewidth]{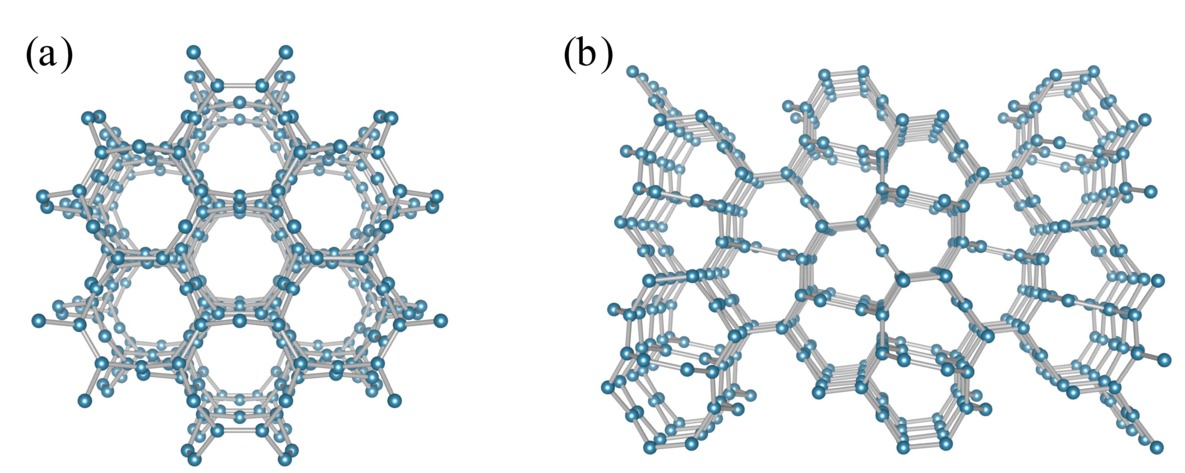}
  \caption{(Color online) The (9,3)a lattice viewed along the (a) crystallographic $\hat c$ axis  and (b)  crystallographic $\hat b$ axis.}
  \label{fig:sym_9_3_a}
\end{figure}

The lattice (9,3)a is depicted along the three-fold rotation axis  in Fig.~\ref{fig:sym_9_3_a}(a) and along $\hat b$ in Fig.~\ref{fig:sym_9_3_a}(b).

%

\subsection{(10,3)a}
Lattice (10,3)a is described by the cubic space group $I4_1 32$ (No. 214).
The Wyckoff positions for the unit cell are $8(a)$ $(\frac{1}{8}\frac{1}{8}\frac{1}{8})$.

This lattice is bipartite with $\kk = (-\bq_2 + \bq_3)/2$, where $\bq_2$ and $\bq_3$ are reciprocal lattice vectors defined in Eq.~\eqref{eq:recip_vec_10_3a}, but lacks inversion symmetry.
All $x$-, $y$- and $z$-bonds are related by a combination of $C_3$ symmetry and a four-fold screw rotation, which is reflected in a phase diagram symmetric in all couplings.

\begin{figure}[h!]
  \includegraphics[width=\linewidth]{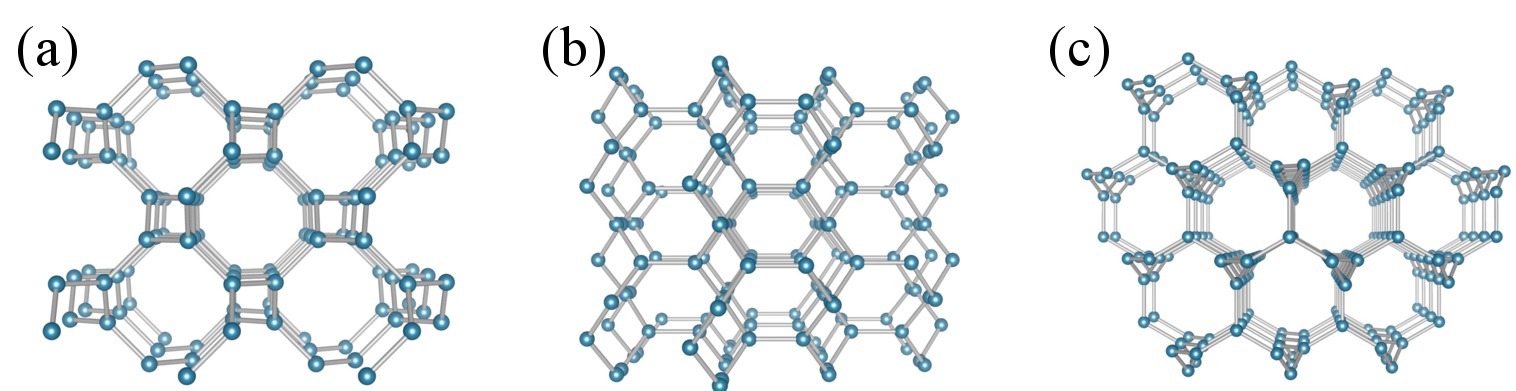}
  \caption{(Color online) The (10,3)a lattice viewed along  the (a) four-fold screw rotation axis $\hat a$, (b) two-fold  rotation axis $\hat a+\hat b$, and (c) three-fold rotation axis $\hat a+\hat b+\hat c$.}
  \label{fig:sym_10_3_a}
\end{figure}

The lattice (10,3)a is depicted in certain high-symmetry projections in Fig.~\ref{fig:sym_10_3_a}:
in (a) the lattice is viewed along a four-fold screw axis and shows the lattice in its square-octagon projection,
(b) shows the lattice along a two-fold rotation axis,
and (c) shows the lattice viewed along a three-fold rotation axis.

The high-symmetry points shown in the dispersion plot of the main text are defined as
\begin{align}
  \Gamma  &= \left( 0, 0, 0 \right)\,,
  &P      &= -\left( \pi, \pi, \pi \right)\,, \nonumber\\
  N       &= -\left( \pi, \pi, 0 \right)\,,
  &H      &= -\left( 0, 2\pi, 0 \right)\,.
\end{align}
%

\subsection{(10,3)b}
Lattice (10,3)b is described by the tetragonal space group $I4_1/amd$ (No. 141) with $c/a=2\sqrt{3}$.
The Wyckoff positions for the unit cell are $8(e)$ $(00z)$ with $z=\frac{1}{12}$.

This lattice is bipartite with vanishing $\kk$ and possesses inversion symmetry with vanishing $\kt$.
All $x$- and $y$-bonds are related by a combination of $C_2$ symmetry and a two-fold screw rotation.
Additionally, all $z$-bonds are related to each other by inversion symmetry, but are not related to any other bond type by lattice symmetries.
The symmetry between $x$- and $y$-bonds is reflected in a phase diagram symmetric under $J_x \leftrightarrow J_y$.

\begin{figure}[h!]
  \includegraphics[width=\linewidth]{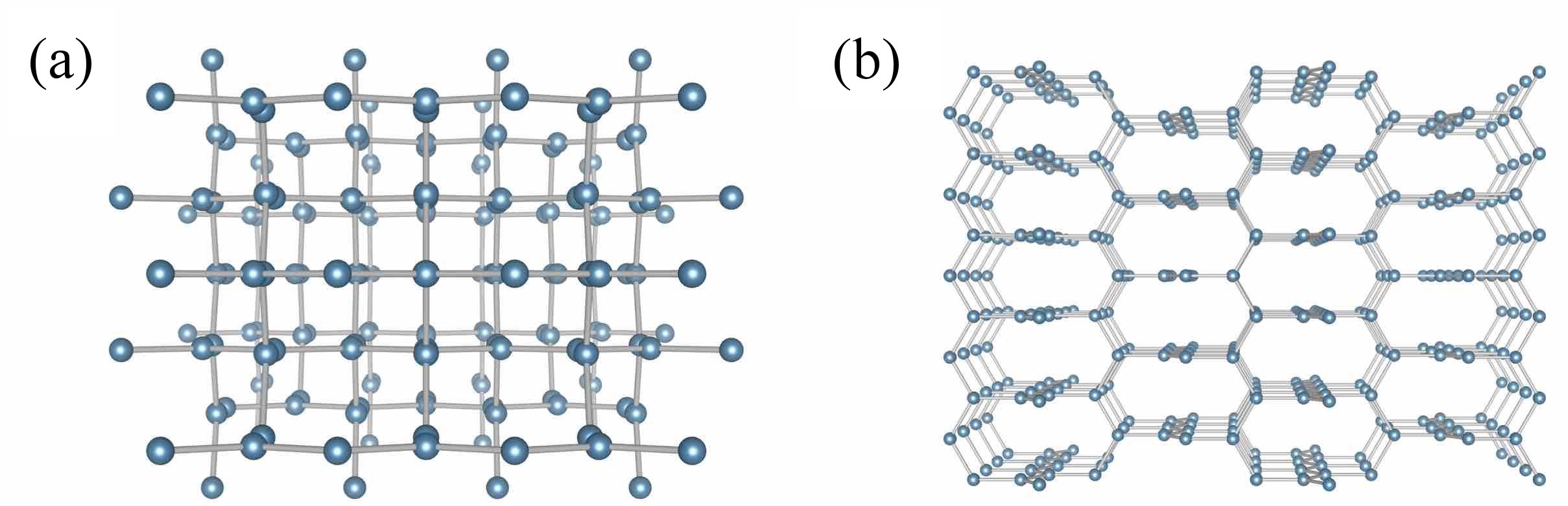}
  \caption{(Color online) The (10,3)b lattice viewed along the (a) crystallographic $\hat c$ axis and (b) crystallographic $\hat b$ axis.}
  \label{fig:sym_10_3_b}
\end{figure}

The lattice (10,3)b is depicted in certain high-symmetry projections in Fig.~\ref{fig:sym_10_3_b} --
in (a) the lattice is viewed along  the four-fold screw rotation axis
and (b) shows the lattice along a two-fold screw axis.

The high-symmetry points shown in the dispersion plot of the main text are defined as
\begin{align}
  \Gamma  &= \Big( 0, 0, 0 \Big)\,,
  &X      &= \left( -\frac{29\pi}{72}, \frac{29\pi}{72}, 0 \right)\,, \nonumber\\
  Y       &= \left( 0, 0, -\frac{\pi}{2} \right)\,,
  &A_1    &= \left( -\frac{11\pi}{72}, \frac{11\pi}{72}, -\frac{\pi}{2} \right)\,, \nonumber\\
  Z       &= \left( \frac{\pi}{6}, \frac{\pi}{6}, 0 \right)\,,
  &T      &= \left( \frac{\pi}{6}, \frac{\pi}{6}, -\frac{\pi}{2} \right)\,.
\end{align}
%

\subsection{(10,3)c}
Lattice (10,3)c is described by the trigonal space group $P3_1 12$ (No. 151) with $c/a = (3\sqrt{3})/2$.
The Wyckoff positions for the unit cell are $6(c)$ with $x=\frac{1}{3}$, $y=\frac{1}{6}$, and $z=\frac{1}{9}$.

This lattice is bipartite with vanishing $\kk$, but lacks inversion symmetry.
All $x$- and $y$-bonds are related by a six-fold screw rotation.
Additionally, all $z$-bonds are related to each other by the same six-fold screw rotation, but are not related to any other bond type by lattice symmetries.
The symmetry between $x$- and $y$-bonds is reflected in a phase diagram symmetric under $J_x \leftrightarrow J_y$.

\begin{figure}[h!]
  \includegraphics[width=\linewidth]{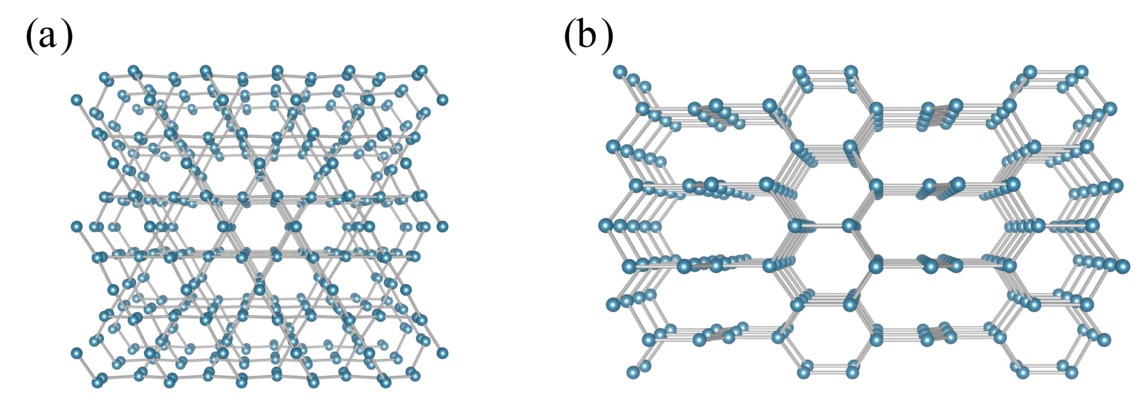}
  \caption{(Color online) The (10,3)c lattice viewed along the (a) crystallographic $\hat c$ axis  and (b) crystallographic $\hat b$ axis .}
  \label{fig:sym_10_3_c}
\end{figure}

The lattice (10,3)c is depicted in certain high-symmetry projections in Fig.~\ref{fig:sym_10_3_c} --
in (a) the lattice is viewed along a three-fold screw axis and shows the lattice in its kagome projection
and (b) shows the lattice along a two-fold screw axis.

The high-symmetry points shown in the dispersion plot of the main text are defined as
\begin{align}
  \Gamma  &= \Big( 0, 0, 0, \Big)\,,
  &M      &= \left( \pi, \frac{\pi}{\sqrt{3}}, 0 \right)\,, \nonumber\\
  X       &= \Big( \pi, 0, 0 \Big)\,,
  &R      &= \left( \pi, 0 ,\frac{2\pi}{3\sqrt{3}} \right)\,, \nonumber\\
  A       &= \left( \pi, \frac{\pi}{\sqrt{3}}, \frac{2\pi}{3\sqrt{3}} \right)\,.
\end{align}
%

%
%
%
\section{Three-dimensional Kitaev lattices}
\label{sec:supplemental}

In this second appendix, we provide additional information on the 3D Kitaev models for the tricoordinated lattices reviewed in the first appendix. 
Our focus here is on a detailed expos\'e of the gauge structure of the ground state of the Kitaev model defined on one of these lattices.
In particular, we give an explicit expression for the Kitaev Hamiltonian in its Majorana representation in this gauge.

%
\subsection{(8,3)a}
\label{ssec:supplemental_8_a}
\begin{figure}[h]
  \includegraphics[width=\linewidth]{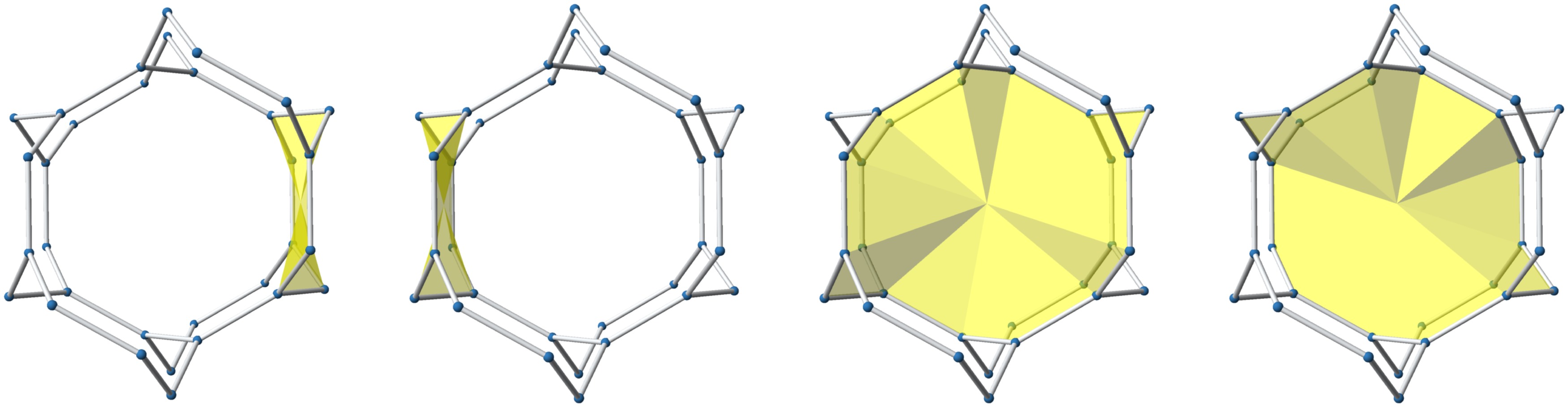}
  \caption{(Color online) Loop operators of the lattice (8,3)a forming a volume constraint.}
  \label{fig:volumes_8_3a}
\end{figure}

\begin{figure}[h]
  \includegraphics[width=\linewidth]{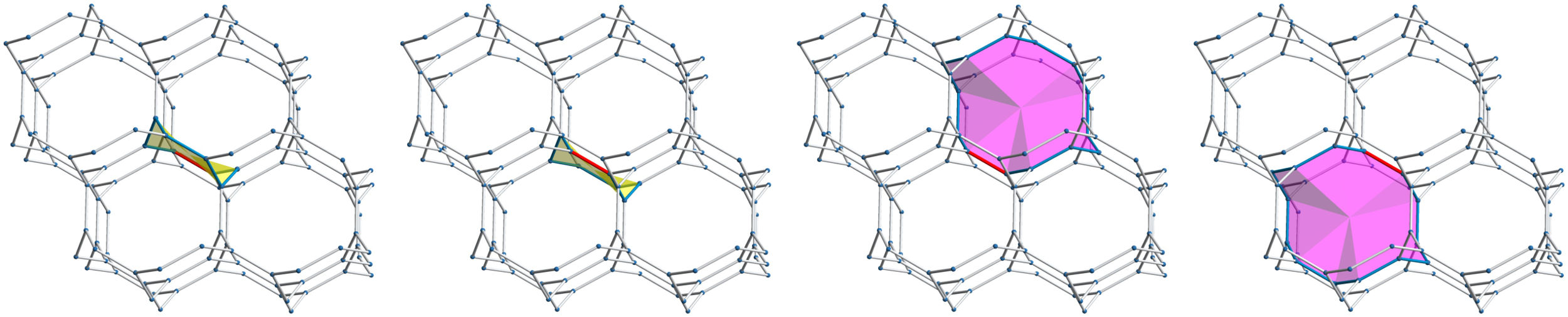}
  \caption{(Color online) Vison excitation threading two plaquettes of length 8 in the lattice (8,3)a, as depicted on the left. In addition, it threads several plaquettes of length 14, two examples of which are depicted on the right shaded in magenta. The flipped bond operator is pictured in red.}
  \label{fig:vison_8_3a}
\end{figure}

The lattice (8,3)a possesses three loop operators of length 8 and three of length 14 per unit cell.
These six loop operators can be combined to form three closed volumes, each of which must have vanishing total flux, resulting in only three linearly independent loop operators per unit cell.
One of these closed volumes is illustrated in Fig.~\ref{fig:volumes_8_3a}.
The remaining two volumes are related by a three-fold screw rotation.
The smallest vison loop in this lattice threads two plaquettes of length 8, as well as several plaquettes of length 14, as visualized in Fig.~\ref{fig:vison_8_3a}. 

The calculations for lattice (8,3)a were performed using the following reference gauge:
\begin{align}
  u^x_{21}  &= +1,  & u^y_{14}  &= +1,  & u^z_{15}  &= +1,  \nonumber\\
  u^x_{34}  &= +1,  & u^y_{26}  &= +1,  & u^z_{24}  &= +1,  \nonumber\\
  u^x_{56}  &= +1,  & u^y_{35}  &= +1,  & u^z_{36}  &= +1.
\end{align}
In this gauge, the momentum space Hamiltonian reads as
\begin{equation}
  H(\bk) =
  \begin{pmatrix}
    0            & -i A_3       & 0      & i J_y & i A_2        & 0     \\
    i A^\star_3  & 0            & 0      & i J_z & 0            & i A_1 \\
    0            & 0            & 0      & i J_x & i J_y        & i J_z \\
    -i J_y       & -i J_z       & -i J_x & 0     & 0            & 0     \\
    -i A^\star_2 & 0            & -i J_y & 0     & 0            & i A_3 \\
    0            & -i A^\star_1 & -i J_z & 0     & -i A^\star_3 & 0     \\
  \end{pmatrix},
\end{equation}
where
\begin{align}
  A_1 &= e^{2 i k_1 \pi } J_y , \nonumber\\
  A_2 &= e^{-2 i k_2 \pi } J_z , \nonumber\\
  A_3 &= e^{-2 i k_3 \pi } J_x.
\end{align}
The gauge-fixed matrix representations of the symmetry operators relevant to our classification scheme are
\begin{equation}
  U_{\mbox{\tiny SLS}} = U_{\mbox{\tiny T}} =
  \begin{pmatrix}
    \id_{3\times 3} & 0                \\
    0               & -\id_{3\times 3}
  \end{pmatrix}.
\end{equation}
%

%
\subsection{(8,3)b}
\label{ssec:supplemental_8_b}
\begin{figure}[h]
  \includegraphics[width=\linewidth]{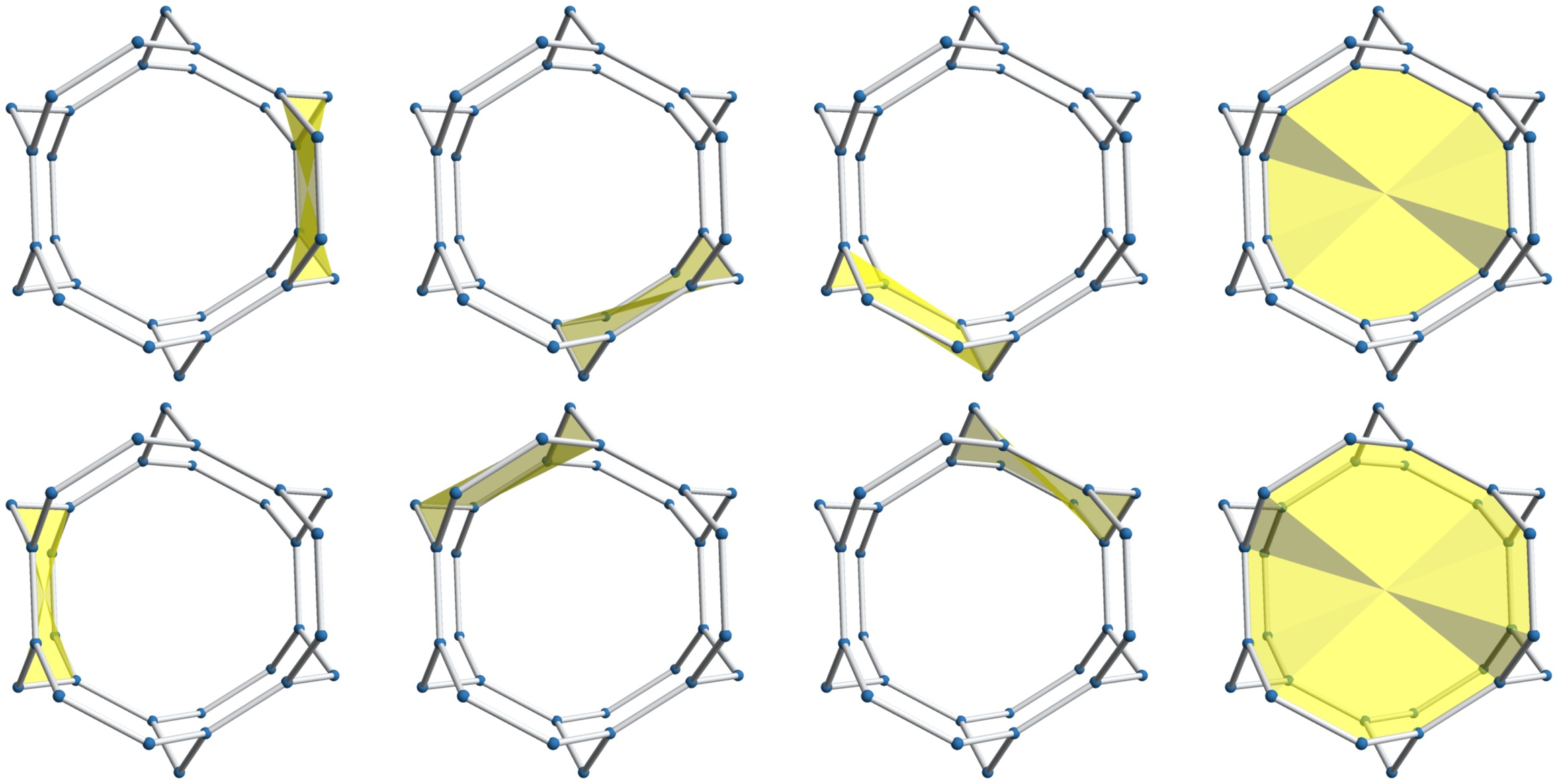}
  \caption{(Color online) Loop operators of the lattice (8,3)b forming a volume constraint. The loop operators in the bottom row are related to those in the top row by lattice translation vectors.}
  \label{fig:volumes_8_3b}
\end{figure}

\begin{figure}[h]
  \includegraphics[width=\linewidth]{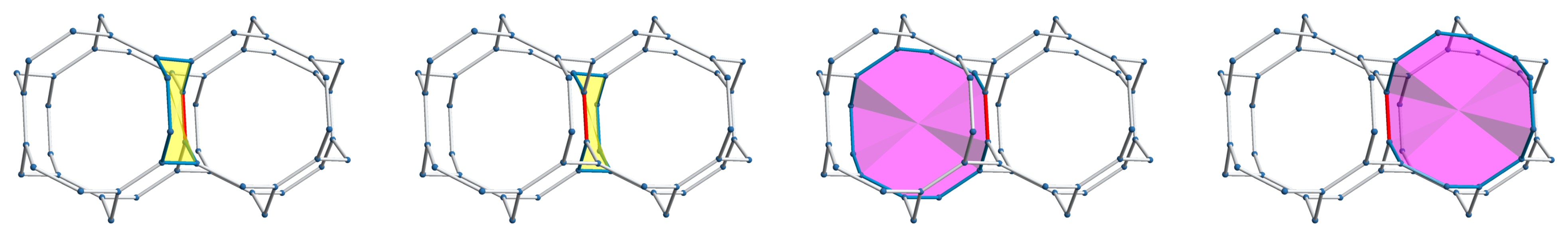}
  \caption{(Color online) Vison excitation threading four plaquettes in the lattice (8,3)b -- two of length 8 shaded in yellow and two of length 12 shaded in magenta. The flipped bond operator is pictured in red.}
  \label{fig:vison_8_3b}
\end{figure}

The lattice (8,3)b possesses three loop operators of length 8 and one of length 12 per unit cell.
These four loop operators can be combined with four loop operators from neighboring unit cells to form a closed volume which must have vanishing total flux, resulting in only three linearly independent loop operators per unit cell.
This closed volume is illustrated in Fig.~\ref{fig:volumes_8_3b}.
The smallest vison loop in this lattice threads two  plaquettes of length 8 and two of length 12 and is visualized in Fig.~\ref{fig:vison_8_3b}.

The calculations for the lattice (8,3)b were performed using the following reference gauge:
\begin{align}
  u^x_{43}  &= +1,  & u^y_{42}  &= +1,  & u^z_{14}  &= +1,  \nonumber\\
  u^x_{21}  &= +1,  & u^y_{53}  &= +1,  & u^z_{25}  &= +1,  \nonumber\\
  u^x_{56}  &= +1,  & u^y_{61}  &= +1,  & u^z_{36}  &= +1.
\end{align}
In this gauge, the momentum space Hamiltonian reads as
\begin{equation}
  H(\bk) =
  \begin{pmatrix}
    0              & -i A_3 & 0           & i J_z  & 0            & -i A_{13} \\
    i A^\star_3    & 0      & 0           & -i J_y & i J_z        & 0         \\
    0              & 0      & 0           & -i A_2 & -i J_y       & i J_z     \\
    -i J_z         & i J_y  & i A^\star_2 & 0      & 0            & 0         \\
    0              & -i J_z & i J_y       & 0      & 0            & i A_3     \\
    i A^\star_{13} & 0      & -i J_z      & 0      & -i A^\star_3 & 0         \\
  \end{pmatrix},
\end{equation}
where
\begin{align}
  A_{13} &= e^{-2 i (k_1+k_3) \pi } J_y ,\nonumber\\
  A_2    &= e^{2 i k_2 \pi } J_x ,\nonumber\\
  A_3    &= e^{-2 i k_3 \pi } J_x.
\end{align}
The gauge-fixed matrix representations of the symmetry operators relevant to our classification scheme are
\begin{equation}
  U_{\mbox{\tiny SLS}} = U_{\mbox{\tiny T}} =
  \begin{pmatrix}
    \id_{3\times 3} & 0                \\
    0               & -\id_{3\times 3}
  \end{pmatrix}
\end{equation}
and
\begin{equation}
  U_{\mbox{\tiny I}} =
  \begin{pmatrix}
    0  & 0  & 0  & 0 & 0 & 1 \\
    0  & 0  & 0  & 0 & 1 & 0 \\
    0  & 0  & 0  & 1 & 0 & 0 \\
    0  & 0  & -1 & 0 & 0 & 0 \\
    0  & -1 & 0  & 0 & 0 & 0 \\
    -1 & 0  & 0  & 0 & 0 & 0
    \end{pmatrix}.
\end{equation}

\begin{figure}[h]
  \includegraphics[width=\linewidth]{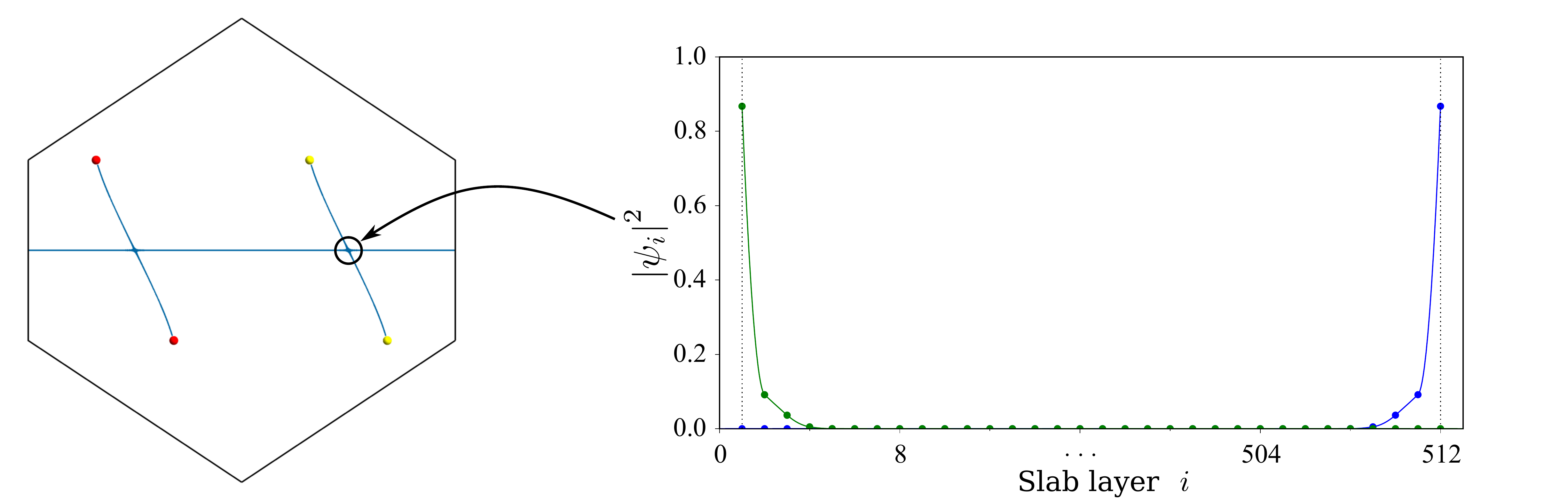}
  \caption{(Color online) \textit{(left)} Fermi arc surface states in the [001]-surface Brillouin zone of lattice (8,3)b for a slab geometry 512 layers thick. \textit{(right)} Probability densities as functions of slab layer of the zero energy states with momentum corresponding to one of the crossing points of the Fermi arcs.}
  \label{fig:surface_crossing}
\end{figure}

As mentioned in Sec.~\ref{ssec:8b}, the exchange couplings of the Kitaev Hamiltonian can be tuned such that the surface Fermi arcs touch one another.
As the exchange couplings are tuned away from this point, the Fermi arcs split once more, connecting the same pairs of Weyl points as before, but winding differently around the surface Brillouin zone (see Fig.~\ref{fig:fermi_arcs_8_3b} in the main text).
This crossing of Fermi arcs is pictured in Fig.~\ref{fig:surface_crossing} with $J_x=J_y=(1-J_z)/2$ and $J_z \approx 0.201$ for a slab geometry with 512 layers.
Also pictured in Fig.~\ref{fig:surface_crossing} are the probability densities as functions of slab layer of the zero energy states with momentum corresponding to one of the crossing points of the Fermi arcs.
The zero modes continue to be highly localized to the surface as the Fermi arcs cross, and the splitting/reconnecting of the Fermi arcs appears to be a purely surface effect with no corresponding bulk effects.
%

%
\subsection{(8,3)c}
\label{ssec:supplemental_8_c}
\begin{figure}[h]
  \includegraphics[width=\linewidth]{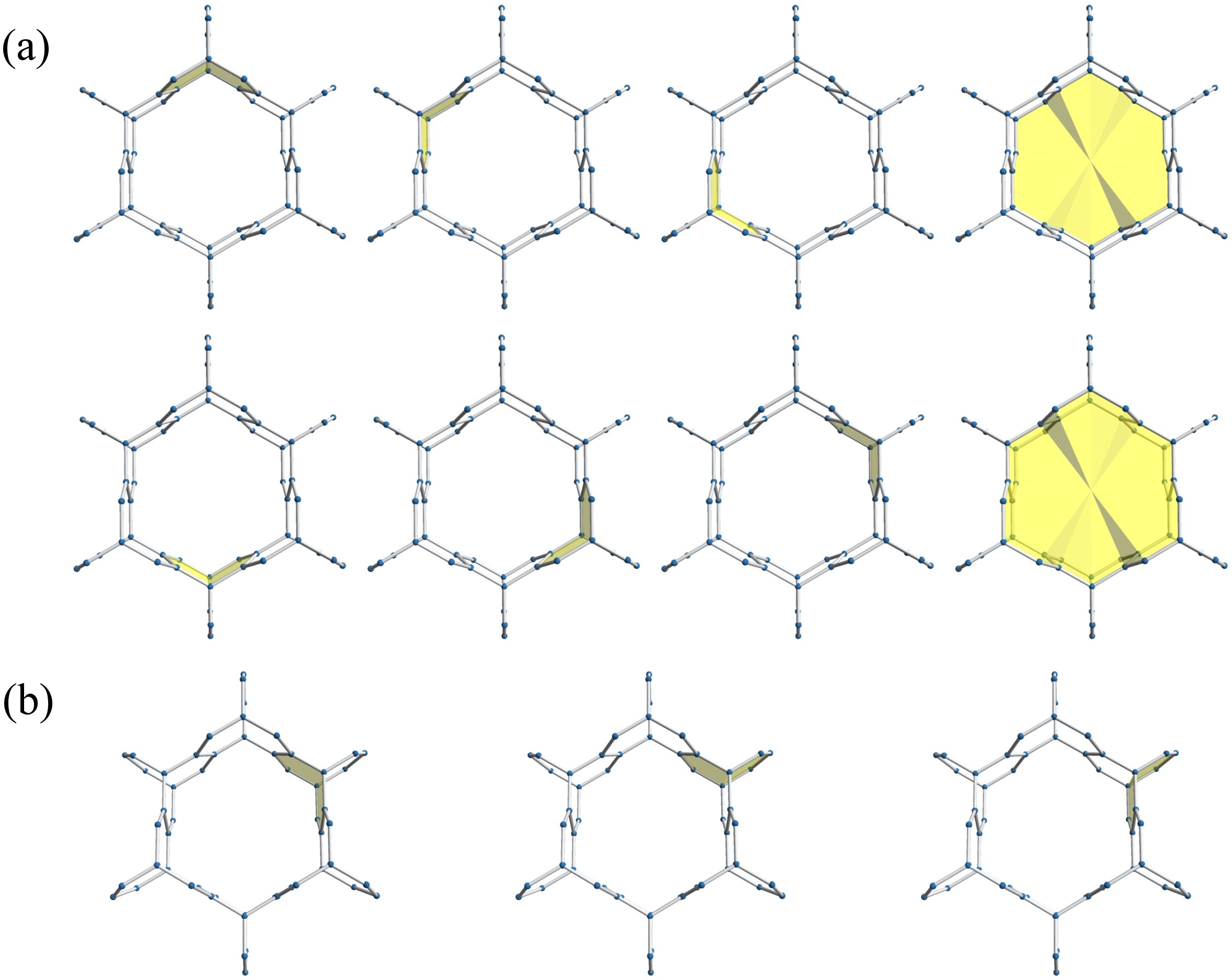}
  \caption{(Color online) Loop operators of the lattice (8,3)c forming two unique volume constraints in (a) and (b).}
  \label{fig:volumes_8_3c}
\end{figure}

\begin{figure}[h]
  \includegraphics[width=\linewidth]{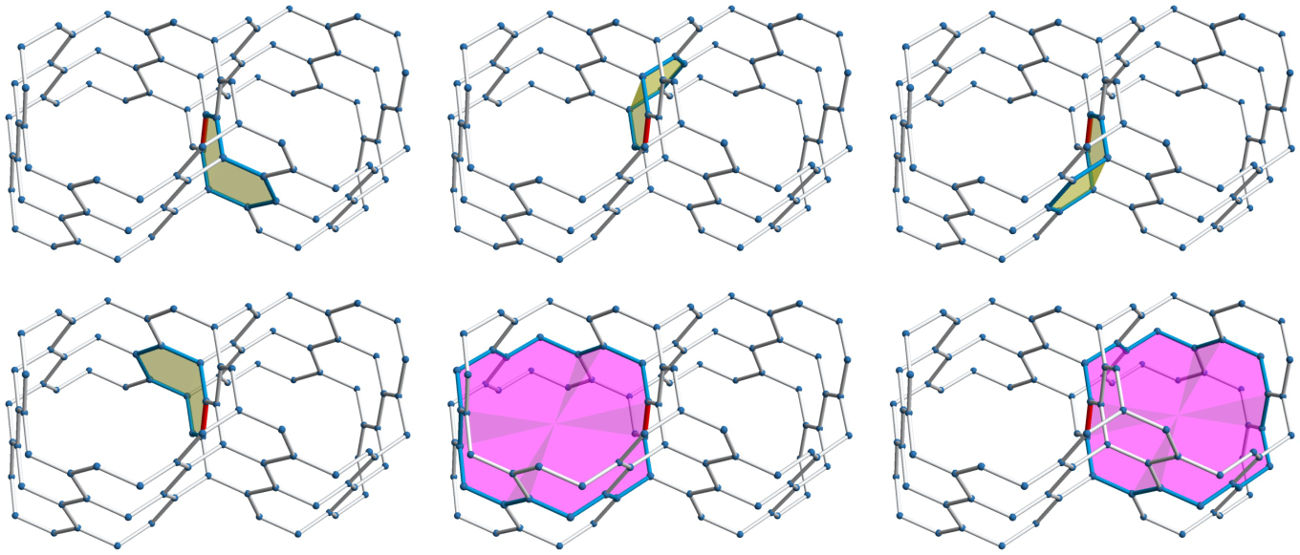}
  \caption{(Color online) Vison excitation threading four plaquettes of length 8 (shaded in yellow) in the lattice (8,3)c.  The flipped bond operator is pictured in red.}
  \label{fig:vison_8_3c}
\end{figure}

The lattice (8,3)c possesses six loop operators of length 8 and one of length 18 per unit cell.
These seven loop operators can be combined to form three closed volumes, each of which must have vanishing total flux, resulting in only four linearly independent loop operators per unit cell.
Two of these volumes are constructed only from loops of length 8 and are related to each other by a six-fold screw rotation.
The remaining volume is constructed from six loops of length 8 and two of length 18.
This larger closed volume and one of the smaller closed volumes are illustrated in Fig.~\ref{fig:volumes_8_3c}.
The smallest vison loop in this lattice threads four plaquettes of length 8, visualized in Fig.~\ref{fig:vison_8_3c} shaded in yellow.
A number of plaquettes of length 18  are also excited -- two examples of such plaquettes (shaded in magenta) are visualized in the second row in Fig.~\ref{fig:vison_8_3c}. 

When considering the assignment of fluxes to the loop operators in this lattice, one might be guided by Lieb's theorem and wish to assign $\pi$ flux through the plaquettes of length 8.
However, such a flux assignment is frustrated due to the volume constraints discussed above.
In Fig.~\ref{fig:volumes_8_3c}(b) one sees that it takes three such loop operators to form a closed volume.
Such a volume must have vanishing total flux, thus making it impossible to assign $\pi$ flux through all plaquettes of length 8. In the main text, we thus considered the zero flux sector, the only remaining flux sector that obeys all the lattice symmetries. 

The calculations for lattice (8,3)c were performed using the following reference gauge
\begin{align}
  u^x_{15}  &= +1  & u^y_{18}  &= +1  & u^z_{18}  &= +1  \nonumber\\
  u^x_{27}  &= +1  & u^y_{27}  &= +1  & u^z_{25}  &= +1  \nonumber\\
  u^x_{36}  &= +1  & u^y_{35}  &= +1  & u^z_{36}  &= +1  \nonumber\\
  u^x_{48}  &= +1  & u^y_{46}  &= +1  & u^z_{47}  &= +1.
\end{align}
In this gauge, the momentum space Hamiltonian reads
\begin{equation}
  H(\bk) =
  \begin{pmatrix}
    0              & A(\bk) \\
    A^\dagger(\bk) & 0 \\
  \end{pmatrix},
\end{equation}
where the matrix $A(\bk)$ is given by
\begin{equation}
  A(\bk) =
  \begin{pmatrix}
    i J_x & 0       & 0        & i A_{13} \\
    i J_z & 0       & i A_{23} & 0        \\
    i J_y & i A_{3} & 0        & 0        \\
    0     & i J_y   & i J_z    & i J_x    \\
  \end{pmatrix}
\end{equation}
and
\begin{align}
  A_{13} &= e^{-2 i k_1 \pi} J_y + e^{-2 i (k_1 + k_3) \pi} J_z,\nonumber\\
  A_{23} &= e^{2 i k_2 \pi} J_x + e^{2 i (k_2 - k_3) \pi} J_y,\nonumber\\
  A_3    &= e^{-2 i k_3 \pi} J_x+J_z.
\end{align}
The gauge-fixed matrix representations of the symmetry operators relevant to our classification scheme are
\begin{equation}
  U_{\mbox{\tiny SLS}} = U_{\mbox{\tiny T}} =
  \begin{pmatrix}
    \id_{4\times 4} & 0                \\
    0               & -\id_{4\times 4}
  \end{pmatrix}
\end{equation}
and
\begin{equation}
  U_{\mbox{\tiny I}} =
  \begin{pmatrix}
    0  & 0  & 0  & 0  & 0 & 0 & 0 & 1 \\
    0  & 0  & 0  & 0  & 0 & 0 & 1 & 0 \\
    0  & 0  & 0  & 0  & 0 & 1 & 0 & 0 \\
    0  & 0  & 0  & 0  & 1 & 0 & 0 & 0 \\
    0  & 0  & 0  & -1 & 0 & 0 & 0 & 0 \\
    0  & 0  & -1 & 0  & 0 & 0 & 0 & 0 \\
    0  & -1 & 0  & 0  & 0 & 0 & 0 & 0 \\
    -1 & 0  & 0  & 0  & 0 & 0 & 0 & 0
    \end{pmatrix}.
\end{equation}
%

%
\subsection{(8,3)n}
\label{ssec:supplemental_8_n}
\begin{figure}[h]
  \includegraphics[width=\linewidth]{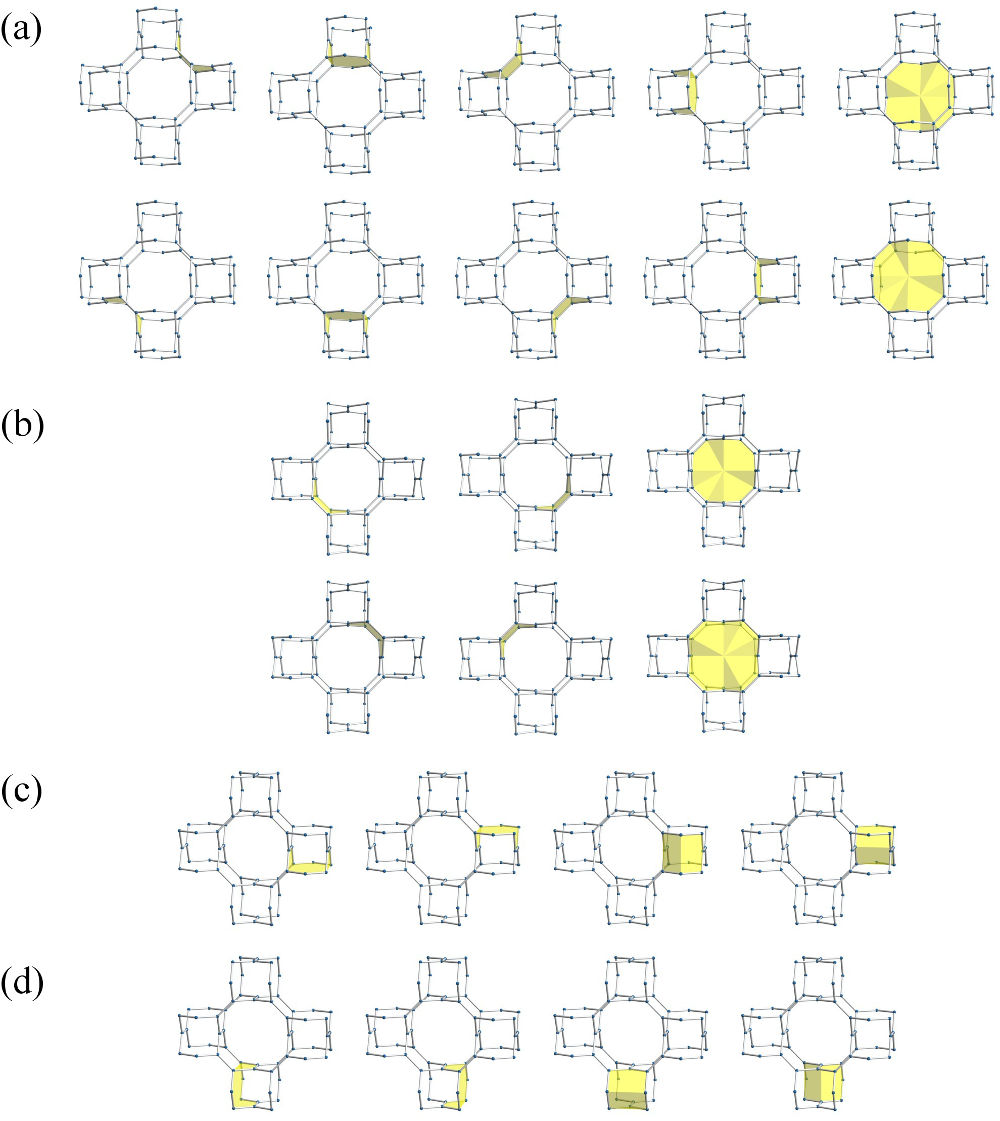}
  \caption{(Color online) Loop operators of the lattice (8,3)n forming four unique volume constraints in (a), (b), (c) and (d).}
  \label{fig:volumes_8_3n}
\end{figure}

\begin{figure}[h]
  \includegraphics[width=\linewidth]{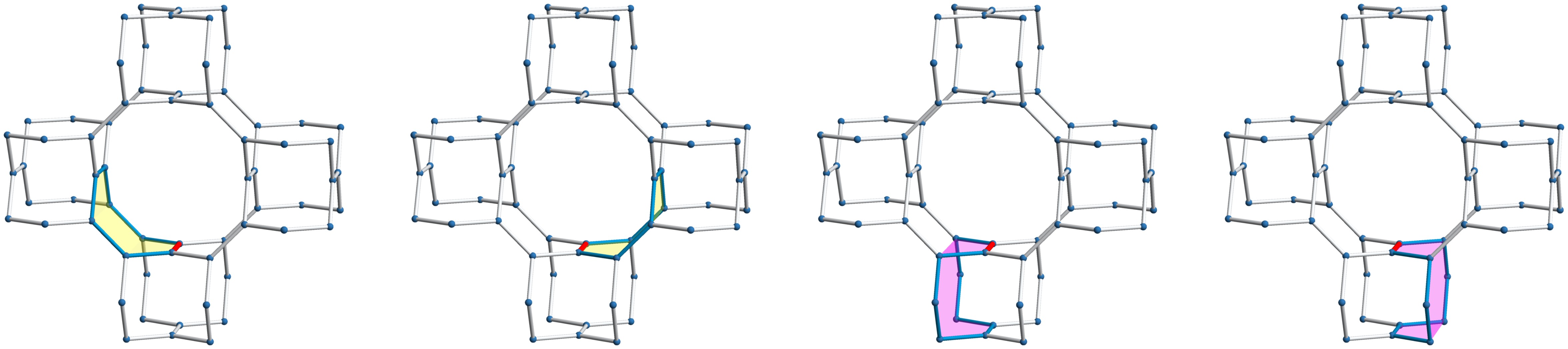}
  \caption{(Color online) Vison excitation threading two plaquettes of length 8 (yellow) and two plaquettes of length 10 (magenta) in the lattice (8,3)n. The flipped bond operator is pictured in red.}
  \label{fig:vison_8_3n}
\end{figure}

The lattice (8,3)n possesses six loop operators of length 8, four of length 10, and two of length 12 per unit cell.
These twelve loop operators can be combined to form four closed volumes, each of which must have vanishing total flux, resulting in only eight linearly independent loop operators per unit cell.
These closed volumes are illustrated in Fig.~\ref{fig:volumes_8_3n}.
The smallest vison loop in this lattice threads two plaquettes of length 8 and two of length 10 and is visualized in Fig.~\ref{fig:vison_8_3n}.

The calculations for lattice (8,3)n were performed using the following reference gauge
\begin{align}
  u^x_{8,9}  &= -1 & u^y_{7,14} &= +1  & u^z_{5,14} &= +1  \nonumber\\
  u^x_{16,2} &= +1 & u^y_{15,1} &= +1  & u^z_{9,1}  &= -1  \nonumber\\
  u^x_{14,6} &= +1 & u^y_{6,9}  &= +1  & u^z_{2,11} &= +1  \nonumber\\
  u^x_{1,10} &= +1 & u^y_{10,2} &= +1  & u^z_{12,4} &= +1  \nonumber\\
  u^x_{11,3} &= +1 & u^y_{3,12} &= +1  & u^z_{13,8} &= +1  \nonumber\\
  u^x_{4,13} &= +1 & u^y_{13,5} &= +1  & u^z_{3,15} &= -1  \nonumber\\
  u^x_{12,7} &= +1 & u^y_{11,8} &= +1  & u^z_{6,16} &= +1  \nonumber\\
  u^x_{5,15} &= +1 & u^y_{4,16} &= -1  & u^z_{10,7} &= -1.
\end{align}
In this gauge, the momentum space Hamiltonian reads
\begin{equation}
  H(\bk) =
  \begin{pmatrix}
    0              & A(\bk) \\
    A^\dagger(\bk) & 0 \\
  \end{pmatrix},
\end{equation}
where the matrix $A(\bk)$ is given by
\begin{widetext}
  \begin{equation}
    A(\bk) =
    \begin{pmatrix}
      i J_z  & i J_x        & 0          & 0          & 0                & 0      & -i J_y     & 0 \\
      0      & -i J_y       & i J_z      & 0          & 0                & 0      & 0          & -i J_x \\
      0      & 0            & -i J_x     & i J_y      & 0                & 0      & -i A_3 J_z & 0 \\
      0      & 0            & 0          & -i J_z     & i J_x            & 0      & 0          & -i  A_2 J_y \\
      0      & 0            & 0          & 0          & -i J_y           & i J_z  & i A_2 J_x  & 0 \\
      i J_y  & 0            & 0          & 0          & 0                & -i J_x & 0          & i A_{13} J_z \\
      0      & i A_{23} J_z & 0          & -i A_1 J_x & 0                & i J_y  & 0          & 0 \\
      -i J_x & 0            & -i A_1 J_y & 0          & -i A_3^\star J_z & 0      & 0          & 0 \\
    \end{pmatrix}
  \end{equation}
\end{widetext}
and
\begin{align}
  A_1    &= e^{-2 i k_1 \pi } ,\nonumber\\
  A_{13} &= e^{-2 i (k_1-k_3) \pi } \nonumber\\
  A_2    &= e^{2 i k_2 \pi } ,\nonumber\\
  A_{23} &= e^{2 i (k_2-k_3) \pi } \nonumber\\
  A_3    &= e^{2 i k_3 \pi }. 
\end{align}
The gauge-fixed matrix representations of the symmetry operators relevant to our classification scheme are
\begin{equation}
  U_{\mbox{\tiny SLS}} = U_{\mbox{\tiny T}} =
  \begin{pmatrix}
    \id_{8\times 8} & 0                \\
    0               & -\id_{8\times 8}
  \end{pmatrix}
\end{equation}
and
\begin{equation}
  U_{\mbox{\tiny I}} =
  \begin{pmatrix}
    0       & B(\bk) \\
    -B(\bk) & 0
  \end{pmatrix},
\end{equation}
where
\begin{equation}
  B(\bk) =
  \begin{pmatrix}
    B_3 & 0    & 0       & 0       & 0       & 0      & 0    & 0       \\
    0   & 0    & -B_{13} & 0       & 0       & 0      & 0    & 0       \\
    0   & 0    & 0       & 0       & 0       & 0      & 0    & -B_{13} \\
    0   & 0    & 0       & B_{123} & 0       & 0      & 0    & 0       \\
    0   & 0    & 0       & 0       & 0       & B_{23} & 0    & 0       \\
    0   & 0    & 0       & 0       & 0       & 0      & -B_3 & 0       \\
    0   & 0    & 0       & 0       & -B_{23} & 0      & 0    & 0       \\
    0   & -B_3 & 0       & 0       & 0       & 0      & 0    & 0       \\
  \end{pmatrix}
\end{equation}
and
\begin{align}
  B_{123} &= e^{-2 i (k_1+k_2-k_3) \pi } ,\nonumber\\
  B_{13}  &= e^{-2 i (k_1-k_3) \pi } ,\nonumber\\
  B_{23}  &= e^{-2 i (k_2-k_3) \pi } ,\nonumber\\
  B_3     &= e^{2 i k_3 \pi }.
\end{align}
%

%
\subsection{(9,3)a}
\label{ssec:supplemental_9_a}
\begin{figure}[h]
  \includegraphics[width=\linewidth]{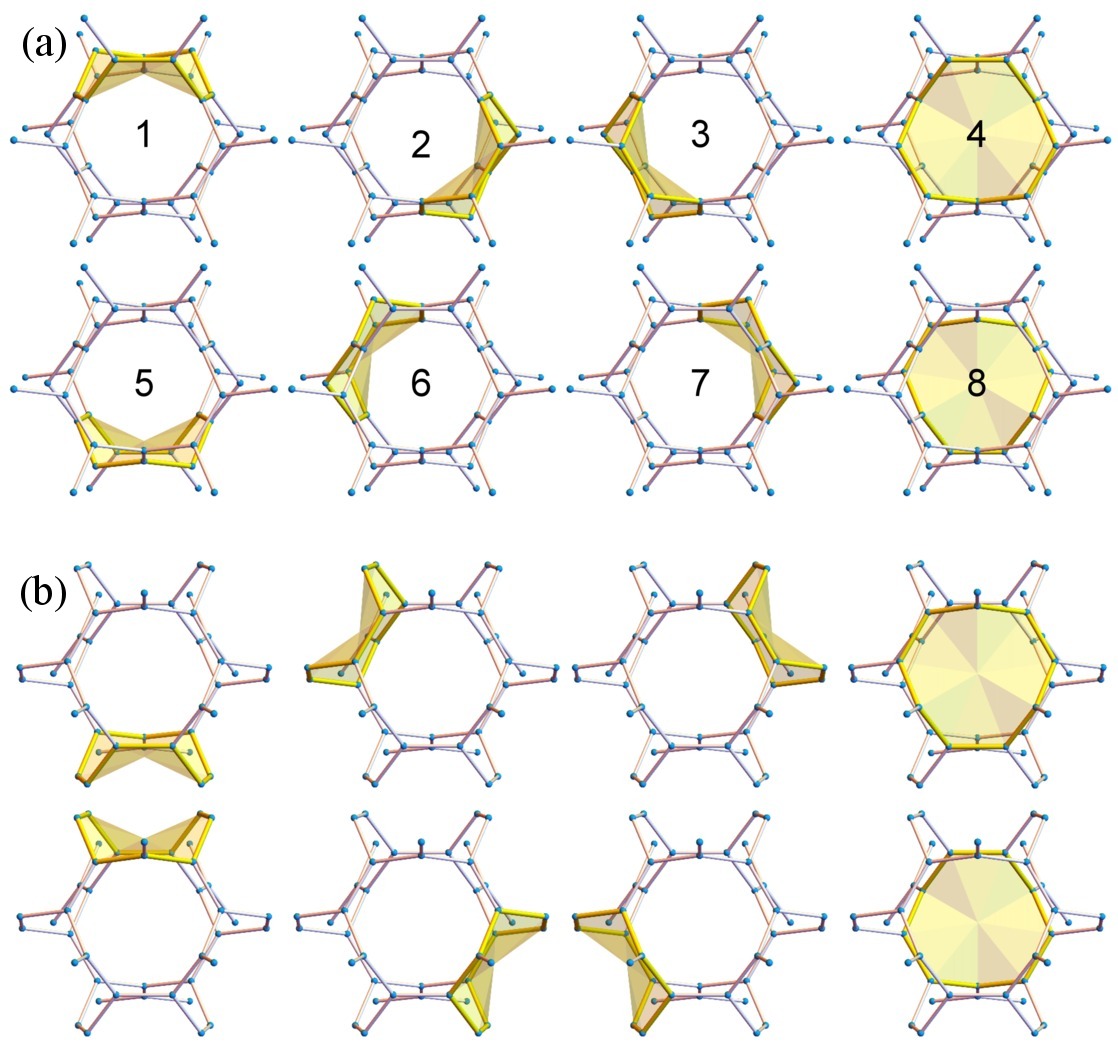}
  \caption{(Color online) The lattice (9,3)a has eight loops of length 9 per unit cell -- labeled by 1,$\ldots,$ 8 in (a) -- that are subject to two volume constraints shown in (a) and (b).}
  \label{fig:volumes_9_3a}
\end{figure}
\begin{figure}[h]
  \includegraphics[width=\linewidth]{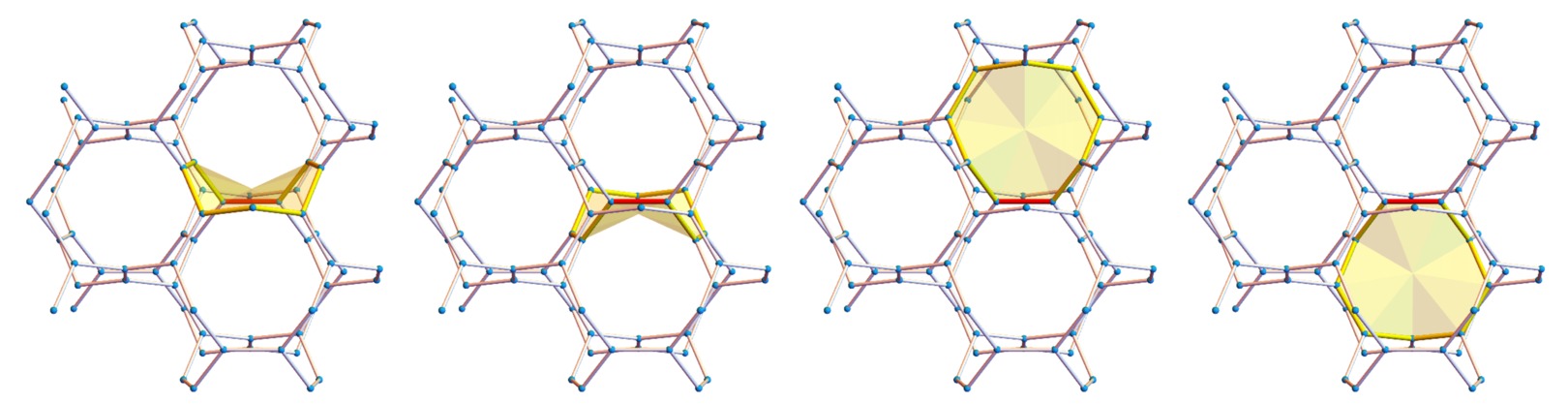}
  \caption{(Color online) Vison excitation threading four plaquettes of length 9 in the lattice (9,3)a. The flipped bond operator is pictured in red.}
  \label{fig:vison_9_3a}
\end{figure}

The lattice (9,3)a has 8 loops per unit cell, all of length 9.
These form two distinct volumes as shown in Fig.~\ref{fig:volumes_9_3a} where, for the sake of clarity, we only show a small section of the lattice.  
Flipping a $z$-type bond creates a vison loop of length four, visualized in Fig.~\ref{fig:vison_9_3a}.

Even though the flux sector considered in the main text is likely not the ground-state sector,  note that we can always stabilize this sector as the ground state by adding additional, local terms to the Hamiltonian. 
To that end, we define six 12-loops per unit cell by using the combinations $W_4(\mathbf R)\cdot W_j(\mathbf R)$ with $j=1,2,3$ and $W_8(\mathbf R)\cdot W_j(\mathbf R)$ with $j=5,6,7$ [see Fig.~\ref{fig:volumes_9_3a}(a) for the definition of the loops].
Assigning a negative energy to  $\pi$-flux for each of these loops stabilizes the flux sector considered in the main text.
Note that this necessarily implies that the flux through the remaining 12-loop in Eq.~\eqref{eq:12-loop} vanishes.
Thus, assigning $\pi$ flux through all 12-loops is prohibited, \ie, the Z$_2$ gauge theory in the (9,3)a lattice is "frustrated", similarly as for the lattice (8,3)c.

%
\subsection{(10,3)a}
\label{ssec:supplemental_10_a}
\begin{figure}[h]
  \includegraphics[width=\linewidth]{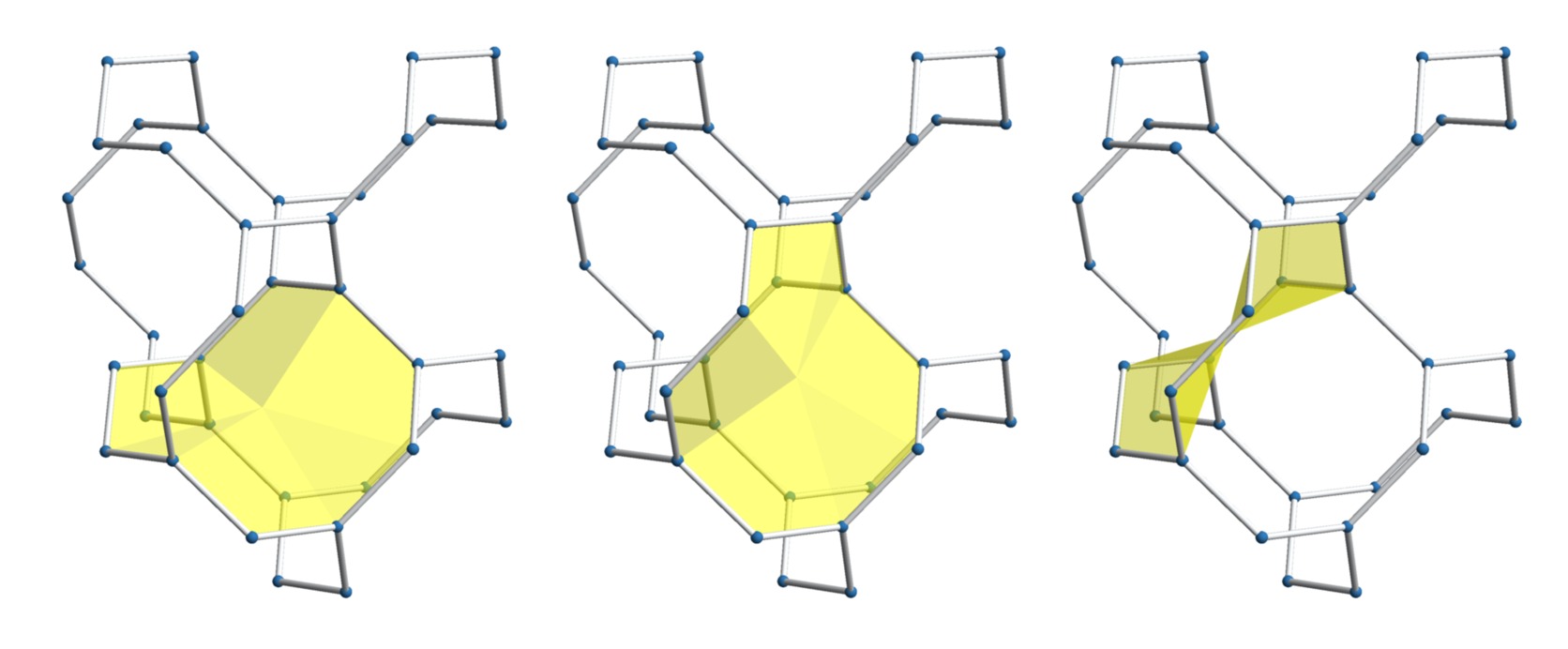}
  \caption{(Color online) Loop operators of the lattice (10,3)a forming a volume constraint.}
  \label{fig:volumes_10_3a}
\end{figure}

\begin{figure}[h]
  \includegraphics[width=\linewidth]{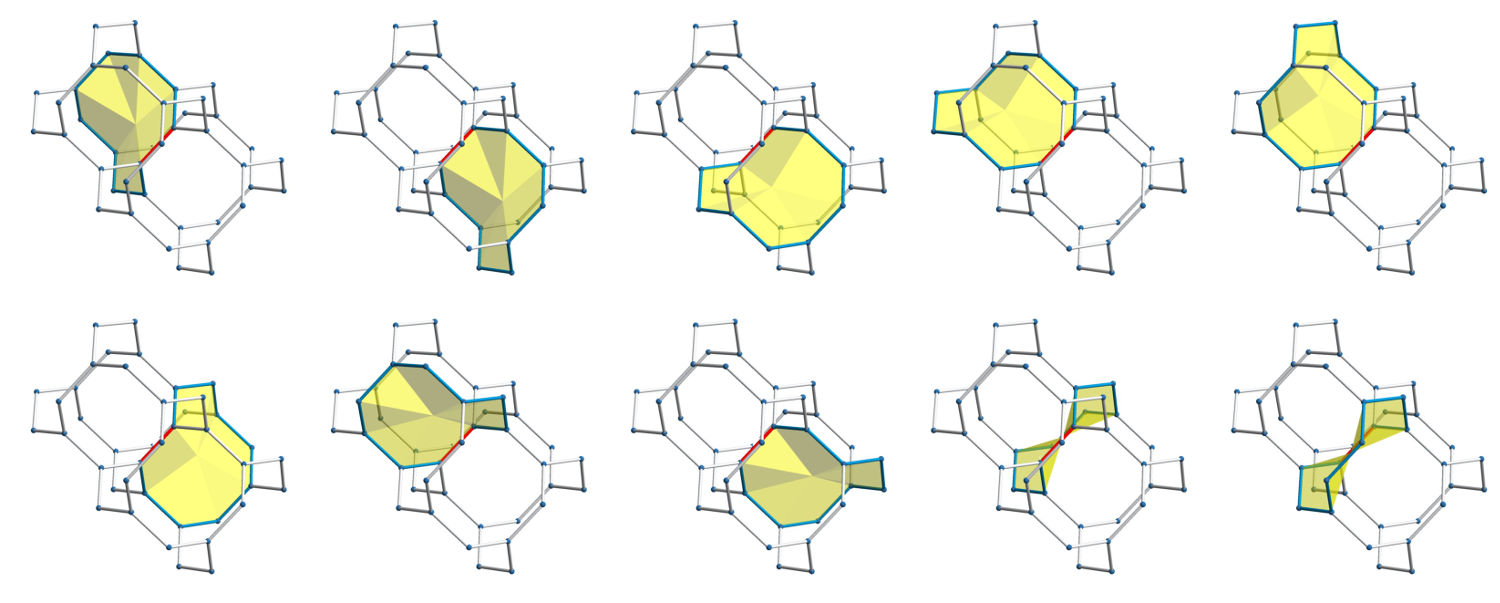}
  \caption{(Color online) Vison excitation threading ten plaquettes of length 10 in the lattice (10,3)a. The flipped bond operator is pictured in red.}
  \label{fig:vison_10_3a}
\end{figure}

The lattice (10,3)a possesses six loop operators of length 10 per unit cell.
These six loop operators can be combined into four closed volumes, each of which must have vanishing total flux, resulting in only two independent loop operators per unit cell.
One of these closed volumes is illustrated in Fig.~\ref{fig:volumes_10_3a}.
The remaining three volumes are related by a four-fold screw rotation.
The smallest vison loop in this lattice threads ten such plaquettes and is visualized in Fig.~\ref{fig:vison_10_3a}.

The calculations for lattice (10,3)a were performed using the following reference gauge:
\begin{align}
  u^x_{12}  &= -1,  & u^y_{13}  &= -1,  & u^z_{32}  &= +1,  \nonumber\\
  u^x_{34}  &= -1,  & u^y_{24}  &= +1,  & u^z_{14}  &= -1.
\end{align}
In this gauge, the momentum space Hamiltonian reads as
\begin{equation}
  H(\bk) = 
  \begin{pmatrix}
    0           & -i A_2 & -i J_y      & -i A_1 \\
    i A_2^\star & 0      & -i J_z      & i J_y  \\
    i J_y       & i J_z  & 0           & -i A_3 \\
    i A_1^\star & -i J_y & i A_3^\star & 0      \\
  \end{pmatrix},
\end{equation}
where
\begin{align}
  A_1 &= e^{-2 i k_1 \pi} J_z ,\nonumber\\
  A_2 &= e^{-2 i k_2 \pi} J_x ,\nonumber\\
  A_3 &= e^{-2 i k_3 \pi} J_x.
\end{align}
The gauge-fixed matrix representations of the symmetry operators relevant to our classification scheme are
\begin{equation}
  U_{\mbox{\tiny SLS}} = U_{\mbox{\tiny T}} = 
  \begin{pmatrix}
    \id_{2\times 2} & 0                \\
    0               & -\id_{2\times 2} \\
  \end{pmatrix}.
\end{equation}
%

%
\subsection{(10,3)b}
\label{ssec:supplemental_10_b}
\begin{figure}[h]
  \includegraphics[width=\linewidth]{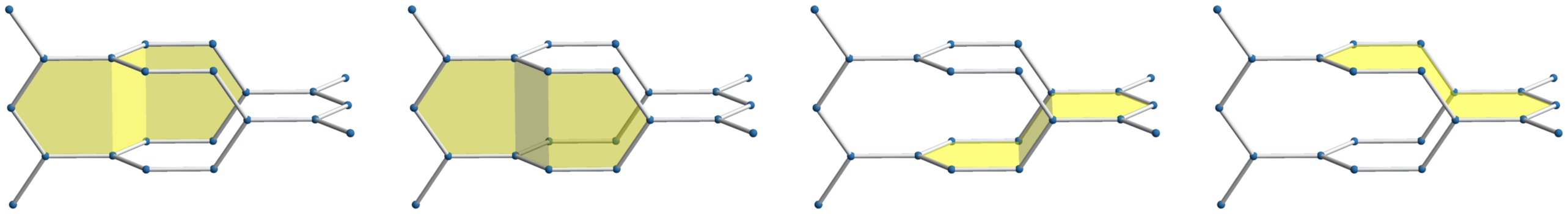}
  \caption{(Color online) Loop operators of the lattice (10,3)b forming a volume constraint.}
  \label{fig:volumes_10_3b}
\end{figure}

\begin{figure}[h]
  \includegraphics[width=\linewidth]{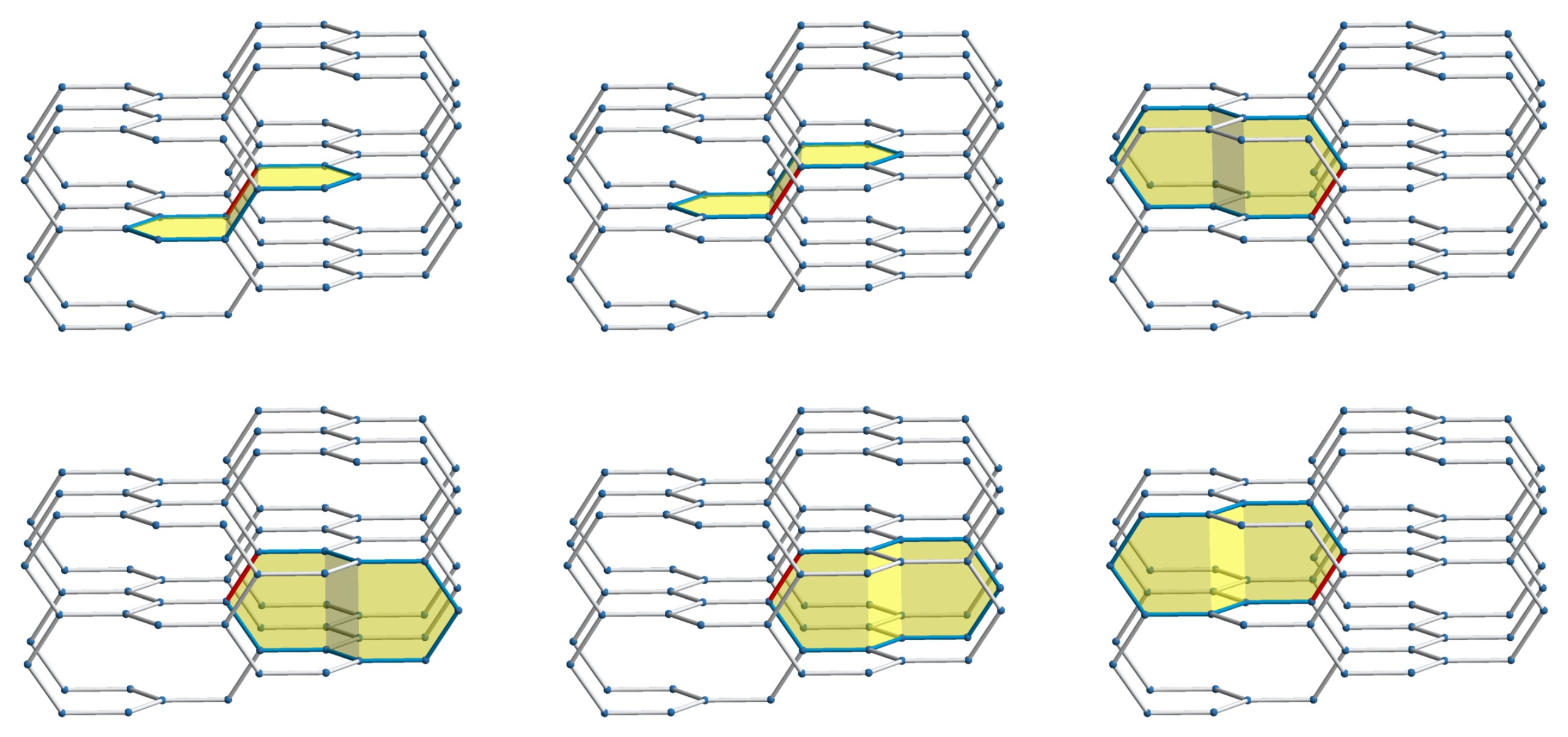}
  \caption{(Color online) Vison excitation threading six plaquettes of length 10 in the lattice (10,3)b. The flipped bond operator is pictured in red.}
  \label{fig:vison_10_3b}
\end{figure}

The lattice (10,3)b possesses four loop operators of length 10 per unit cell.
These four loop operators can be combined to form two closed volumes, each of which must have vanishing total flux, resulting in only two independent loop operators per unit cell.
One of these closed volumes is illustrated in Fig.~\ref{fig:volumes_10_3b}.
The remaining volume is related by a two-fold screw rotation.
The smallest vison loop in this lattice threads six such plaquettes and is visualized in Fig.~\ref{fig:vison_10_3b}.

The calculations for lattice (10,3)b were performed using the following reference gauge
\begin{align}
  u^x_{23}  &= +1  & u^y_{14}  &= +1  & u^z_{13}  &= +1  \nonumber\\
  u^x_{14}  &= +1  & u^y_{23}  &= +1  & u^z_{24}  &= +1.
\end{align}
In this gauge, the momentum space Hamiltonian reads
\begin{equation}
  \begin{pmatrix}
    0               & 0            & i J_z & i A_{13} \\
    0               & 0            & i A_2 & i J_z    \\
    -i J_z          & -i A^\star_2 & 0     & 0        \\
    -i A^\star_{13} & -i J_z       & 0     & 0        \\
  \end{pmatrix},
\end{equation}
where
\begin{align}
  A_{13} &= e^{-2 i k_3 \pi } \left(J_x+e^{2 i k_1 \pi } J_y\right) ,\nonumber\\
  A_2    &= J_x+e^{2 i k_2 \pi } J_y.
\end{align}
The gauge-fixed matrix representations of the symmetry operators relevant to our classification scheme are
\begin{equation}
  U_{\mbox{\tiny SLS}} = U_{\mbox{\tiny T}} =
  \begin{pmatrix}
    \id_{2\times 2} & 0                \\
    0               & -\id_{2\times 2}
  \end{pmatrix}
\end{equation}
and
\begin{equation}
  U{\mbox{\tiny I}} =
  \begin{pmatrix}
    0  & 0  & 0 & 1 \\
    0  & 0  & 1 & 0 \\
    0  & -1 & 0 & 0 \\
    -1 & 0  & 0 & 0
  \end{pmatrix}.
\end{equation}
%

%
\subsection{(10,3)c}
\label{ssec:supplemental_10_c}
\begin{figure}[h]
  \includegraphics[width=\linewidth]{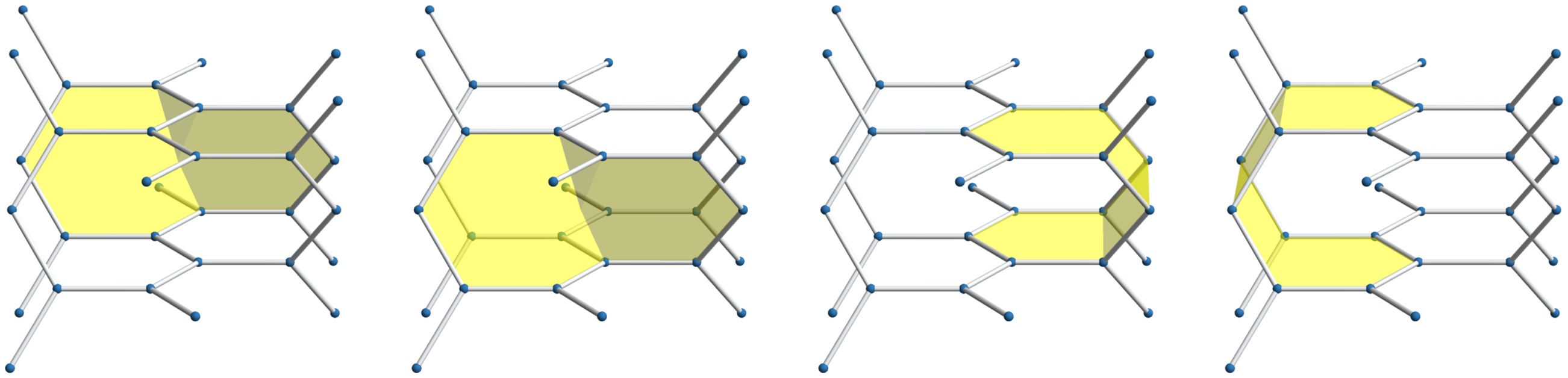}
  \caption{(Color online) Loop operators of the lattice (10,3)c forming a volume constraint.}
  \label{fig:volumes_10_3c}
\end{figure}

\begin{figure}[h]
  \includegraphics[width=\linewidth]{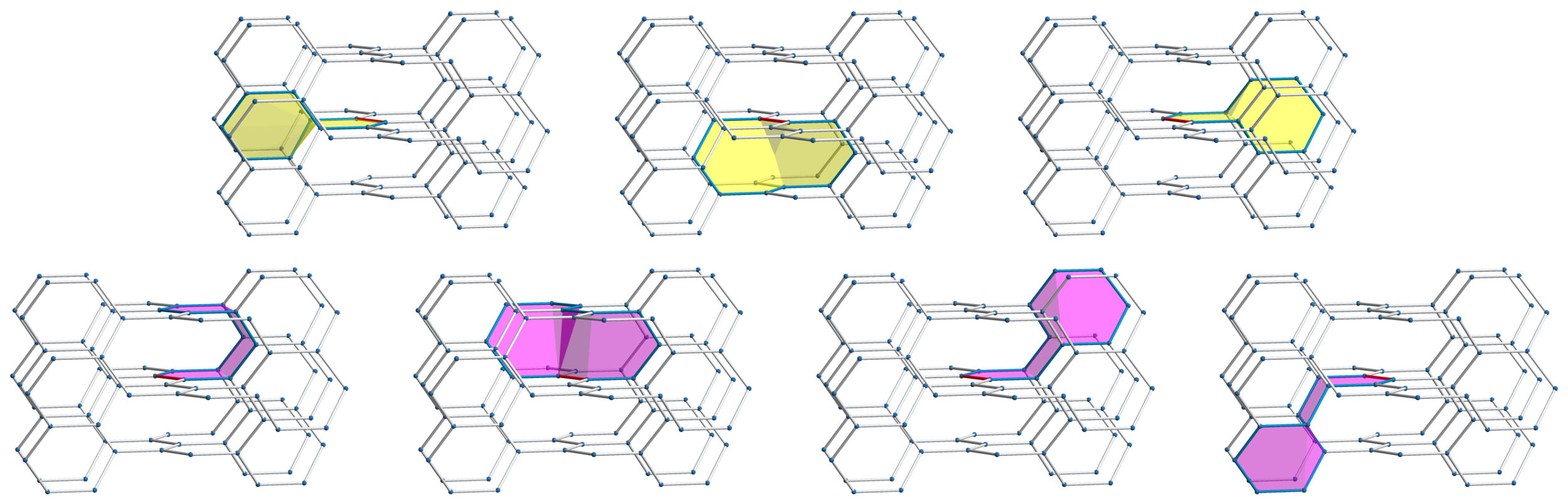}
  \caption{(Color online) Vison excitation threading three plaquettes of length 10 (yellow) in the lattice (10,3)c. In addition, 11 plaquettes of length 12 are excited; four examples of such plaquettes (shaded in magenta) are shown in the second row.  The flipped bond operator is pictured in red.}
  \label{fig:vison_10_3c}
\end{figure}

The lattice (10,3)c possesses three loop operators of length 10 and three of length 12 per elementary 6-site unit cell.
These six loop operators can be combined to form three closed volumes, each of which must have vanishing total flux, resulting in only three linearly independent loop operators per unit cell.
One of these closed volumes is illustrated in Fig.~\ref{fig:volumes_10_3c}: note that this particular visualization obscures the fact that the loop operators of length 12 are symmetry-related.
The remaining two volumes are related by a six-fold screw rotation.
The smallest vison loop in this lattice threads three plaquettes of length 10 and 11 of length 12 and is visualized in Fig.~\ref{fig:vison_10_3c}.
\begin{figure}[h]
  \includegraphics[width=.75\linewidth]{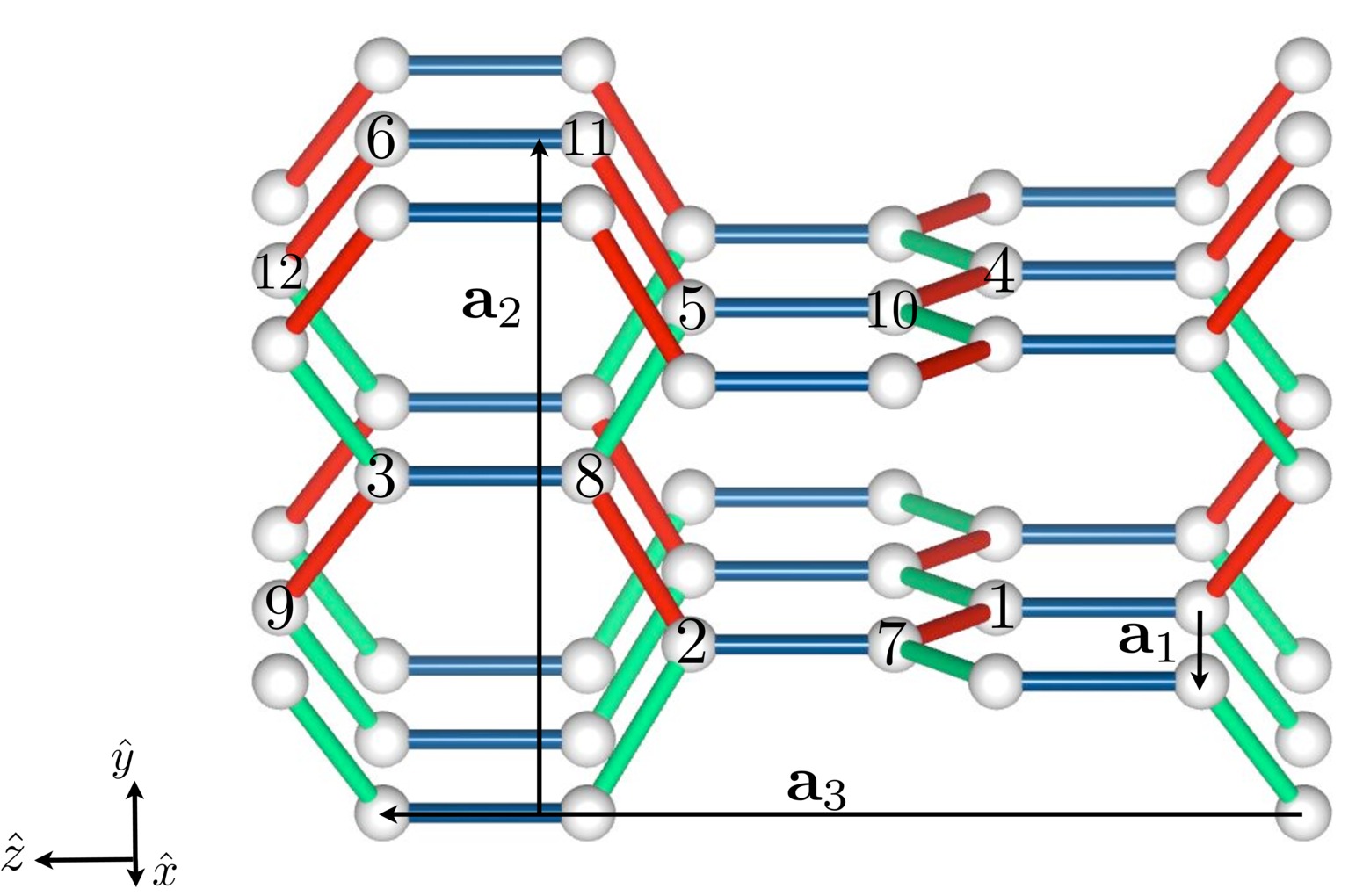}
  \caption{(Color online) Visualization of the Kitaev couplings, the unit cell and the translation vectors for the lattice (10,3)c with enlarged unit cell.}
  \label{fig:10c_enlarged}
\end{figure}

As mentioned in Sec.~\ref{ssec:10c}, in order to accommodate the ground-state flux configuration, the unit cell must be enlarged in the 010-direction to a 12 site unit cell.
The sites have been relabeled for the enlarged unit cell, which is depicted in Fig.~\ref{fig:10c_enlarged}.
The calculations for lattice (10,3)c were performed using the following reference gauge:
\begin{align}
  u^x_{1,7}  &= +1,  & u^y_{1,7}  &= +1,  & u^z_{7,2}  = -1,  \nonumber\\
  u^x_{2,8}  &= -1,  & u^y_{8,5}  &= -1,  & u^z_{8,3}  = -1,  \nonumber\\
  u^x_{3,9}  &= -1,  & u^y_{3,12} &= +1,  & u^z_{9,1}  = +1,  \nonumber\\
  u^x_{4,10} &= -1,  & u^y_{4,10} &= +1,  & u^z_{10,5} = -1,  \nonumber\\
  u^x_{5,11} &= +1,  & u^y_{11,2} &= -1,  & u^z_{11,6} = -1,  \nonumber\\
  u^x_{6,12} &= -1,  & u^y_{6,9}  &= +1,  & u^z_{12,4} = +1.
\end{align}
In this gauge, the momentum space Hamiltonian reads as
\begin{equation}
  H(\bk) =
  \begin{pmatrix}
    0              & A(\bk) \\
    A^\dagger(\bk) & 0      \\
  \end{pmatrix},
\end{equation}
where
\begin{equation}
  A(\bk) =
  \begin{pmatrix}
    i A_1 & 0      & -i A_2 & 0     & 0     & 0      \\
    i J_z & -i J_x & 0      & 0     & i A_3 & 0      \\
    0     & i J_z  & -i J_x & 0     & 0     & i A_4  \\
    0     & 0      & 0      & i A_5 & 0     & -i A_2 \\
    0     & i J_y  & 0      & i J_z & i J_x & 0      \\
    0     & 0      & i A_6  & 0     & i J_z & -i J_x \\
  \end{pmatrix},
\end{equation}
with
\begin{align}
  A_1    &= J_x+e^{-2 i k_1 \pi } J_y ,\nonumber\\
  A_2    &= e^{-2 i k_3 \pi } J_z ,\nonumber\\
  A_3    &= e^{-2 i k_2 \pi } J_y ,\nonumber\\
  A_4    &= e^{2 i k_1 \pi } J_y ,\nonumber\\
  A_5    &= -J_x+e^{-2 i k_1 \pi } J_y ,\nonumber\\
  A_6    &= e^{2 i (k_1+k_2) \pi } J_y.
\end{align}
The gauge-fixed matrix representations of the symmetry operators relevant to our classification scheme are
\begin{equation}
  U_{\mbox{\tiny SLS}} = U_{\mbox{\tiny T}} =
  \begin{pmatrix}
    \id_{6\times 6} & 0                \\
    0               & -\id_{6\times 6}
  \end{pmatrix}.
\end{equation}


\begin{thebibliography}{00}
\bibitem{equidistantbond}
None of these criteria are essential for the physical properties, which are solely determined by the topology of the lattice. However, these criteria select a subset of lattices that exhibit particularly simple, regular structures and that already showcase all the different spin liquid phases that are allowed by our symmetry analysis.
\bibitem{FootnoteGappedVisons}
For the lattices (9,3)a and (8,3)c, the flux sector we consider is not the ground state sector for the pure Kitaev model. However, one can always stabilize it as the ground state sector by adding additional terms to the Kitaev Hamiltonian that favor/penalize flux through the plaquettes. For such an augmented Hamiltonian, also the lattices (9,3)a and (8,3)c have fully gapped vison excitations throughout the phase diagram.
\bibitem{FootnoteChernNumberArgument}
As the nodal structure of the (8,3)c lattice is rather non-generic at the isotropic point, one should consider anisotropic couplings --- where the lines form contractable loops --- for this argument.
\bibitem{Footnote:Scaling}
As the unit cell for this flux configuration contains already 96 sites, a thorough finite-size scaling analysis is extremely hard.
\bibitem{FootNoteSym10b}
While the most symmetric incarnation of the (10,3)b lattice exhibits 90 degree angles between the zig-zag chains (see Fig.~\ref{fig:comparison_10_3_b_c}), we use a smooth deformation of the lattice in our calculations which corresponds to the hyperhoneycomb lattice structure studied in the context of $\beta$-Li$_2$IrO$_3$ \cite{beta}.
\end{thebibliography}
\end{document}